\documentclass[%
 reprint,
nofootinbib,
 amsmath,amssymb,
 aps,
prd,
]{revtex4-2}

\usepackage{booktabs}
\usepackage{tabularx}
\usepackage[hidelinks,colorlinks=true,linkcolor=blue,citecolor=blue]{hyperref}
\newcommand\aap{A\&A}                
\newcommand\aj{AJ}                   

\newcommand\apjl{ApJ}                

\newcommand\mnras{MNRAS}             
\newcommand\na{New~Astron.}          
\newcommand\pasa{Publ. Astron. Soc. Australia}  
\newcommand\planss{Planet. Space~Sci.} 

\usepackage{tabu}
\usepackage{array}
\usepackage{multirow}
\usepackage{CJK}
\usepackage{graphicx}
\usepackage{dcolumn}
\usepackage{bm}
\usepackage[normalem]{ulem}

\begin{document}

\title{The construction and use of dephasing prescriptions for environmental effects in gravitational wave astronomy}

\author{János Takátsy$^{1,2}$}
\email{janos.takatsy@nbi.ku.dk}
\author{Lorenz Zwick$^{1,2}$}
\author{Kai Hendriks$^{1,2}$}
\author{Pankaj Saini$^{1,2}$}
\author{Gaia Fabj$^{1,2}$}
\author{Johan Samsing$^{1,2}$}
\affiliation{%
$^{1}$Niels Bohr International Academy, The Niels Bohr Institute, Blegdamsvej 17, DK-2100, Copenhagen, Denmark\\
$^{2}$Center of Gravity, Niels Bohr Institute, Blegdamsvej 17, 2100 Copenhagen, Denmark.}%


%


\date{\today}

\newcommand{\cyan}{\color{cyan}}
\newcommand{\PS}[1]{{\cyan [PS: #1]}}             

\begin{abstract}
In the first part of this work, we provide a curated overview of the theoretical framework necessary for incorporating dephasing due to environmental effects (EE) in gravitational wave (GW) templates. We focus in particular on the relationship between orbital perturbations in the time-domain and the resulting dephasing in both time and frequency domain, elucidating and resolving some inconsistencies present in the literature. We discuss how commonly studied binary environments often result in several sources of dephasing that affect the GW signal at the same time. This work synthesizes insights from two decades of literature, offering a unified conceptual narrative alongside a curated reference of key formulas, illustrative examples and methodological prescriptions. It can serve both as a reference for researchers in the field as well as a modern introduction for those who wish to enter it. In the second part, we derive novel aspects of dephasing for eccentric GW sources and lay the foundations for consistently treating the full problem. Importantly, we demonstrate that the detectability of EEs can be significantly enhanced in the presence of eccentricity, even for $e_\mathrm{10Hz}\lesssim0.2$, substantially increasing the prospects for detection in ground based detectors. Our results highlight the unique potential of modeling and searching for EE in eccentric binary sources of GWs.
\end{abstract}

\maketitle

\section{Introduction}
\label{sec:introduction}
One of the most intriguing open problems in the field of gravitational wave (GW) astronomy is the impact of so called ``environmental effects” (EE) on GW signals. This term refers to the myriad of astrophysical and fundamental processes that can influence a GW source, alter its evolution and perturb its GW emission \citep{1993chakrabarti,1995ryan,2008barausse,2011PhRvD..83d4030Y,2007levin,kocsis,2014barausse,inayoshi2017,2017meiron,2017Bonetti,2019alejandro,2019randall,2020cardoso,DOrazioGWLens:2020,2022liu,2022xuan,garg2022,2022cole,2022chandramouli,2022sberna,2023zwick,2023Tiede,2024dyson,2022destounis,2022cardoso,2020caputo,2024zwicknovel,Derdzinksi:2021}. The study of EE is already considered  a priority in the mission statement of the recently adopted Laser Interferometer Space Antenna (LISA) and the complementary proposed TianQin, since the majority of targeted GW sources, such as massive black hole (BH) binaries and extreme mass ratio in-spirals (EMRIs), are likely to merge deep within the accretion discs or nuclear clusters hosted in the centre of galaxies \citep{2017lisa,2019lisa,2022lisaastro,2024lisa,TianQin,2021tian,tianqin2024}. Additionally, recent work is strongly indicating that EE will leave detectable imprints in a significant fraction of stellar mass binary sources of GW for next generation ground based detectors \citep{2025zwick}.

The potential presence of EE in GW signals has profound implications on the science objectives of upcoming detectors, both in terms of fundamental physics and astrophysics. On the one hand, current waveform generation models are exclusively based on \textit{vacuum solutions} of general relativity. If not accounted for, environmental influences can potentially confound results based on vacuum expectations, limiting the effectiveness of parameter estimation and introducing spurious biases, severely hinder the purported scientific goals of placing extremely stringent constraints on the strong field regime of gravity \citep{2014barausse,2020cardoso,2020chen,2025delgrosso}. On the other hand, environmental signatures in GWs can be used to probe the astrophysical surroundings of the sources they originate from. Prospects range from measuring properties of accretion discs \citep{2007levin,Derdzinksi:2021,garg2022,2022speri,2024duque,DuttaRoy:2025gnu} or dark matter profiles \citep{2013kazunari,2015kazunari,2022coogan,2023figureido,2022cole}, to centre of mass accelerations \citep{2017meiron,Tiwari:2023cpa,Vijaykumar:2023tjg,Tiwari:2024pvb,2024samsing,kai2024,kai22024,2018PhRvD..98f4012R} and even tidal effects related to neutron star interiors \citep{LIGOScientific:2018hze,LIGOScientific:2018cki,De:2018uhw,Annala:2021gom,Takatsy:2023xzf,Takatsy:2024sin}. Many of these measurements are beyond what is possible with other messengers, such as light or neutrinos, even in principle.

Most models of EE are based on the concept of dephasing, i.e. a phase shift between the vacuum signal and the signal including environmental perturbations. The standard procedure is to relate an environmental perturbation to a modification of the binary's evolution in frequency, which in turn implies a change to the total accumulated phase of the GW signal. If the total dephasing is large enough, typically $\sim \pi$ or $\sim \pi/$SNR \citep[without accounting for degeneracies, see e.g.][]{kocsis,2023zwick,2024zwick}, where SNR is the total signal-to-noise ratio of the signal, then the EE is said to be potentially detectable. Note that the magnitude of a detectable phase shift is also detector dependent. Most commonly, dephasing prescriptions reduce to power laws of the GW frequency $f$, in a manner analogous to the various post-Newtonian (PN) corrections to the binary's GW phase \citep{1994cutler,blanchet2014,2018maggiore}. However, EE most typically induce dephasing with a much steeper frequency scaling, corresponding to the fact that they are more relevant for wider binary separations, where the intrinsic gravitational force of the binary is weaker. For this reason EE are said to appear at ``negative PN orders", i.e. they accumulate with a steeper frequency scaling than even the leading order GW phase $\propto f^{-5/3}$.

\begin{figure}[!t]
    \centering
    \includegraphics[width=0.35\textwidth]{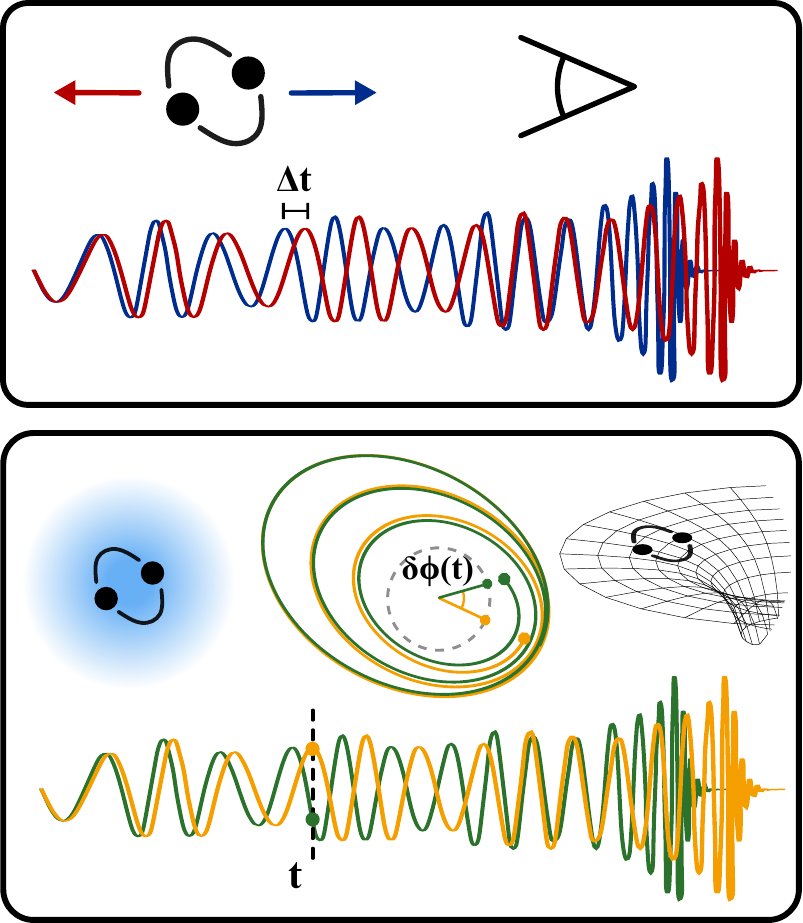}
    \caption{Depiction of the various aspects analysed in this work. Dephasing in GWs can be produced by both observer dependent redshift-like effects (top-panel) as well as intrinsic perturbations to the binary source secular evolution (bottom panel). These manifest as observables in Fourier domain waveforms in two different ways. They are best thought of as a time delay $\Delta t$ in the former case, and as a binary phase shift $\delta \phi$ at a given time $t$ in the latter case. Dephasing in eccentric binaries is mostly unexplored, and is a significantly richer problem than the corresponding circular case.}
    \label{fig:Illustration}
\end{figure}

The detailed theory of building up and including PN corrections in GW templates -both in the time and the frequency domain- has been worked out in many seminal works in the 90s, for circular binaries \citep[see, e.g.][]{1985damour,1991damour,1998jaranowski,1999jaranowski}. Based on these works, the first investigations of EE have followed over the following decades, focusing in particular on neutron star (NS) tidal deformations \citep{Flanagan:2007ix,Hinderer:2016eia,Schmidt:2019wrl,Steinhoff:2021dsn} and the effect of gas drag on extreme mass ratio inspirals (EMRIs) \citep{1993chakrabarti,2007levin,kocsis}. Now, interest in the topic is rising due to the adoption of LISA and the advent of 3rd generation ground based detectors. More and more research on the astrophysical modelling side is demonstrating the relevance of EE in a plethora of GW sources, and more and more works are suggesting dephasing and other EE as an important observable to distinguish between binary formation channels and as a novel way to probe matter distributions in highly curved backgrounds \citep{2024basu,2024santoro,2024samsing,kai22024,2025delgrosso,2024soumen,2024dyson,2025dyson,2025zwick}. Furthermore, this trend is in concomitance with the increasing relevance of residual eccentricity in gravitational wave sources. Weak evidence for eccentricity is already present in a handful of LIGO-Virgo-KAGRA (LVK) signals \citep{2021ApJ...921L..31R,2024gupte}, and is considered a smoking gun for the dynamical and AGN formation channels for stellar mass binaries \citep[e.g.][]{2006ApJ...640..156G, 2014ApJ...784...71S, 2017ApJ...840L..14S, Samsing18a, Samsing2018, Samsing18, 2018ApJ...855..124S,
2018MNRAS.tmp.2223S, 2018PhRvD..98l3005R, 2019ApJ...881...41L,2019ApJ...871...91Z, 2019PhRvD.100d3010S, 2019arXiv190711231S}. Similarly, EMRIs are most likely formed with high eccentricities \citep{amaro-seoane2007}, and recent research is showing evidence that even massive BH binaries may retain some eccentricity as they enter the mHz band \citep[see e.g.][]{2024garg}. All in all, the emerging picture is that a large quantity of signals in upcoming GW detectors will showcase complex features such as eccentricity, EE or most likely both. If the goal is to maximally benefit from such observations, it will be necessary to develop waveform templates capable of modelling these aspects to sufficient precision.

With this goal in mind, we present an overview of the fundamental concepts of how to include dephasing in GW waveforms, going from a description of EE in the time domain to a description of dephasing in the frequency domain, useful for GW data analysis. Our aim is to provide a comprehensive conceptual overview on the topic, as well as provide a useful reference for formulas and other prescriptions. We achieve this by revisiting and collecting the insight gained in many publications over the last two decades, as well as clear up some conceptual misunderstandings that are present in several more recent publications.  Additionally, we lay some initial foundations for the analysis of dephasing due to EE in eccentric GW sources, which is a much richer problem with many open avenues of investigation. A depiction of the various aspects touched on by our work is presented in Fig. \ref{fig:Illustration}.

The paper is structured as follows. In section~\ref{sec:SPA_rev} we review how the Fourier domain waveform is obtained through the stationary phase approximation for circular binary inspirals, and interpret the connection between the time- and Fourier domain phases. In section~\ref{sec:circ_deph} we provide the formulas and emphasize the important distinctions between different types of phase-shifts in the time- and Fourier domain. We also categorize the main types of environmental phase-shifts relevant for circular binaries, i.e. dephasing due to additional energy fluxes, radial potentials and Doppler shifts. In section~\ref{sec:circ_examp} we go through some common astrophysical examples, in which several sources of dephasing act at the same time and compare their relative importance. In section~\ref{sec:ecc_deph} we extend the discussion of dephasing due to environmental effects to eccentric binaries, introduce novel ways in which energy and angular momentum fluxes can indirectly affect the GW phase, and illustrate the importance of these indirect effects through simple examples. In section~\ref{sec:SNR} we close our overview with a brief discussion of signal-to-noise ratios related to the detectability of environmental dephasing. We finally summarize the main take-aways of our paper in section~\ref{sec:conclusion}.

\section{The stationary phase approximation - revisited}
\label{sec:SPA_rev}
The analysis of GW signals can be performed in either the time domain -where the relevant quantity is the GW strain time series $h(t)$- or the frequency domain -where the relevant quantity is the Fourier transform of the strain $\tilde{h}(f)\equiv{\rm FT}(h)$. While both representations are used in different applications, the numerical integration of GW signals is typically much more efficient in the frequency domain. In particular, the analysis of dephasing due to EE is almost exclusively performed in the frequency domain, by employing either simple estimates (see section~\ref{sec:SNR}), Fisher matrix methods \citep{1994cutler,2008PhRvD..77d2001V}, or Bayesian inference methods \citep{Veitch:2009hd, Veitch:2014wba, 2019PASA...36...10T}. The contrary is true when treating most perturbative effects on the binary inspiral. As an example, additional energy and angular momentum fluxes that modify the binary inspiral are typically given as functions of the binary parameters \textit{in time}. Similarly, any other time-varying perturbations such as centre of mass accelerations, accretion, or other perturbative forces acting on the GW source are easy to model in the time domain. For these reasons, it is essential to be able to relate GW signals and dephasing prescriptions in the time domain to a representation of the perturbed signal in frequency space. The following ingredients are required: (i) the GW signal in the time domain (ii) a strategy to perform the Fourier transform analytically (iii) the relations between time and Fourier domain dephasing and (iv) a method to relate the dephasing to different physical scenarios. Let us first consider the first two points.

The detailed analysis of observed GW signals requires complex waveform models that contain all the relevant PN effects \citep[e.g.][]{owen23} and detector characteristics \citep{2019robson}. However, in order to assess the detectability of environmental effects it is often sufficient to incorporate dephasing prescriptions in the leading-order Newtonian GW waveform, neglecting here the problem of degeneracies. This is commonly done in phenomenological works, and is in fact more instructive as it allows to retain physical intuition about the GW signal and about the environmental dephasing itself. Note that this choice does not affect dephasing formulas in subsequent sections, where the relations $t(f)$ and $\phi(t_f)$ are not explicitly written out. Therefore, in this section we start by briefly summarising the formulas and the main takeaways of the seminal analysis of Newtonian frequency-domain GW waveforms, performed by Cutler \& Flanagan \citep{1994cutler}. The conclusions drawn are also valid for the general case. We use geometrical units with $G=c=1$.

Let us also note that the dephasing prescriptions we provide here are valid when EEs introduce a small perturbation on the system, the evolution of which is driven mainly by GW radiation. It is also assumed here that the timescale of the frequency evolution, or "chirp" of the binary is short compared to the observation timescale. These conditions are not satisfied, for example, for wide binaries in triple systems, where the detectability of EEs requires a different treatment \citep[see e.g.][]{gupta2020,Deme:2020ewx,2022chandramouli}. Additionally, other effects such as turbulent drag \citep{Trani:2025edb} or abrupt resonant excitations can instead introduce stochastic evolution or discontinuities, leading in the frequency domain to phase noise, spectral amplitude changes, or other distortions beyond a simple phase shift.

\subsection{From time domain waveforms...}
In the Newtonian description, binary BHs are considered as two point particles on a Keplerian orbit, which is secularly evolving due to quadrupolar GW radiation. The instantaneous orbital frequency is then $\Omega=M^{1/2}/r^{3/2}$, where $M = m_1 + m_2$ is the total mass. Then the evolution of the leading order quadrupolar GW frequency $f\equiv f_\mathrm{GW} = \Omega/\pi$ reads:
\begin{equation}
    \frac{\mathrm{d}f}{\mathrm{d}t} = \frac{96}{5}\pi^{8/3} \mathcal{M}^{5/3}f^{11/3} \: ,
    \label{eq:dfvac}
\end{equation}
where $\mathcal{M} \equiv \mu^{3/5} M^{2/5}$ is the chirp mass and $\mu = m_1 m_2/M$ is the reduced mass. Integrating this equation, we can obtain a closed solution for $f(t)$. Then the observed GW strain becomes
\begin{equation}
    h(t) = \mathcal{A}(t)\cos\left(\int^t 2\pi f(t')\, \mathrm{d}t'\right) \: ,
    \label{eq:ht}
\end{equation}
where
\begin{equation}
    \mathcal{A}(t)= \frac{(384/5)^{1/2}\pi^{2/3} Q\mu M}{D_Lr(t)}
\end{equation}
is the GW strain amplitude, which is a function of the component masses, the separation of the binary $r(t)$, the luminosity distance $D_L$ and the relative orientation of the binary and the detector, characterized by the scalar factor $Q$ of $\mathcal{O}(1)$. The prefactor in this equation is defined such that it is cancelled after a Fourier transformation.

Inverting $f(t)$ we acquire the relation $t(f)$ describing the time at which the GW signal reaches the frequency $f$. This reads:
\begin{equation}
    t(f) = -5(8\pi f)^{-8/3} \mathcal{M}^{-5/3} + t_c \: ,
    \label{eq:tf}
\end{equation}
where $t_c$ is an integration constant having the meaning of a fictitious collision time, i.e. $r(t_c) = 0$. In reality, the inspiral turns into a merger before the separation $r$ can become formaly zero, producing a sharp cut-off in the GW signal. The time domain phase of the GW signal is given by the integral in the parenthesis of Eq.~\eqref{eq:ht}. Evaluating the integral yields:
\begin{equation}
    \phi(t) = -2\left( \frac{t_c-t}{5\mathcal{M}} \right)^{5/8} + \phi_c \: ,
\end{equation}
where $\phi_c$ is a second integration constant corresponding to the fictitious phase at the collision time. Expressing this as a function of the GW frequency we get yet another important relation:
\begin{equation}
\label{eq:time_phase_freq}
    \phi(t(f)) = -2(8\pi\mathcal{M} f)^{-5/3} + \phi_c \: 
\end{equation}
Eq.~\eqref{eq:time_phase_freq} denotes the time-domain phase, expressed as a function of frequency.

\subsection{...to Fourier domain waveforms}
Now that we have the time domain waveform, we can relate it to its representation in the frequency domain via a Fourier transformation:
\begin{equation}
    \tilde{h}(f) \equiv \int\limits_{-\infty}^{\infty} e^{2\pi ift} h(t) \, \mathrm{d}t \: .
    \label{eq:FT}
\end{equation}
While the latter equation is formally correct, the integral cannot be evaluated analytically, even for the simplest circular Newtonian waveform. Therefore, some approximation scheme -the stationary phase approximation- is required to simplify this integral.

\begin{figure}[!t]
    \centering
    \includegraphics[width=0.48\textwidth]{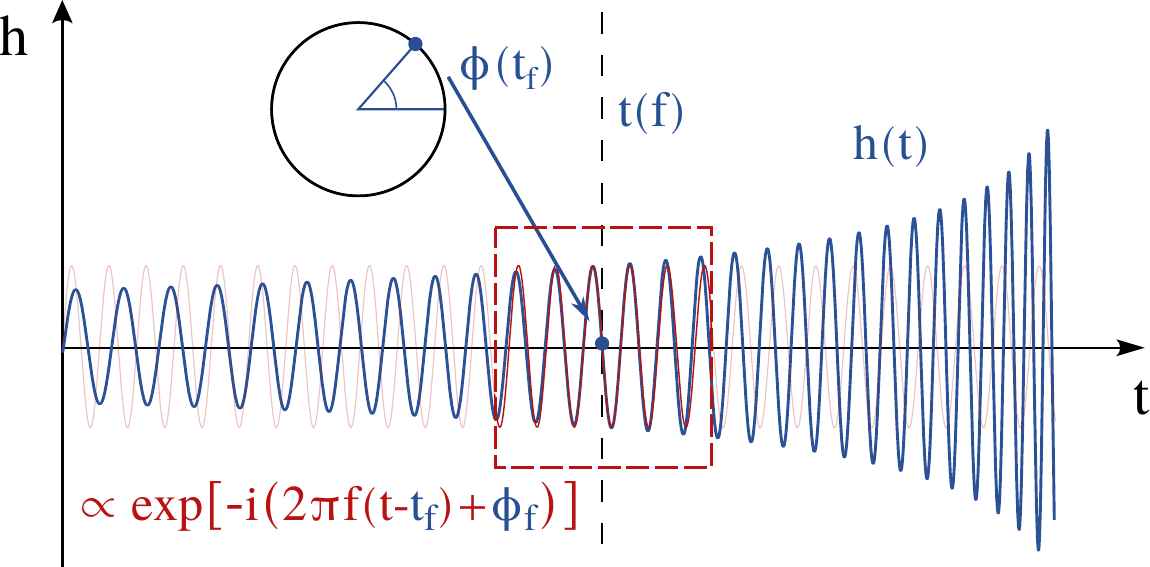}
    \caption{Illustration of the stationary phase approximation. The main contribution to the Fourier domain amplitude for a given frequency $f$ is given by the part of the signal where $\mathrm{d}\phi/\mathrm{d}t=2\pi f$. In the close vicinity of this point the signal can be approximated by a sinusoidal of that frequency (red). The sinusoidal's phase has to be matched to the GW phase $\phi(t_{f})$ at the time $t(f)$ at which the GW frequency is $f$ (blue dot). Thus, the Fourier representation contains information on both $\phi(t_{f})$ and $t(f)$.}
    \label{fig:SPA}
\end{figure}

Circular binaries evolve slowly due to GW emission, with respect to the orbital timescale. Therefore, we can use that the modulation of the amplitude is slow with respect to the accumulation of the GW phase, i.e. $\mathrm{d}(\ln \mathcal{A})/\mathrm{d}t \ll \mathrm{d}\phi/\mathrm{d}t$. Similar considerations apply to the GW frequency i.e. $ \mathrm{d}^2\phi/\mathrm{d}t^2 \ll (\mathrm{d}\phi/\mathrm{d}t)^2$. Under such conditions the integral can be evaluated using the stationary phase approximation, as follows: Given a time domain function $B(t) = A(t) \exp[-i\phi(t)]$, we use the fact that the Fourier amplitude at a frequency $f$ is mainly accumulated when the phase of the integrand in Eq.~\eqref{eq:FT} becomes stationary, i.e. when $\mathrm{d}\phi/\mathrm{d}t = 2\pi f$. This condition assures that the exponent in Eq.~\eqref{eq:FT} is close to 1, such that the integral can accumulate rather than being dominated by the superposition of incoherent phases. Given the stationary phase condition, the phase in the integrand can be expanded as
\begin{equation}
\label{eq:SPA_phase}
    2\pi f t - \phi(t) \approx 2\pi f\, t_f - \phi_f - \frac{1}{2}\ddot{\phi}(t_f)(t-t_f)^2 \: ,
\end{equation}
where $t_f \equiv t(f)$ and $\phi_f \equiv \phi(t_f)$, and the linear term vanishes by construction. Then the integral can be evaluated analytically as a Gaussian integral, yielding:
\begin{align}
    \tilde{B}(f) &\approx A(t_f) \left[ \frac{-i}{\dot{f}(t_f)} \right]^{1/2} \exp\left[ i(2\pi f t_f - \phi_f) \right] \nonumber \\
    &= \frac{A(t_f)}{{\dot{f}(t_f)}^{1/2}} \exp\left[ i\left(2\pi f t_f - \phi_f - \frac{\pi}{4}\right) \right] \: .
\end{align}
The factor in front of the exponential can be understood intuitively. The Fourier domain amplitude of the waveform is proportional to the time domain amplitude, multiplied by the time the GW spends in that given frequency bin, which is proportional to $\dot{f}(t_f)^{1/2}$. Thus the Fourier domain waveform in the Newtonian approximation becomes
\begin{equation}
    \tilde{h}(f) \approx \frac{Q}{D_L} \mathcal{M}^{5/6} f^{-7/6} \exp[i\psi(f)] \: ,
\end{equation}
where $f \geq 0$ and
\begin{equation}
    \psi(f) = 2\pi f t(f) - \phi(t_f) -\frac{\pi}{4} \: .
    \label{eq:psif}
\end{equation}
We see that the Fourier phase $\psi(f)$ is composed of two separate components. It contains information about the corresponding time domain GW phase $\phi(t_f)$, but also about the time $t(f)$ at which the binary reaches a given GW frequency. This will be a crucial insight when attempting to add perturbations of different types to the GW signal. Fig.~\ref{fig:SPA} illustrates how we obtain the Fourier transform of the waveform in the stationary phase approximation. The waveform at a given time instance is approximated by a sinusoidal with the respective frequency, which has information about both $t(f)$ and $\phi(t_f)$.

\subsection{General considerations}
We now conclude our discussion of time and frequency domain waveforms with some additional important insights. While in the Newtonian example above we have given the exact functional forms of $t(f)$ and $\phi(t_f)$, Eq.~\eqref{eq:psif} is true in general in the stationary phase approximation. Given a general prescription for $\dot{f}$ we have
\begin{align}
    \label{eq:tf}
    t(f) &= \int^f \dot{f}^{-1} \, \mathrm{d}f' \: , \\
    \label{eq:phif}
    \phi(t_f) &= 2\pi\int^f f'\,\dot{f}^{-1} \, \mathrm{d}f' \: .
\end{align}
As $\psi(f)$ contains two integration constants, let us take its derivative. Due to cancellations we have:
\begin{equation}
    \frac{\mathrm{d}\psi}{\mathrm{d}f} = 2\pi t(f) \: .
\end{equation}
Note that as $\mathrm{d}\phi/\mathrm{d}t = 2\pi f(t)$, this makes $t$ and $f$ (or more precisely $\omega=f/2\pi$) conjugate variables and $\psi(f)$ and $\phi(t)$ related by Legendre transformations:
\begin{align}
    \psi(f) &= \frac{\mathrm{d\phi}}{\mathrm{d}t}\bigg|_{t(f)} t(f) - \phi(t(f)) \: , \\
    \phi(t) &= \frac{\mathrm{d\psi}}{\mathrm{d}f}\bigg|_{f(t)} f(t) - \psi(f(t)) \: . 
\end{align}
This insight will be useful when interpreting the connection between dephasings in the time and Fourier domain.

Differentiating $\psi(f)$ once again we obtain:
\begin{equation}
    \frac{\mathrm{d}^2\psi}{\mathrm{d}f^2} = \frac{2\pi}{\dot{f}} = \frac{8\pi^2}{\dot{E}_\mathrm{orb}} \frac{\mathrm{d}E_\mathrm{orb}}{\mathrm{d}\Omega}\: ,
    \label{eq:dpsidf2}
\end{equation}
where in the second step we transformed the relation into a form used by multiple other studies \citep[e.g.]{Tichy:1999pv,Flanagan:2007ix}. Crucially, Eq.~\eqref{eq:dpsidf2} contains the full physical information about the Fourier phase evolution of the binary, while $\psi(f)$ contains two additional degrees of freedom from the two integration constants.

Finally, we briefly note that the validity of the stationary phase approximation has been thoroughly studied in \cite{1999PhRvD..59l4016D}. Corrections to the waveform appear in two forms: Firstly, the cubic term in the expansion used in Eq.~\eqref{eq:SPA_phase} produces an additional phase term $\delta\psi_{\rm SPA}^{\rm error}$ which is suppressed by 5PN orders:
\begin{align}
   \delta\psi_{\rm SPA}^{\rm error} \propto \psi\times\left(\frac{v}{c} \right)^{10}.
\end{align}
Secondly, the finite observation time of real signals induces windowing effects that modify both the phase and the amplitude of the signal.
\section{Dephasing in circular waveforms}
\label{sec:circ_deph}
We can now approach the topic of dephasing, i.e. of modelling perturbations on the GW signal related to some external influence acting on the binary inspiral. Firstly, note that deviations in the amplitude of the GW signal are extremely hard to produce, as they require a substantial instantaneous variation in the binary separation. On the contrary, even small instantaneous changes in the GW phase can accumulate over the entire inspiral and cause a significant difference in the overall waveform. There are many physical mechanisms that can induce phase shifts in the observed GW signals. In particular, perturbations to the orbital evolution induce a change in the binary's chirp rate $\dot{f}(f)$, which ultimately determines $\psi(f)$. This can be realised as a result of additional energy fluxes that remove or inject energy from the orbit, or alternatively from the presence of external radial potentials, which change the binding energy of the orbit and the GW radiation at a given GW frequency. These two effects are usually called "dissipative" and "conservative" effects in GW modeling literature. Finally, observational effects, in particular Doppler shifts, can induce an apparent dephasing in the observed signal. In this section we discuss these different types of perturbations for circular inspirals, while also emphasizing and explaining the difference between time- and frequency-domain phase shifts.

The results of this section are summarised in Table \ref{tab:summary}, which lists the various relations between frequency chirp, time domain and Fourier domain dephasing for circular binary sources of GW.

\subsection{Time-domain and Fourier domain dephasing}
\label{ssec:deph_circ_tF}

We start our discussion of dephasing with some general considerations that are independent of the cause of dephasing and the exact functional form of the perturbation on the binary. The only required simplification is for any perturbation to the binary inspiral to be small. Then, regardless of the specifics, we can express the binary's total chirp rate as a sum of the vacuum chirp and a small perturbation $\delta\dot{f}$:
\begin{equation}
    \dot{f}_\mathrm{tot}(f) = \dot{f}_\mathrm{vac}(f) + \delta \dot{f}(f) \: .
\end{equation}

To obtain the dephasing in Fourier space we follow the same steps as in the vacuum case, but replace $\dot{f}_\mathrm{vac}$ with $\dot{f}_\mathrm{tot}$. Then $t(f)$ and $\phi(t_f)$ becomes:
\begin{align}
    t(f) &= \int^f {\dot{f}_\mathrm{tot}}^{-1} \, \mathrm{d}f' = t_\mathrm{vac}(f) + \delta t(f) \: , \\
    \phi(t_f) &= 2\pi \int^f \frac{f'}{\dot{f}_\mathrm{tot}} \, \mathrm{d}f' =\phi_\mathrm{vac}(t_f) + \delta\varphi(f) \: ,
\end{align}
where
\begin{align}
    \delta t(f) &\approx- \int^f\frac{\delta\dot{f}}{{\dot{f}_\mathrm{vac}}^2} \, \mathrm{d}f' \: , \\
    \delta\varphi(f) &= \phi(t_f) - \phi_\mathrm{vac}(t_{f,\mathrm{vac}}) \approx - 2\pi \int^f\frac{f'\delta\dot{f}}{{\dot{f}_\mathrm{vac}}^2} \, \mathrm{d}f' \: .
\end{align}
Note that we have introduced the notation $\delta \varphi(f)$ for the difference in the time-domain phases of the perturbed and vacuum signals at a given frequency to emphasize its distinction from the difference in the time-domain phases at a given time, $\delta \phi(t)$. This makes the perturbation of the Fourier phase:
\begin{equation}
\label{eq:dphasing_fourier}
    \delta\psi(f) = 2\pi f\delta t(f) - \delta\varphi(f) \: .
\end{equation}
Thus, the Fourier domain dephasing also contains two contributions, one from the binary dephasing at a fixed frequency, and one from the difference in time at which the binary reaches a given frequency. 
We note here that the additional contribution from the time shift $\delta t(f)$ is not accounted for in multiple studies regarding dephasing in astrophysical scenarios \citep[see e.g.][]{Derdzinksi:2021,garg2022,2023zwick,2023chatterjee,2023cole}\footnote{This list includes work by the authors.}, or at least it is unclear methodologically whether it is \citep[e.g.][]{kavanagh2020,2022coogan,2024santoro,2024rivera}. Nevertheless, the qualitative results of these works are most likely not affected, as the differences for power law dephasing prescriptions only account to pre-factors of order unity. We note that even though the time-domain phase shift contains all the necessary information about the EE, using the frequency-domain waveform usually simplifies the analysis, is more natural from a data-analysis perspective, and provides a better detectability criterion \citep{Lindblom:2008cm}.

Differentiating Eq.~\eqref{eq:dphasing_fourier} twice we obtain:
\begin{equation}
    \frac{\mathrm{d}^2(\delta\psi)}{\mathrm{d}f^2} = - 2\pi \frac{\delta \dot{f}}{\dot{f}_\mathrm{vac}^2} \: ,
\end{equation}
in analogy to Eq.~\eqref{eq:dpsidf2}.

\begin{figure}[!t]
    \centering
    \includegraphics[width=0.48\textwidth]{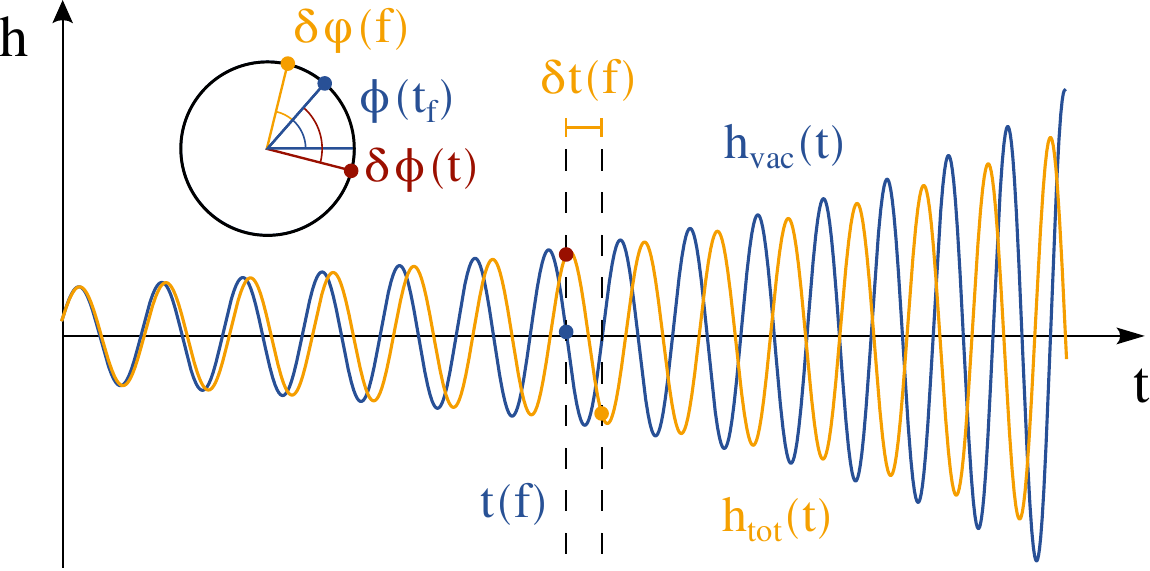}
    \caption{Illustration of the different types of dephasing addressed in section \ref{ssec:deph_circ_tF}. The coloured dots on the phase circle correspond to the same dots on the GW signals and represent the different phases. The perturbed waveform $h_{\rm tot}$ (yellow) has a different phase from the vacuum waveform $h_{\rm vac}$ at a given reference time, yielding a dephasing $\delta \phi(t)$. It may also have a different phase at a given reference frequency, yielding a dephasing $\delta \varphi (f)$. The two signals reach a given reference frequency with a time difference $\delta t(f)$. The Fourier dephasing is then related to these quantities by $\delta \psi = 2\pi f \delta t(f) - \delta\varphi(f) \approx -\delta \phi(t)$.}
    \label{fig:dphidt}
\end{figure}

Figure~\ref{fig:dphidt} illustrates the various perturbed quantities, where we again highlight the difference between $\delta\varphi(f)$ and $\delta\phi(t)$. The perturbed orbit reaches the frequency $f$ at a time $t+\delta t$, meaning that:
\begin{align}
    \delta\varphi(f) &= \phi_\mathrm{tot}(t+\delta t) - \phi_\mathrm{vac}(t) \nonumber \\
    &\approx \phi_\mathrm{tot}(t) - \phi_\mathrm{vac}(t) + 2\pi f\delta t \: ,
\end{align}
where we have assumed that $\delta t$ is small. Thus:
\begin{equation}
    \delta\phi(t_f) \approx \delta\varphi(f) - 2\pi f \delta t(f) = - \delta\psi(f) \: .
    \label{eq:dphi_dpsi}
\end{equation}
The above expression has important practical implications: It states that in order to obtain the magnitude of the Fourier domain phase shift of the GW signal it is sufficient to evaluate the time-domain phase difference at a specific time. This can expedite the evaluation of Fourier dephasing in binaries analysed in the time domain, as it allows to skip over a potentially costly Fourier transformation. In particular, such a relation may be useful in population synthesis codes, where large numbers of binaries are followed from formation to merger. Note also that since $\phi(t)$ and $\psi(f)$ are connected by a Legendre-transformation, Eq.~\eqref{eq:dphi_dpsi} is analogous to saying that a Lagrangian perturbation $\delta L$ generates a Hamiltonian perturbation $-\delta H$.

Summarizing once again the different time and phase shifts for clarity:
\begin{itemize}
    \item \emph{$\delta t (f)$:} time difference of the vacuum and perturbed signal reaching the reference frequency,
    \item \emph{$\delta \varphi (f)$:} time-domain phase difference between the vacuum and the perturbed signal at a given reference frequency $f$,
    \item \emph{$\delta \phi (t)$:} time-domain phase difference between the vacuum and the perturbed signal at a given reference time $t$,
    \item \emph{$\delta \psi (f)$:} frequency-domain phase difference between the vacuum and the perturbed signal at a given reference frequency $f$.
\end{itemize}

\subsection{Dephasing due to additional energy fluxes}
\label{ssec:deph_circ_Edot}

We now start our discussion of individual mechanisms that induce dephasing in GW signals from inspiralling binaries, starting with the category of external perturbations. The simplest case is to directly change the binary inspiral rate by extracting or adding energy to it. In order to investigate these it is worth expressing $\dot{f}$ in terms of the flux of orbital energy and the external perturbative energy flux $\delta P$:
\begin{equation}
\label{eq:enerflux}
    \dot{f}_\mathrm{tot} = \dot{E}_\mathrm{tot}\left(\frac{\mathrm{d}E_\mathrm{tot}}{\mathrm{d}f}\right)^{-1} = \dot{f}_\mathrm{vac} + \delta \dot{f}
\end{equation}
Therefore, a perturbation $\delta P$ or $\delta \dot{a}$ can be directly translated into $\delta \dot{f}$:
\begin{align}
\label{eq:enerflux2}
    \dot{f}_\mathrm{tot} &= \left(\dot{E}_\mathrm{vac}+\delta P \right)\left(\frac{\mathrm{d}E_\mathrm{vac}}{\mathrm{d}f}\right)^{-1} \nonumber \\
    &\approx\dot{f}_\mathrm{vac} \left[ 1 + \frac{\delta P}{\dot{E}_\mathrm{vac}} \right] \: ,
\end{align}
thus:
\begin{equation}
    \delta \dot{f} \approx \dot{f}_\mathrm{vac}\frac{\delta P}{\dot{E}_\mathrm{vac}} \: .
    \label{eq:dfdP}
\end{equation}
Note however, that between Eqs.~\eqref{eq:enerflux} and \eqref{eq:enerflux2}--\eqref{eq:dfdP} we have substituted the binary vacuum binding energy $E_{\rm vac}$ for the binary total binding energy. This substitution is only valid in the case the external perturbation does not also induce a change in the binding energy, which is the topic of the following section.

\subsection{Dephasing due to radial potentials}
\label{ssec:cons_pot}

Beyond fluxes, the orbit of binaries can be perturbed by the presence of an external conservative potential $\delta U$, as can be the case with three-body tidal interactions or the gravitational potential of a disc or dark matter spike. We show here that this will have multiple effects on the chirp of the signal. The clearest effect is the addition of an extra binding energy to the binary potential, which will have to be radiated away by GW emission. Additionally, the presence of an external conservative potential will modify the equilibrium point for circular binary orbits, i.e. binaries at a given radius will orbit with a different velocity in order to maintain a circular orbit. In turn, this induces a shift in the binary semi-major axis $\delta a(f)$ that corresponds to a given GW frequency $f$. The consequences are to modify the Keplerian binding energy, as well as implicitly change the GW radiation flux at a given GW frequency. In other words, one can understand this as the external perturbation causing a mismatch between the actual Keplerian binding energy or the GW radiation flux, and what an observer would expect by only considering the GW frequency $f$. Thus, the relations $E_\mathrm{vac}(a_\mathrm{vac}(f),f)$ and $\dot{E}_\mathrm{vac}(a_\mathrm{vac}(f),f)$ no longer hold as the Keplerian relation $a_\mathrm{vac}(f)$ does not hold.
Here we consider the specific case of radial potentials $\delta U(r)$. Then the total binding energy will read:
\begin{equation}
    E_\mathrm{tot}(f) \approx E_\mathrm{vac}(f) + \frac{\partial E_\mathrm{vac}}{\partial a}\Bigg|_f \delta a(f) + \delta U \: ,
\end{equation}
where $E_\mathrm{vac}(f)\equiv E_\mathrm{vac}(a_\mathrm{vac}(f),f)$. The GW radiation flux will also become:
\begin{equation}
    \dot{E}_\mathrm{tot}(f) \approx \dot{E}_\mathrm{vac}(f) + \frac{\partial \dot{E}_\mathrm{vac}}{\partial a}\Bigg|_f \delta a(f) \: ,
\end{equation}
with $\dot{E}_\mathrm{vac}(f)\equiv \dot{E}_\mathrm{vac}(a_\mathrm{vac}(f),f)$. This means that
\begin{align}
    \frac{\delta \dot{f}}{\dot{f}_\mathrm{vac}} &\approx \dot{E}^{-1}_\mathrm{vac}\frac{\partial \dot{E}_\mathrm{vac}}{\partial a}\Bigg|_f \delta a(f) - \left( \frac{\mathrm{d}E_\mathrm{vac}}{\mathrm{d}f} \right)^{-1} \nonumber \\
    &\times\left[ \frac{\mathrm{d(}\delta U)}{\mathrm{d}f} + \frac{\mathrm{d}}{\mathrm{d}f}\left(\frac{\partial E_\mathrm{vac}}{\partial a}\Bigg|_f \delta a(f)\right) \right] \: .
    \label{eq:delf_cons}
\end{align}
All of these different contributing terms must be considered when an external potential $\delta U$ is present.

Let us try to simplify these relations by utilizing the Keplerian energy and 2.5PN GW radiation formulas. Since $\dot{E}_\mathrm{vac}\propto a^4 f^6$, the first term becomes:
\begin{equation}
    \dot{E}^{-1}_0\frac{\partial \dot{E}_0}{\partial a}\Bigg|_f \delta a(f) = 4\frac{\delta a(f)}{a_0(f)} \: ,
    \label{eq:delf_cons_rad}
\end{equation}
where we introduced the lower index $0$ as shorthand for the vacuum definitions. From now on, for simplicity we also omit dependencies on $f$ and $a_0(f)$, where these are implied by the equations. Considering the force balance equations and defining $\delta F \equiv-\mathrm{d(\delta U)/\mathrm{d}r}$ we get:
\begin{equation}
    \frac{\delta a}{a_0} \approx \frac{1}{6} \frac{\delta F}{E_0}a_0 \: .
\end{equation}
Let us now consider the second line in Eq.~\eqref{eq:delf_cons}, where we have:
\begin{equation}
    \frac{\mathrm{d(}\delta U)}{\mathrm{d}f} \approx -\delta F \frac{\mathrm{d}a_0}{\mathrm{d}f}
\end{equation}
For the partial derivative we have:
\begin{equation}
    \frac{\partial E_0(a,f)}{\partial a}\Bigg|_f = -4 \frac{E_0}{a_0} \: .
\end{equation}
Note the difference between this and the total derivative:
\begin{equation}
    \frac{\mathrm{d} E_0(a,f(a))}{\mathrm{d} a} = - \frac{E_0}{a_0} \: ,
\end{equation}
where the frequency is not fixed but is also changing according to Kepler's law. Using these the second line in Eq.~\eqref{eq:delf_cons}, we continue the derivation:
\begin{align}
    &\frac{\mathrm{d(}\delta U)}{\mathrm{d}f} + \frac{\mathrm{d}}{\mathrm{d}f}\left(\frac{\partial E_0}{\partial a}\Bigg|_f \delta a(f)\right) \nonumber \\
    \approx& -\frac{5}{3} \delta F \frac{\mathrm{d}a_0}{\mathrm{d}f} - \frac{2}{3} a \frac{\mathrm{d(\delta F)}}{\mathrm{d}r} \frac{\mathrm{d} a_0}{\mathrm{d}f} \: .
\end{align}
Dividing this by $\mathrm{d}E_\mathrm{vac}/\mathrm{d}f$ and further manipulating the equations we finally get:
\begin{align}
    &- \left( \frac{\mathrm{d}E_0}{\mathrm{d}f} \right)^{-1} \left[ \frac{\mathrm{d(}\delta U)}{\mathrm{d}f} + \frac{\mathrm{d}}{\mathrm{d}f}\left(\frac{\partial E_0}{\partial a}\Bigg|_f \delta a(f)\right) \right] \nonumber \\
    &\approx - \frac{10}{3} \frac{\delta U'}{V_g'}\Bigg|_{a_0(f)} + \frac{8}{3} \frac{\delta U''}{V_g''}\Bigg|_{a_0(f)} \: ,
\end{align}
where $V_g(r)$ is the gravitational potential and the prime denotes differentiation with respect to $r$. After similar manipulations on Eq.~\eqref{eq:delf_cons_rad} we get
\begin{equation}
    \dot{E}^{-1}_0\frac{\partial \dot{E}_0}{\partial a}\Bigg|_f \delta a \approx \frac{4}{3} \frac{\delta U'}{V_g'}\Bigg|_{a_0(f)} \: .
\end{equation}
Thus the total deviation in the binary's frequency chirp becomes:
\begin{equation}
    \frac{\delta \dot{f}}{\dot{f}_\mathrm{vac}} \approx - 2 \frac{\delta U'}{V_g'}\Bigg|_{a_\mathrm{vac}(f)} + \frac{8}{3} \frac{\delta U''}{V_g''}\Bigg|_{a_\mathrm{vac}(f)} \: .
    \label{eq:Deph_U}
\end{equation}
It is worth further discussing the different terms in this equation. First of all, all the terms are proportional to only derivatives of the perturbing potential $\delta U$. This is expected, as a constant shift in the potential should have no physical implications. The first term is the same as what we would get by only considering the effect of the extra binding energy from $\delta U$. This is due to some accidental cancellations between the modification of the binding energy and the 2.5PN radiation power due to the fact that the orbit has a different semi-major axis at a given GW frequency. The second term, interestingly, is proportional to the second derivative of the perturbing potential and is due to the fact that $\delta a$, which is proportional to the first derivative of the potential, is itself also changing as the orbit shrinks.

This equation has an additional interesting implication. We can make $\delta \dot{f}/\dot{f}_\mathrm{vac}$ constant by introducing a potential with the same functional dependence on $r$ as $V_g$, i.e. $\delta U\propto r^{-1}$. This, however, would be equivalent to increasing the total mass of the binary while keeping $\mu$ constant. Note how the coefficient $2/3$ we obtain from Eq.~\eqref{eq:Deph_U} in this case matches with the power-law dependence of $\dot{f}_\mathrm{vac}$ on $M$. On the other hand, the extra chirp can also be non-trivially cancelled up to first order by introducing a perturbing potential $\delta U \propto r^{5/8}$. The dephasing due to e.g. a tidal potential linear in $r$ would result in the same dephasing as what we would naively get from only taking into account the extra binding energy $\delta U$.

\subsection{Dephasing due to Doppler shift}

The interpretation of the Fourier phase shift requires a slightly different treatment when considering a Doppler-shifted source. It is well-known that the time-domain strain, $h(t)$, is invariant under the transformation \citep{2018maggiore}:
\begin{equation}
    (f,\mathcal{M},\mu,r,D_L,t) \rightarrow (f/\lambda,\mathcal{M}\lambda,\mu\lambda,r\lambda,D_L\lambda,t\lambda) \: .
\end{equation}
This means that a GW signal shifted by a constant Doppler shift $(1+z)$ would be indistinguishable from a non-Doppler-shifted signal with redshifted masses $\mathcal{M}_\mathrm{obs} = (1+z)\mathcal{M}$ and $\mu_\mathrm{obs} = (1+z)\mu$. However, time-dependent Doppler shifts (e.g. due to non-zero accelerations) cannot be modeled as a constant change in the binary parameter, but appear as frequency-dependent phase shifts, as we are about to show in this section. In order to get a more intuitive insight about this statement let us look at the Fourier phase of the observed signal of a binary source receding at a constant velocity $v$. When the intrinsic GW frequency of the binary is $f$, the observed frequency is Doppler-shifted by
\begin{equation}
    f_\mathrm{obs} = \frac{f}{1+v/c} \equiv \frac{f}{\lambda} \: .
\end{equation}
This part of the signal will also be observed at a different, Doppler-shifted time:
\begin{equation}
    t_\mathrm{obs}(f_\mathrm{obs}) = t(f) + \frac{v}{c}t(f) = \lambda t(f_\mathrm{obs}/\lambda) \: .
\end{equation}
Then, since for the time-domain phase $\phi_\mathrm{obs}(t_\mathrm{obs}(f_\mathrm{obs})) = \phi(t(f))$, we have
\begin{align}
    \psi_\mathrm{obs}(f_\mathrm{obs}) &= 2\pi f_\mathrm{obs} t_\mathrm{obs}(f_\mathrm{obs}) - \phi_\mathrm{obs}(t_\mathrm{obs}(f_\mathrm{obs})) \nonumber \\
    &= 2\pi \lambda f_\mathrm{obs} t(f_\mathrm{obs}/\lambda) - \phi(t(f_\mathrm{obs}/\lambda)) \nonumber \\
    &= \lambda^{-5/3} \psi^{\mathcal{M}}(f_\mathrm{obs}) = \psi^{\lambda\mathcal{M}}(f_\mathrm{obs})
\end{align}

Let us now relax the condition of constant velocity and investigate the phase shift for small perturbations. Then the binary signal with intrinsic frequency $f$ is observed at a time:
\begin{equation}
    t_\mathrm{obs} = t(f) + \int^{t(f)} \frac{v(t')}{c}\mathrm{d}t' = t(f) +\Delta t(f)
\end{equation}
Then using that $f=f_\mathrm{obs} + \Delta f$ we obtain the following for the Fourier phase:
\begin{align}
    \psi&_\mathrm{obs}(f_\mathrm{obs}) = 2\pi f_\mathrm{obs} \left[t(f)+\Delta t(f)\right] - \phi(t(f)) \nonumber \\
    \approx &2\pi f_\mathrm{obs}\left[ t(f_\mathrm{obs}) + \Delta f \frac{\mathrm{d}t(f)}{\mathrm{d}f}\Bigg|_{f_\mathrm{obs}} +\Delta t(f) \right] \nonumber\\
    &- \phi(t(f_\mathrm{obs})) - \Delta f \frac{\mathrm{d}\phi(t(f))}{\mathrm{d}f}\Bigg|_{f_\mathrm{obs}} \: .
\end{align}
Using the definition of $t(f)$ and $\phi(f)$ in Eqs.~\eqref{eq:tf} and \eqref{eq:phif}, we can see that the derivative terms cancel out. Then formally replacing $f_\mathrm{obs}$ with $f$ we get:
\begin{equation}
    \psi_\mathrm{obs}(f) \approx \psi(f) + 2\pi f\Delta t(f) \: ,
\end{equation}
where we used that $\Delta t(f_\mathrm{obs}) \approx \Delta t(f)$ up to first order. Thus, we see that in contrast to our previous analysis of intrinsic dephasing, the Fourier domain dephasing of Doppler shifted GW sources is only due to the observational time difference between the vacuum and shifted binary, i.e.
\begin{equation}
    \delta \psi(f) = 2\pi f \Delta t(f) \: .
\end{equation}
This can also be shown by the equivalence relation in Eq.~\eqref{eq:dphi_dpsi}. Since the time by which the observed signal reaches the phase $\phi(f)$ is delayed by $\Delta t(f)$, the phase of the Doppler-shifted signal lags behind by:
\begin{equation}
    \delta \phi(t_f) \approx -2\pi f_\mathrm{obs} \Delta t(f) \approx -2\pi f \Delta t(f) \: ,
\end{equation}
which agrees with our previous result.

Also note that since a constant velocity simply appears as a change in the chirp mass (as shown in the beginning of the section), the actual observable dephasing would be due to a deviation from this constant velocity, i.e. an acceleration of the center of mass. Denoting our reference time for the constant recession as $t_0$, we have:
\begin{equation}
    \Delta t(f) = \int\limits_{t_0}^{t(f)} \frac{v(t')-v(t_0)}{c} \mathrm{d}t'
\end{equation}

This formula is also applicable to redshifts not originating from proper motion, i.e. gravitational or cosmological redshift. In this case the time difference translates to a difference in proper and observer time and the formula becomes:
\begin{equation}
    \Delta t(f) = \int\limits_{t_0}^{t(f)} \left[ z(t')-z(t_0) \right] \mathrm{d}t'.
\end{equation}

\renewcommand{\arraystretch}{1.5}
\begin{table*}[]
\caption{Relations between frequency chirp, time domain and Fourier domain dephasing for circular binary sources of GW.
\label{tab:summary}}
\begin{tabular}{ccccc}
\toprule
Dephasing cause & $\delta\dot{f}/\dot{f}_\mathrm{vac}$                                & $\delta \varphi(f)$                                                                              & $\delta t(f)$                                                                             & $\delta \psi(f) = -\delta \phi (t_f)$                   \\ \midrule\midrule
Flux $\delta P$     & $\delta P/ \dot{E}_\mathrm{vac}$                             & \multirow{2}{*}{$\quad- 2\pi {\displaystyle\int^f} \left(f'\delta\dot{f}/{\dot{f}_\mathrm{vac}^2}\right) \, \mathrm{d}f'\quad$} & \multirow{2}{*}{$ - {\displaystyle\int^f} \left( \delta\dot{f}/{\dot{f}_\mathrm{vac}^2} \right) \, \mathrm{d}f' $} & \multirow{4}{*}{$\quad2\pi f\delta t(f) - \delta\varphi(f)\quad$} \\ \cmidrule{1-2}

Potential $\delta U$     & $- 2 \delta U'/V_g' + \frac{8}{3} \delta U''/V_g''$ &  &   &  \\ \cmidrule{1-4}

\multirow{2}{*}{Doppler}        & \multirow{2}{*}{-}                                                                   & \multirow{2}{*}{$2\pi f\left( zf/\dot{f}_\mathrm{vac} \right) $} & \multirow{2}{*}{$z f/\dot{f}_\mathrm{vac}+{\displaystyle\int_{t_0}^{t(f)}} \left[ z(t')-z(t_0) \right] \mathrm{d}t'$} &  \\
 & & & & \\ \bottomrule
\end{tabular}
\end{table*}

\section{Illustrative Examples}
\label{sec:circ_examp}
In realistic astrophysical environments phase shifts typically originate from multiple sources. For example, perturbations from a third body will induce a center-of-mass acceleration, create an additional conservative perturbing potential, and induce secular changes in the orbital parameters. Drag forces from the surrounding gas will act as a source of extra energy flux, induce an additional conservative potential, while asymmetric forces on the two components will also accelerate the center of mass of the binary \citep{2020cardosoself}.

We now go through some simple analytic examples in astrophysics, in which several sources of dephasing act at the same time. We split the discussion into two separate astrophysical sources of GW, i.e. extreme mas ratio inspirals (EMRIs) in the first part and stellar mass binary sources in the second. For the former, we analyse the relative importance of drag vs. conservative radial potentials in typical matter backgrounds. For the latter we compare the importance of Doppler shifts due to accelerations vs. gravitational redshifts in the presence of a third massive body.

\subsection{Drag vs potentials in EMRI inspirals}
EE in EMRI sources are the subject of many studies \citep{2014barausse,2020cardoso,2020chen,2025delgrosso,2007levin,Derdzinksi:2021,garg2022,2022speri,2024duque,2013kazunari,2015kazunari,2022coogan,2023figureido,2022cole}. The two most commonly studied effects are 1) the presence of gas torques for EMRIs embedded in accretion discs, and 2) the effect of dynamical friction in dark matter spikes. Here, we compare and discuss the dephasing caused by the two aforementioned astrophysical effects with the dephasing caused by the additional potential $\delta U$ implied by the presence of a gaseous disc or a dark matter distribution (also called spike) around the central massive BH.

Accretion disc potentials and dark matter distributions around massive black holes are a field of active research. Here we employ two simplified models, that capture the basic physics at play. As a proxy for a realistic accretion disc, we use the potential and surface density profile of a rigidly rotating Mestel disk \citep{1952kuzmin,1963mestel,1987Kuzmin}. The surface density $\Sigma$ of a Mestel disk is given by the simple relation:
 \begin{align}
     \Sigma = \frac{M_{\rm d}}{2 \pi R_{\rm d}}\frac{1}{r}
 \end{align}
where $M_{\rm d}$ is the disc's total mass, $R_{\rm d}$ its truncation radius and $r$ is a radial coordinate. The potential in the midplane is given by:
\begin{align}
    \delta U_{\rm d}(R) = -\frac{GM_{\rm d}}{R_{\rm d}} \ln\left(r/R_{\rm d} \right).
\end{align}
For sufficiently large radii, the potential experienced by the EMRI is therefore well approximated by the Newtonian result:
\begin{align}
    U_{\rm tot} \approx -\frac{G M_{\bullet}}{r} -\frac{GM_{\rm d}}{R_{\rm d}} \ln\left(r/R_{\rm d} \right),
\end{align}
where $M_{\bullet}$ is the central BH mass. As shown before, the presence of this additional potential would induce a perturbation in the frequency chirp. We evaluate Eq.~\eqref{eq:Deph_U} to find:

\begin{align}
    \delta \dot{f}_{\rm pot} \approx q\frac{64 \pi^2}{5 c^5}\frac{G^3 M_{\bullet} M_{\rm d}}{R_{\rm d}}f^{3},
\end{align}
where $q <<1$ is the EMRI mass ratio. This additional chirp contribution increases rapidly as a function of frequency. This corresponds to the fact that, as the EMRI chirps faster and faster radially due to GW radiation, it is experiencing more variation in the perturbative potential as a function of time.

We now compare the dephasing due to the perturbative potential to the effect of gas torques. Small orbiters embedded in thin accretion discs excite density perturbations, which can induce a back reaction on the mass. This effect is known as migration torque, or type I torque, and is commonly invoked as an important EE. The magnitude of type I migration torques is \citep{GoldreichTremain:1980}:
\begin{align}
    \Gamma_{\rm I} = -q^2\Sigma r^4 \Omega_{\rm K}^2 \mathcal{M}_{\rm a}^2
\end{align}
where here $\Omega_{\rm K}$ is the Keplerian angular frequency and $\mathcal{M}_{\rm a}$ is the disc's Mach number. Here we assume a constant Mach number, which corresponds to isothermality for a rigidly rotating Mestel disc. Type I torques induce an additional energy flux onto a circular orbit:
\begin{align}
    \delta P_{\rm I}= \Gamma_{\rm I} \Omega_{\rm K}.
\end{align}
As shown in Eq.~\eqref{eq:dfdP}, this implies a corresponding change in the frequency chirp:
\begin{align}
    \delta \dot{f}_{\rm I}= 3q \frac{\mathcal{M}_{\rm a}^2 G^{2/3}\Sigma}{\pi ^{1/3} M_\bullet^{1/3}}   f^{2/3}. 
\end{align}
Evaluated for our simple disc model, we find:
\begin{align}
    \delta \dot{f}_{\rm I}= q \frac{3\mathcal{M}_{\rm a}^2 G^{1/3} M_{\rm d}}{2\pi ^{2/3} M_\bullet^{2/3} R_{\rm d}}   f^{4/3}. 
\end{align}

The ratio $\mathcal{R}_{\rm d}$ between the induced frequency chirps from  a Mestel disc reveals some insight:
\begin{align}
   \mathcal{R}_{\rm d}= \frac{\delta \dot{f}_{\rm pot}}{\delta \dot{f}_{\rm I}} = \frac{128\pi^{8/3}}{15 \mathcal{M}_{\rm a}^2} \left(\frac{G M_{\bullet}}{c^3} f\right)^{5/3},
\end{align}
which rescaled in terms of the ISCO frequency for a non-rotating BH is:
\begin{align}
   \mathcal{R}_{\rm d}= \frac{1}{\mathcal{M_{\rm a}}^2} \frac{16 \pi\sqrt{2}  }{135\sqrt{3}}\left(\frac{f}{f_{\rm ISCO}} \right)^{5/3}. 
\end{align}
We see that even for highly relativistic EMRIs, the dephasing caused by the disc's potential is suppressed with respect to the gas torque effect by a factor $\mathcal{M}_{\rm a}^{-2}$. Thus, for any but the hottest of discs it is entirely neglectable, and we recover the conclusions discussed in \cite{kocsis,2014barausse}.

We now turn our attention to the case of dark matter spikes. Here we model the density of the spike as a singular isothermal sphere:
\begin{align}
    \rho = \rho_0 \left(\frac{r}{r_{\rm ISCO}}\right)^{-2}
\end{align}
where $\rho_0$ is typically $10^{18}$ GeV/cm$^3$ to $10^{20}$ GeV/cm$^3$, scaled to the radius of the ISCO \citep{2022coogan}. The enclosed mass of the spike is:
\begin{align}
    M_{\rm encl} = r\times 4 \pi \rho_0 r_{\rm ISCO}^2
\end{align}
which implies a gravitational potential:
\begin{align}
    \delta U_{\rm DM} =  -4 \pi G\rho_0 r_{\rm ISCO}^2 \ln(r/r_{\rm ISCO}).
\end{align}
As we can see, this has an identical functional form as the Mestel disc we previously discussed. In dark matter spikes, a small orbiter experiences a drag force due to dynamical friction. In the supersonic limit, dynamical friction is well approximated by the following formula \citep{chandrasekhar1943}:
\begin{align}
    \lvert \mathbf{F}_{\rm dyn} \rvert = q^2\frac{4 \pi G^2M_{\bullet}^2}{v^2}\rho\ln(\Lambda),
\end{align}
where $v$ is the orbital velocity and  $\ln(\Lambda)$ is the Coulomb logarithm. This induces a modification to the EMRI chirp, in complete analogy to the disc embedded case. The energy flux is $-\lvert \mathbf{F}_{\rm dyn} \rvert  v$ and the frequency chirp is: 
\begin{align}
    \delta\dot{f}_{\rm DF}= 12\, q\,   G \rho(f) \ln(\Lambda).
\end{align}
We compare this with the potential induced chirp, which we compute by evaluating Eq.~\eqref{eq:Deph_U}:

\begin{align}
     \delta\dot{f}_{\rm DM} = - 2304\pi^3  \frac{G^2 M_\bullet\rho_0 r_{\rm ISCO}^2}{5 c^5}qf^3.
\end{align}
Once again, the ratio $\mathcal{R}_{\rm DM}$ of the frequency chirps induced by the presence of a dark matter spike is informative:
\begin{align}
   \mathcal{R}_{\rm DM} = \frac{\delta \dot{f}_{\rm DM}}{\delta \dot{f}_{\rm DF}} &= 192 \pi^{5/3}\left( \frac{G M_{\bullet}}{c^3} f\right)^{5/3} \nonumber \\
   &= \frac{8}{15}\sqrt{\frac{2}{3}} \left( \frac{f}{f_{\rm ISCO}}\right)^{5/3}.
\end{align}
where we neglect the Coulomb logarithm. We see here that for highly relativistic EMRIs, both contributions to the dephasing can be of comparable order. This implies that the two contributions to dephasing in such systems must be inferred jointly, in order to properly recover the value of the dark matter density.

Note how the conclusions for Mestel discs and isothermal sphere dark matter mostly differ by a factor of the Mach number squared, which is purely a measure of the geometrical flattening of the matter distribution. This hints at a broader connection between dynamical friction -- caused by the gravitational pull of wake-like overdensities -- and type I torques, caused by the gravitational pull of interference patterns between wake-like overdensities.

\subsection{Doppler and gravitational redshifts}
Here we compare two types of redshifts that can modify the observed frequency of GW sources. As discussed previously, dephasing due a Doppler shift of the GW signal can be observed if the inspiralling binary is accelerated along the line of sight. We consider here the acceleration caused by a third massive object with mass $m_3$ and at a distance $r$. The change in redshift caused by the line-of-sight acceleration $a_{\rm ls}$ is:
\begin{equation}
    \dot{z}_{\rm D} = \frac{a_{\rm ls}}{c}  = \frac{Gm_3 }{r^2c} \times P \: ,
\end{equation}
where $P$ is the projection factor onto the line of sight and we assume that the acceleration is constant. Thus the phase shift is:
\begin{equation}
    \delta \psi_\mathrm{D} = 2\pi f \frac{Gm_3}{r^2c} \frac{t^2}{2} P \: .
\end{equation}
Using Eq.~\eqref{eq:tf} and $c=G=1$ we get at leading order (cf. Eq.~(9) in \citep{Tiwari:2023cpa} or Eq.~(4) in \citep{Vijaykumar:2023tjg})
\begin{equation}
    \delta \psi_\mathrm{D} = \frac{25 M}{65536 \eta^2} a_\mathrm{ls} (\pi M f)^{-13/3} \: , 
\end{equation}
where $\eta=\mu/M$ is the symmetric mass ratio.

Thus, a Doppler induced dephasing arises as a consequence of the acceleration induced by the third body. However, the presence of a third body has the additional implication of changing the gravitational potential in the vicinity of the GW source, inducing a gravitational redshift \citep[e.g.][]{2017meiron,2022sberna,Kuntz:2022juv}. Taking into account the time dilation effect due to the proper motion of the center of mass as well, we get the following expression for weak gravitational fields \citep[e.g.][]{Kuntz:2022juv}:
\begin{align}
    1+z_\mathrm{G} = \left( -g_{\mu\nu} V^\mu V^\nu \right)^{-1/2} \approx 1 + \frac{Gm_3}{c^2 r} + \frac{V^2}{2c^2} \: ,
\end{align}
where $g_{\mu\nu}$ is the Schwarzschild metric \citep{1916SPAW189S}, $V^\mu = (1,\boldsymbol{V})$ with $\boldsymbol{V}$ being the center-of-mass velocity of the binary, and $r$ is the distance from the third object with mass $m_3$. Then for a binary moving radially:
\begin{align}
    \dot{z}_\mathrm{G} \approx - \frac{Gm_3}{c^2r^2} V_r + \frac{a_r V_r}{c^2} = -2\frac{Gm_3}{c^2r^2} V_r \: ,
\end{align}
where $V_r$ is the radial component of the center-of-mass velocity and we have used the fact that center of mass only experiences a radial acceleration due to the third body. Thus we get
\begin{equation}
    \delta \psi_\mathrm{G} \approx -2\frac{Gm_3}{r^2c} \frac{V_r}{c} \frac{t^2}{2} (2\pi f) \: .
\end{equation}
The ratio $\mathcal{R}_{\rm DG}$ between gravitational redshift and Doppler induced phase shifts :
\begin{align}
   \lvert\mathcal{R}_{\rm DG}\rvert  \approx 2\frac{V_{\rm r}}{c P},
\end{align}
where we neglected a factor of order unity. We see that the Doppler shift due to gravitational redshift is formally suppressed compared to the Doppler shift due to line-of-sight motion by a factor $V_{\rm r}/c$. The magnitude of the radial velocity of binary on an eccentric orbit around the third body scales approximately as:
\begin{align}
    V_{\rm r} \sim c \left(\frac{r_{\rm S}}{r_0} \right)^{1/2}e_{3},
\end{align}
where $e_3$ is the orbital eccentricity and $r_{\rm S}$ is the Schwarzschild radius of the third body. For binaries around massive black holes, such radial velocities could therefore easily reach order $0.01c$. Therefore, while we expect Doppler dephasing to dominate, dephasing due to radial motion in the third body potential may dominate for small values of the projection factor.

Before moving on to eccentric waveforms let us note that analytic formulas for the magnitude of Doppler-dephasing in eccentric binaries have already been worked out in some recent studies \citep[see][]{2024samsing,kai22024,kai2024} and further elaborated on by Ref.~\citep{2025zwick}.

\section{Dephasing in eccentric waveforms}
\label{sec:ecc_deph}
On a conceptual level, the presence of orbital eccentricity adds several new aspects to the analyisis of dephasing perfomred in previous sections. Firstly, eccentricity influences the functional dependence of the GW phase, and therefore of any dephasing, on the GW frequency. Additionally, other ways in which perturbations can modify the orbital evolution emerge, related to the extra degrees of freedom the binary orbit has. We list here a few considerations:
\begin{itemize}
    \item An external angular momentum flux can cause a secular perturbation of the binary eccentricity in addition to affecting the semi-major axis and the orbital frequency itself. This can indirectly induce a modification to the chirp of the signal.
    \item  Any precession of the argument of pericentre induced by a perturbing force directly influences the GW phase and modulates the amplitudes of the GW polarisations. Additionally, changes in the binary orientation due to precession will also in turn change the coupling between the binary and the perturbing force. In fact, 1PN forces will already necessarily make the orbit precess, which precession can then interact with perturbing forces in non-trivial ways. Furthermore, the misalignment of spin angular momentum of the binary components with the orbital angular momentum of the binary leads to the precession of the orbital plane, further modulating the GW signal and making off-plane forces interact with the orbit.
    \item The presence of external perturbative potientials breaks the Keplerian definitions of both semi-major axis and eccentricity in non trivial ways, similarly to the discussion in section \ref{ssec:cons_pot}.
\end{itemize}
Here we do not aim to cover all of these options in full. Instead, we focus on how a generic dephasing prescription can be translated between time and frequency domains, analogously to the case for cricular orbits in section \ref{ssec:deph_circ_tF}. Additionally, we analyse the behaviour of dephasing prescriptions that arise from energy and angular momentum fluxes, i.e. a generalised version of section \ref{ssec:deph_circ_Edot}.

\subsection{Basics of eccentric waveforms}
\begin{figure}[!t]
    \centering
    \includegraphics[width=0.48\textwidth]{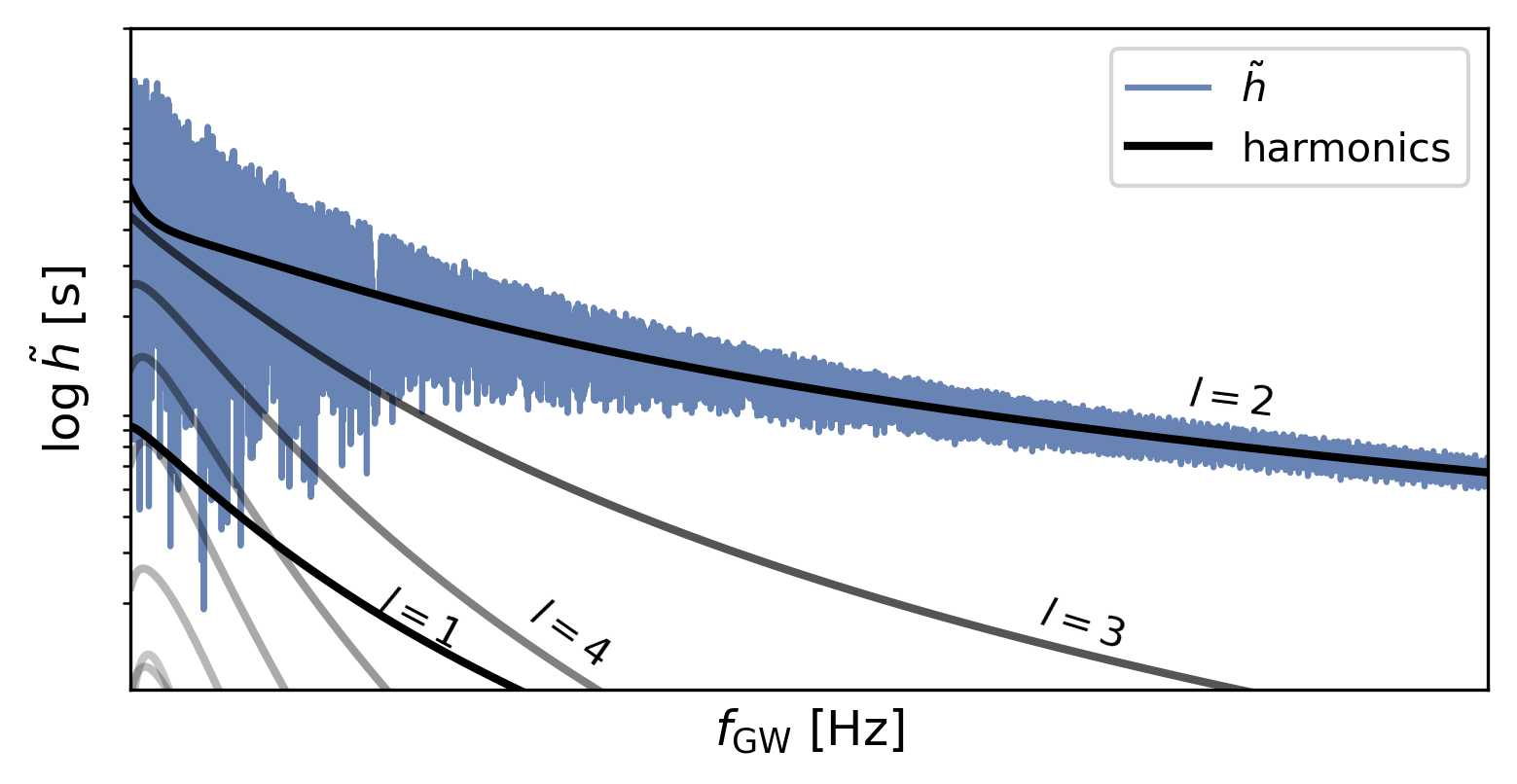}
    \caption{Illustration of an eccentric waveform in Fourier space (computed with the Newtonian model described in \cite{2009yunes}) and the harmonics it is composed of. Note how the waveform appears noisy, while in reality it is composed of a sum of smooth harmonics, the phase of which adds incoherently. As the binary evolves in frequency, its eccentricity decreases and the GW waveform is better described by the $\ell=2$ mode. In this example, the minimum eccentricity of the binary is $e_0 = 0.04$ and the maximum is $e_0 = 0.29$. The amplitude and frequency can be rescaled arbitrarily by changing binary parameters.}
    \label{fig:ecc_harmonics}
\end{figure}

We start our discussion with a review of the basics of eccentric waveforms, based on the seminal work of Yunes et al. \citep{2009yunes}. While at a given time instance circular binaries only emit quadrupolar GWs at a singular frequency, i.e. twice the orbital frequency, eccentric binaries have a wide spectrum corresponding to the various epicyclic frequencies of the orbit, i.e. multiples of the orbital frequency. This makes the GW strain at a given time a sum of of different harmonics:
\begin{equation}
    h(t) = \sum_{\ell=1}^{\infty} \mathcal{A}_\ell(t) \exp\left[ -i\ell \Phi(t) \right] \: ,
    \label{eq:Ecc_waveform}
\end{equation}
where $\Phi$ is the mean anomaly, while $\ell$ corresponds to the epicyclic expansion. Fig~\ref{fig:ecc_waveform} illustrates this expansion for a single period of an orbit with $e=0.5$.
Note that $\mathcal{A}_\ell$ now is slowly varying complex quantity, the phase of which depends on the projection of the elliptic orbit on the observer's frame and thus can vary due to the precession of the orbit. Such an expansion for non-precessing binaries was carried out for low eccentricities by \cite{2009yunes}.
The mean anomaly in Eq.~\eqref{eq:Ecc_waveform} will evolve as
\begin{equation}
    \Phi(t) = \int^t 2\pi F(t') \, \mathrm{d}t' \: ,
\end{equation}
where $F\equiv \Omega/2\pi$ is the orbital frequency. Then in a similar manner as we did in the circular case we can use the stationary phase approximation to translate Eq.~\eqref{eq:Ecc_waveform} into a Fourier domain waveform. This can be done as long as the two conditions for the approximation are valid, i.e. $\mathrm{d}(\ln \mathcal{A}_\ell)/\mathrm{d}t \ll \ell\mathrm{d}\Phi/\mathrm{d}t$, while the frequency is also evolving slowly, i.e. $ \ell\mathrm{d}^2\Phi/\mathrm{d}t^2 \ll \ell^2(\mathrm{d}\Phi/\mathrm{d}t)^2$, where we replaced $\phi$ by $\phi_\ell = \ell \Phi$. Note that in general, the phase of the Fourier integrand can encounter catastrophes, i.e locations where multiple stationary points coalesce and the second time derivative of the phase of the integrand vanishes together with the first derivative \cite{Loutrel:2018ydu,Loutrel:2023rsl}. Such catastrophes render the stationary phase approximation ill-suited for the evaluation of the Fourier integral, and modifications to this approximation scheme must be employed \cite{Klein:2013qda,Klein:2014bua,BERRY1980257}. In practice, as long as most of the power of the eccentric waveform comes from harmonics below $\ell_\mathrm{max}$, where catastrophes are not present, the stationary phase approximation remains valid. Then we can write:
\begin{figure}[!t]
    \centering
    \includegraphics[width=0.48\textwidth]{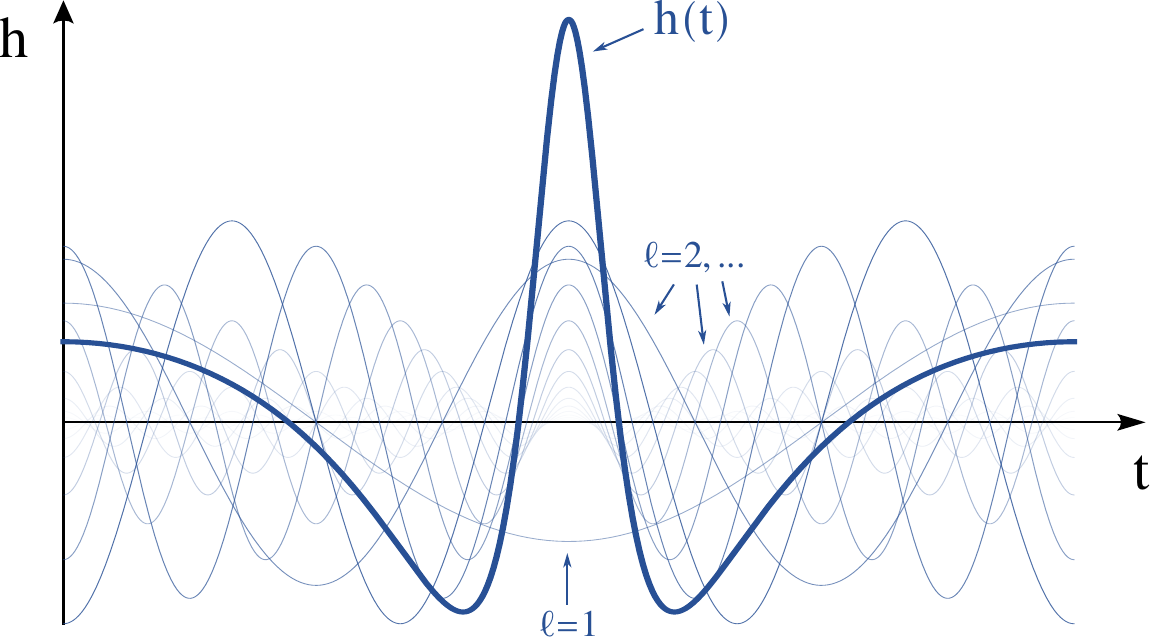}
    \caption{Illustration of the epicyclic expansion of $h(t)$ for a single period of an orbit with $e=0.5$. The time domain waveform $h(t)$ has been rescaled so that its amplitude is comparable to those of the various harmonics. The transparency and amplitude of the harmonics both reflect their relative importance.}
    \label{fig:ecc_waveform}
\end{figure}
\begin{equation}
    \tilde{h}(f) \approx \sum_{\ell=1}^{\ell_\mathrm{max}} \tilde{\mathcal{A}}_\ell\left( f \right) \exp\left[ i\psi_\ell(f) \right] = \sum_{\ell=1}^{\ell_\mathrm{max}} \tilde{h}_\ell(f) \: ,
\end{equation}
where $f=\ell F$ is still the GW frequency. The phases $\psi_\ell(f)$ are just multiples of the Fourier phase:
\begin{equation}
    \psi_\ell (f) = \ell \Psi\left( F=\frac{f}{\ell} \right) \: ,
\end{equation}
where
\begin{equation}
    \Psi(F) =  2\pi F t(F) - \Phi(t_F) \: .
\end{equation}
The formulas for $t(F)$ and $\Phi(t_F)$ are then also analogous to the circular case:
\begin{align}
    t(F) &= \int^F \dot{F}^{-1} \, \mathrm{d}F' \: , \label{eq:tF} \\
    \Phi(t_F) &= 2\pi \int^F F' \dot{F}^{-1} \, \mathrm{d}F' \label{eq:PhiF} \: .
\end{align}
Hence, there is a one-to-one mapping between $\Psi$ and the orbital frequency $F$, independent of $\ell$, which is not the case for $\Psi$ and $f$. The Fourier phase $\Psi$ at a specific $F$ will give contribution to phases at multiple different GW frequencies $f=\ell F$, or equivalently, the Fourier domain waveform and thus its amplitude and phase at $f$ will be accumulated from different stages of the binary evolution for different $\ell$ that correspond to an orbital frequency $F/\ell$. The Fourier domain amplitude of the different harmonics and the total waveform is illustrated in Fig.~\ref{fig:ecc_harmonics}. While the amplitude of each of the harmonics is smooth, note how when added up they create a noisy waveform due to their incoherent phases.

Also in an analogous way to the circular case we have the following relation between Fourier phase and frequency evolution:
\begin{align}
    \frac{\mathrm{d}\Psi}{\mathrm{d}F} &= 2\pi t(F)\: , 
    \label{eq:dPsidF}\\
    \frac{\mathrm{d}^2\Psi}{\mathrm{d}F^2} &= \frac{2\pi}{\dot{F}}\: .
    \label{eq:dPsidF2}
\end{align}

While so far the formulas have been analogous to the circular case, for eccentric binaries the frequency evolution is modified in a non-trivial way:
\begin{align}
    \dot{F} = \dot{F}^{e=0}\times \mathcal{F}(e)
\end{align}
where
\begin{equation}
    \dot{F}^{e=0}(F) = \frac{96}{5}(2\pi)^{8/3} \mathcal{M}^{5/3}F^{11/3} \: ,
    \label{eq:dFvace0}
\end{equation}
and the enhancement function $\mathcal{F}(e)$ reads \citep{peters1964}:
\begin{align}
    \mathcal{F}(e) = \left(1+\frac{73}{24}e^2 + \frac{37}{96}e^4 \right)(1-e^2)^{-7/2} \: .
    \label{eq:Fe}
\end{align}
In order to evaluate the integrals in Eqs.~\eqref{eq:tF} and \eqref{eq:PhiF} we need to have the relation $e(F)$. While $F(e)$ can be given as an analytic function, the inverse formula generally cannot be given in analytic form. However, for small eccentricities, a series expansion can be carried out. Since the aim of this work is only to illustrate different types of dephasings, we only carry out the expansion to leading order in eccentricity, however, expansions of up to $\mathcal{O}(e^9)$ can be found for example in Ref.~\cite{2009yunes}. The relation between the semi-major axis and the eccentricity with the 2.5PN GW radiation term reads:
\begin{equation}
    a(e) = \left( \frac{M}{4\pi^2F^2} \right)^{1/3} = \frac{a_\mathrm{in}}{g(e_\mathrm{in})}g(e) \: ,
    \label{eq:ae}
\end{equation}
where $e_\mathrm{in}$ is the initial eccentricity at the initial semi-major axis $a_\mathrm{in}$, and
\begin{align}
    g(e) = \frac{e^{12/19}}{1 - e^2}\bigg(1 + \frac{121}{304}e^2\bigg)^{870/2299} \: .
\label{eq:ge}
\end{align}
Then to leading order $e(F)$ reads:
\begin{equation}
    e(F) \approx e_\mathrm{in} \left( \frac{F}{F_\mathrm{in}} \right)^{-19/18} \: ,
\end{equation}
where $F_\mathrm{in}$ is the initial orbital frequency.

\subsection{Time-domain and Fourier-domain dephasing}

\begin{figure}[!t]
    \centering
    \includegraphics[width=0.48\textwidth]{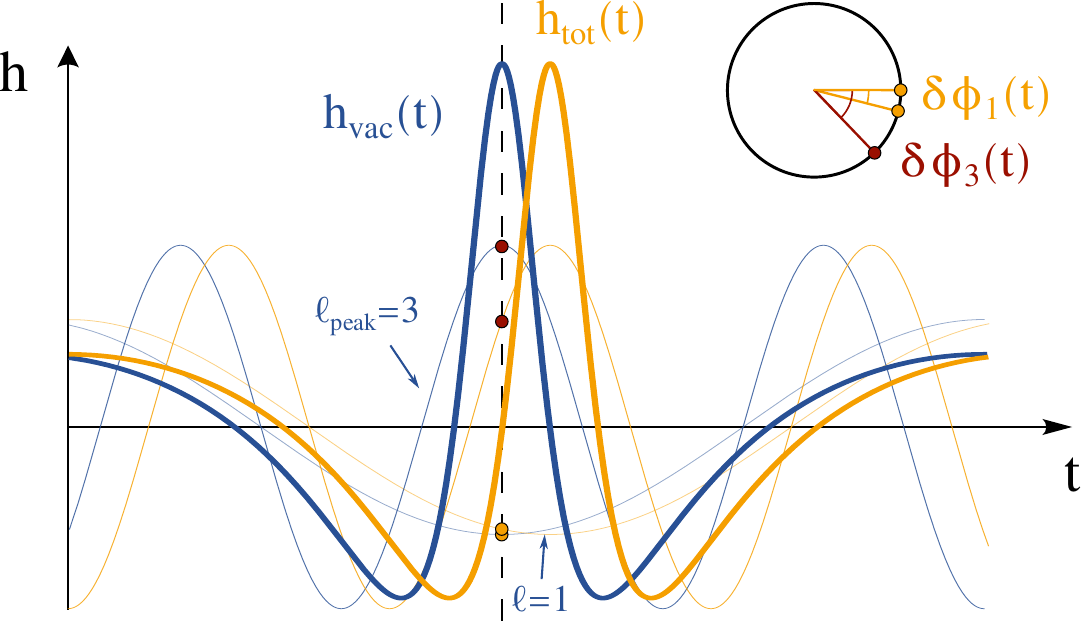}
    \caption{Illustration of a phase shifted GW for an orbit with $e=0.5$. In addition to the $\ell=1$ harmonic we show the peak harmonic as well, i.e. the harmonic with the largest amplitude, which in this case is $\ell_\mathrm{peak}=3$. Note how a constant phase shift is more significant if the waveform presents sharper features. This corresponds to the fact that a given dephasing is boosted in higher harmonics, i.e. $\delta\phi_3 = 3\delta\phi_1 = \delta\Phi$. However, the contribution to the total SNR of higher harmonics decreases exponentially as $\ell\gg1$.}
    \label{fig:ecc_dphi}
\end{figure}

After finding an appropriate representation of the GW signal in the Fourier domain our task then reduces to finding $\delta\Psi(F)$, from which the dephasing for all the different harmonics can be found by $\delta\psi_\ell(f/\ell) = \ell\delta\Psi(F)$ (see illustration in Fig.~\ref{fig:ecc_dphi}). Most of the statements that we found for the GW dephasing in the circular case remains valid for $\delta\Psi(F)$ as well. This means that assuming a perturbation in the chirp of the orbital frequency:
\begin{equation}
    \dot{F}_\mathrm{tot}(F) = \dot{F}_\mathrm{vac}(F) + \delta \dot{F}(F) \: ,
\end{equation}
we get that
\begin{align}
    \delta t(F) &\approx- \int^F\frac{\delta\dot{F}}{{\dot{F}_\mathrm{vac}}^2} \, \mathrm{d}F' \: , \\
    \delta\varphi(F) &= \Phi(t_F) - \Phi_\mathrm{vac}(t_{F,\mathrm{vac}}) \nonumber \\
    &\approx - 2\pi \int^F\frac{F'\delta\dot{F}}{{\dot{F}_\mathrm{vac}}^2} \, \mathrm{d}F' \: .
\end{align}
And thus:
\begin{equation}
    \delta\Psi(F) = 2\pi F\delta t(F) - \delta\varphi(F) \: .
\end{equation}
For the second derivative with respect to frequency we get:
\begin{equation}
    \frac{\mathrm{d}^2(\delta\Psi)}{\mathrm{d}F^2} = - 2\pi \frac{\delta \dot{F}}{\dot{F}_\mathrm{vac}^2} \: .
\end{equation}
Note that in this case as well $\delta\Psi(F)$ itself is free up to two integration constants:
\begin{equation}
    \delta\Psi \rightarrow \delta\Psi + 2\pi F t_c - \Phi_c \: .
\end{equation}
The relation between the time- and Fourier domain dephasings also holds:
\begin{equation}
    \delta \Phi(t_F) \approx -\delta\Psi(F) \: ,
\end{equation}
meaning that in order to get the Fourier dephasing at an orbital frequency $F$, to first order it is sufficient to evaluate the difference in the mean anomalies of the perturbed and unperturbed orbits at $t_F$ when the orbital frequency of the unperturbed orbit reaches $F$. This relation becomes even more useful in the eccentric case, as the analytic evaluation of the exact dephasing can become complicated, and so direct numerical simulations can be used instead to obtain $\delta\Phi(t_F)$. Then $\delta\Psi(F)$ can be used to calculate $\delta$SNR with an analytic waveform, or simply to validate analytic dephasing prescriptions.

All this means that the formulas in columns $3-5$ of Table~\ref{tab:summary} for intrinsic and Doppler dephasings remain valid for eccentric orbits as well by replacing $f$, $\phi$ and $\psi$ with $F$, $\Phi$ and $\Psi$, respectively. The exact form of $\delta\dot{F}(F)$ can however be complicated, as we will see later, due to the mutual dependence of the parameters on each other.

\subsection{Dephasing due to additional energy and angular momentum fluxes}

\begin{figure}[!t]
    \centering
    \includegraphics[width=0.48\textwidth]{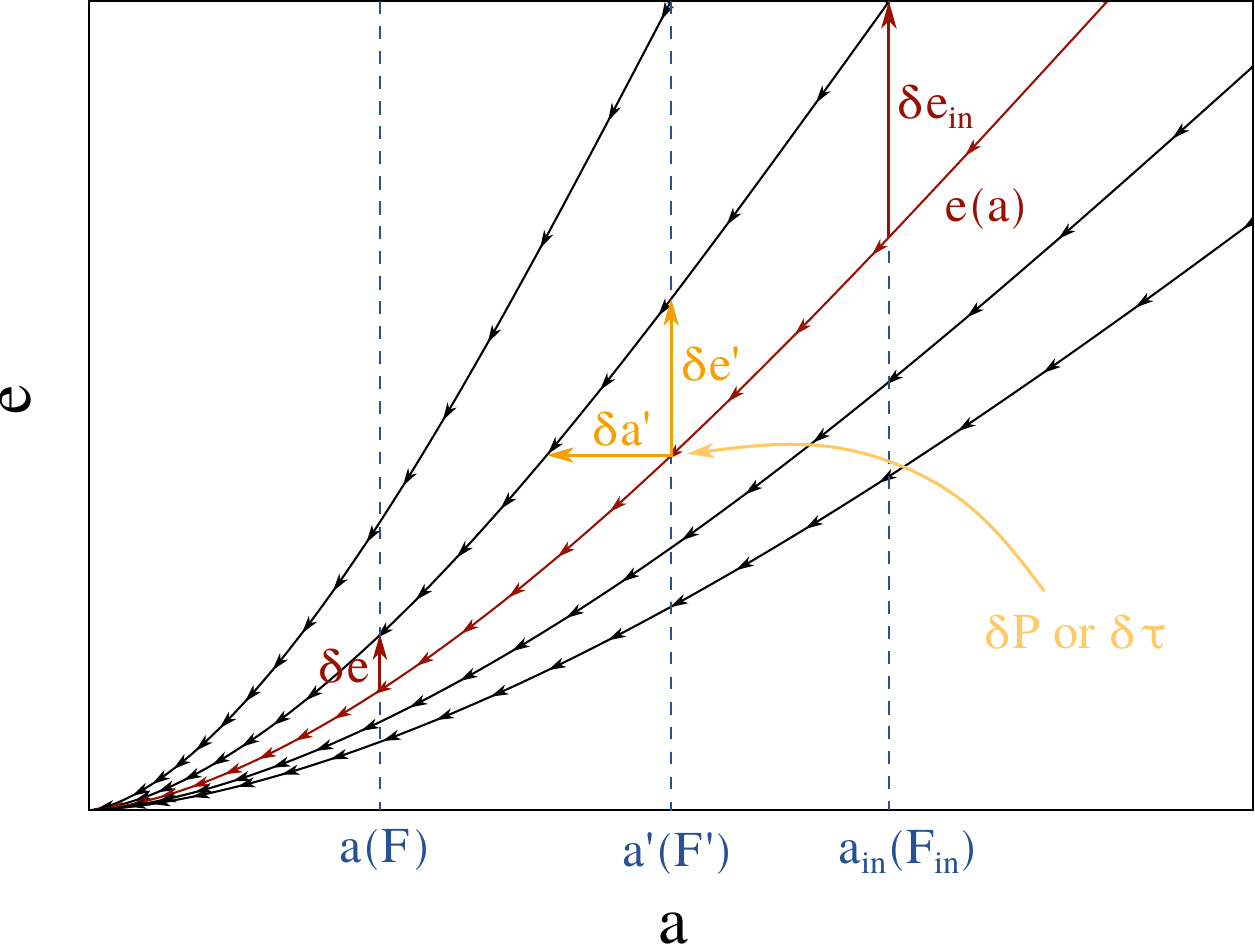}
    \caption{Different evolutionary paths for eccentric binaries on the $a-e$ plane. Both perturbing the semi-major axis and the eccentricity can divert the binary to a different evolutionary path, corresponding to a change $\delta e_\mathrm{in}$ in the initial eccentricity parameter and thus a change $\delta e$ at a semi-major axis $a(F)$. These perturbations can be achieved by additional external energy ($\delta P$) and angular momentum fluxes ($\delta \tau$), and result in different dephasing behaviours as a function of frequency.}
    \label{fig:ecc_evol}
\end{figure}

An additional energy flux generated by an external perturbation has a similar direct effect as in the circular case:
\begin{equation}
    \delta \dot{F} \approx \dot{F}_\mathrm{vac} \frac{\delta P}{\dot{E}_\mathrm{vac}} \: .
\end{equation}
However, in the case of eccentric orbits we have an additional, indirect effect as well. This arises from the fact that the chirp is not only a function of the semi-major axis (or equivalently, the orbital frequency) but the eccentricity as well. Thus if due to an external perturbation the eccentricity at a given orbital frequency $F$ is modified with respect to what the leading order evolution from the 2.5PN radiation would dictate, the chirp will also be modified as:
\begin{equation}
    \delta \dot{F} \approx \frac{\partial \dot{F}_{\rm{vac}} }{\partial e}\delta e \: ,
\end{equation}
where
\begin{align}  
    \frac{\partial \dot{F} }{\partial e} &= \dot{F}^{e=0}\left(\frac{307}{23}e+\frac{9611}{552}e^3+\frac{37}{32}e^5 \right)\left(1-e^2 \right)^{-9/2}
\end{align}
Figure~\ref{fig:ecc_evol} illustrates the cause of this extra dephasing. The solid lines show different evolutionary paths the binary can take depending on the initial eccentricity $e_\mathrm{in}$ at some initial semi-major axis $a_\mathrm{in}$. A perturbation in eccentricity or in semi-major axis can divert the binary from its original trajectory, and put it on a trajectory corresponding to an initial eccentricity $e_\mathrm{in}+\delta e_\mathrm{in}$. Thus the eccentricity at a given orbital frequency will also be modified by $\delta e(F)$. A surprising result is that even though a change in the semi-major axis will not explicitly modify the eccentricity, by changing the orbital frequency while keeping the eccentricity constant it will put the binary on a different trajectory and thus the eccentricity after the frequency change will differ from the expected eccentricity. A perturbation will only keep the binary on the same trajectory if $\delta\dot{e}/\delta\dot{a}\propto \dot{e}_\mathrm{vac}/\dot{a}_\mathrm{vac}$.

Parameterizing the trajectories with $e_\mathrm{in}$ and differentiating Eq.~\eqref{eq:ae} with respect to time we have that for general $\dot{a}$ and $\dot{e}$:
\begin{equation}
    \dot{a} = -\frac{a_\mathrm{in}}{g(e_\mathrm{in})^2}g'(e_\mathrm{in}) g(e) \dot{e}_\mathrm{in} + \frac{a_\mathrm{in}}{g(e_\mathrm{in})} g'(e) \dot{e} \: ,
\end{equation}
where $g'(e)\equiv \mathrm{d}g(e)/\mathrm{d}e$. Note that the 2.5PN radiation terms give $\dot{e}_\mathrm{in}=0$ and thus from the perturbations we get:
\begin{equation}
    \dot{e}_\mathrm{in} = \frac{g(e_\mathrm{in})}{g(e)} \frac{g'(e)}{g'(e_\mathrm{in})} \delta \dot{e} - \frac{g(e_\mathrm{in})^2}{g(e)g'(e_\mathrm{in})} \frac{\delta\dot{a}}{a_\mathrm{in}} \: .
\end{equation}
Then the total change in $e_\mathrm{in}$ parameterizing the trajectory will be:
\begin{equation}
    \delta e_\mathrm{in}(F) = \int^{t_F} \dot{e}_\mathrm{in} \, \mathrm{d}t' \: .
\end{equation}
We then need to determine what this change in the initial eccentricity parameter will mean for $\delta e(F)$. Differentiating the relation $a(e,e_\mathrm{in})$ for fixed semi-major axis to linear order we get that a change $\delta e_\mathrm{in}$ results in
\begin{equation}
    \delta e \approx \frac{g(e)}{g(e_\mathrm{in})} \frac{g'(e_\mathrm{in})}{g'(e)} \delta e_\mathrm{in} \: .
\end{equation}
Combining these relations we can now determine how perturbations in eccentricity and semi-major axis modify the eccentricity at a given orbital frequency. Replacing the time integral with a frequency integral and keeping everything to linear order we have:
\begin{align}
    \delta e(F) = \frac{g(e_0)}{g'(e_0)}\int^F \frac{{\rm d}F'}{\dot{F}_{\rm vac}} \left[ \frac{g'(e_0)}{g(e_0)} \delta\dot{e} - \frac{\delta\dot{a}}{a_0}\right] \: ,
\end{align}
where $e_0 \equiv e_\mathrm{vac}(F)$ and $a_0 \equiv a_\mathrm{vac}(F)$ and $F'$ is a dummy integration variable. At lowest order in eccentricity this reduces to:
\begin{align}
    \delta e(F) = e_0(F)\int^F \frac{\delta\dot{e}(F')}{\dot{F}_{\rm vac}^{{e=0}}(F') e_0(F')} \, {\rm d}F' \: .
\end{align}
Thus, perturbations in the semi-major orbit do not modify the trajectory up to linear order in eccentricity. This can be understood by the fact that the trajectories in Fig.~\ref{fig:ecc_evol} are almost horizontal for small eccentricities, thus changing the the semi-major axis would keep you on the same trajectory to leading order. Along the same argument, a small change in the eccentricity changes the initial eccentricity by an increasingly large amount as the eccentricity decreases.

These eccentricity and semi-major axis perturbations can be induced by external energy and angular momentum fluxes (torques, $\delta\tau$). These quantities are related by the following relations:
\begin{align}
    \delta \dot{e} &= \frac{e^2-1}{e} \left( \frac{\delta\tau}{L_\mathrm{orb}} + \frac{\delta P}{2 E_\mathrm{orb}} \right) \: , \\
    \delta \dot{a} &= -a \frac{\delta P}{E_\mathrm{orb}} \: ,
\end{align}
where $\delta P$ and $\delta \tau$ are negative when the binary is losing energy or angular momentum.

Thus the induced eccentricity finally becomes
\begin{align}
    \delta e(F) = \frac{g(e_0)}{g'(e_0)}&\int^F \frac{{\rm d}F'}{\dot{F}_{\rm vac}} \left[ \frac{g'(e_0)}{g(e_0)} \frac{e_0^2-1}{e_0} \frac{\delta \tau}{L_\mathrm{orb}} \right. \nonumber \\
    &\left. + \left(\frac{g'(e_0)}{g(e_0)} \frac{e_0^2-1}{2e_0}+1\right) \frac{\delta P}{E_\mathrm{orb}} \right] \: .
\end{align}
To lowest order in eccentricity this means
\begin{align}
    &\delta e(F) =\nonumber\\& -e_0(F)\int^F \frac{1}{\dot{F}_{\rm vac}^{{e=0}} e_0^2} \left( \frac{\delta\tau}{L_\mathrm{orb}} + \frac{\delta P}{2 E_\mathrm{orb}} \right) \, {\rm d}F' \: ,
\end{align}
and once again the total induced perturbation to the frequency chirp is\footnote{To our knowledge, this is the first instance of the derivation of the full induced chirp and consequent dephasing due to energy and angular momentum fluxes for eccentric orbits.}: 
\begin{align}
   \delta \dot{F}_{\rm tot} = \dot{F}_{\rm vac}\frac{\delta P}{\dot{E}_{\rm vac}} + \frac{\partial \dot{F}_{\rm vac}}{\partial e}\delta e.
   \label{eq:Fd_ecc_twoterms}
\end{align}
These results of our calculations are summarized in Table~\ref{tab:eccsummary}.
\renewcommand{\arraystretch}{1.5}
\begin{table*}[]
\caption{Relations for dephasing caused by perturbative energy fluxes $\delta P$ and angular momentum fluxes $\delta \tau$ in eccentric GW.
\label{tab:eccsummary}}
\begin{tabular}{cccc}
\toprule
Flux type & \quad $\delta\dot{F}$ \qquad  &  $\delta e(F)$  &  $\delta\Psi(F) = - \delta\Phi(t_F)$ \\ \midrule\midrule
$\delta P$  & \quad $\dot{F}_\mathrm{vac} \,\delta P/ \dot{E}_\mathrm{vac} + {\displaystyle\frac{{\rm d} \dot{F}_{\rm{vac}} }{{\rm d} e}\delta e}$ \qquad &  \qquad ${\displaystyle\frac{g(e_0)}{g'(e_0)} \int^F \frac{{\rm d}F'}{\dot{F}_{\rm vac}} \left[  \left(\frac{g'(e_0)}{g(e_0)} \frac{e_0^2-1}{2e_0}+1\right) \frac{\delta P}{E_\mathrm{orb}} \right]}$ \qquad & \multirow{3}{*}{ $ {\displaystyle - 2\pi \iint {\rm d}F\,\, \frac{\delta \dot{F}}{{\dot{F}_\mathrm{vac}^2}}}$} \\ \cmidrule{1-3}
$\delta \tau$  &  \quad ${\displaystyle\frac{{\rm d} \dot{F}_{\rm{vac}} }{{\rm d} e}\delta e}$ \qquad  &  \qquad ${\displaystyle\frac{g(e_0)}{g'(e_0)} \int^F \frac{{\rm d}F'}{\dot{F}_{\rm vac}} \left[  \frac{g'(e_0)}{g(e_0)} \frac{e_0^2-1}{e_0} \frac{\delta \tau}{L_\mathrm{orb}} \right]}$ \qquad &   \\ \bottomrule
\end{tabular}
\end{table*}

It is interesting to ask which of the two components of the dephasing  dominates. While of course the answer depends on the specifics of the perturbing force, we work through an illustrative example where:
\begin{align}
    \frac{\delta\dot{a}}{a} &= {\rm const.} \equiv K_{\rm a}\\
    \delta\dot{e} &=0.
\end{align}
where one would naively expect no dephasing contribution to dephasing from an eccentricity perturbation. We can calculate the resulting dephasing via the usual prescriptions reported in Table \ref{tab:eccsummary}. here we report the time-domain binary dephasing at a given frequency, which is given by the integral:
\begin{align}
    \delta \varphi(F) = 2 \pi\int {\rm d}F' \left[F'\frac{\delta \dot{F}_{\rm tot}}{\dot{F}_{\rm vac}^2} \right]\equiv \delta \varphi_a + \delta \varphi_e
\end{align}
where $\delta \varphi_a$ comes from the first term in Eq.~\eqref{eq:Fd_ecc_twoterms} and $\delta \varphi_e$ from the second. We can solve the integral analytically in the limit of small eccentricity, finding that the ration of the two induced phase-shifts reads:
\begin{align}
    \frac{\delta \varphi_e}{\delta \varphi_a} \approx - \frac{116337}{14848}e_0(F)^2 \approx- 7.8\times e_0(F)^2.
\end{align}
Therefore, even without any specific eccentricity perturbation $\delta \dot{e}$, the dephasing due to the induced change in eccentricity $\delta e$ can still be dominant for $e_0 \sim 0.4$. Compare now a similar calculation in which $\delta\dot{e} = K_{e}$ is constant. Repeating the integral yields:
\begin{align}
    \frac{\delta \varphi_e}{\delta \varphi_a} \approx - \frac{55107}{5626} \frac{K_{\rm e}}{K_{\rm a}}e_0(F).
\end{align}
That is, the dephasing due to the induced change in eccentricity can already dominate for $e_0\sim 0.1$, provided that $\delta \dot{a}/a$ and $\delta \dot{e}$ are of similar order.

Another interesting and more physically relevant example would be that of a von Zeipel--Kozai--Lidov (ZKL) eccentricity oscillation induced by a tertiary object orbiting around the binary \citep{zeipel1910,koz62,lid62}. In such a case in the secular limit the inner binary exchanges angular momentum without exchanging energy, thus an eccentricity perturbation is induced without perturbing the inner binary's semi-major axis. For a circular outer binary and optimal relative orientation of the inner and outer binaries we have that (see e.g. \citep{2013MNRAS.431.2155N}):
\begin{equation}
    \frac{\delta \dot{e}}{e} \approx \frac{15}{8} \sqrt{1-e^2} \frac{T_\mathrm{in}}{T_\mathrm{out}^2} \: ,
\end{equation}
where $T_\mathrm{in}$ is the orbital period of the inner binary and $T_\mathrm{out}$ is that of the outer binary formed by the tertiary object and the inner binary. In the regime where ZKL oscillations dominate over GW radiation, it has been shown that eccentricity oscillations can cause a gradual phase shift over many oscillation periods, which might be detectable by LISA. In our case, where we are considering perturbations that are small compared to the 2.5PN radiation reaction terms, 1PN precession of the binary will also inevitably become important, which will suppress ZKL oscillations. However, for pedagogical reasons let us now imagine an environmental effect that causes an eccentricity perturbation with the same scaling with eccentricity as the ZKL effect. Then by integrating Eq.~\eqref{eq:Fd_ecc_twoterms} once again we can find that the scaling of the resulting induced dephasing would be:
\begin{equation}
    \delta \varphi_{\rm e} \propto e_{0}(F)^2 F^{-16/3}.
\end{equation}

\section{SNR and $\delta$SNR calculations}
\label{sec:SNR}
We conclude our overview of dephasing with a brief discussion of signal-to-noise (SNR) ratios, which are related to the observability of a GW signal and any potential perturbation. The SNR is calculated from the Fourier domain strain according to:
\begin{equation}
    \mathrm{SNR}^2 = 4\int \frac{|\tilde{h}(f)|^2}{S_n(f)} \mathrm{d}f \: ,
\end{equation}
where $S_n(f)$ is the spectral noise density curve of the detector. The interpretation of this integral is that of summing the excess power of the GW signal with respect to the detector noise, in every frequency bin.  Similarly, the observability of a perturbation is usually estimated by the SNR of the difference signal \citep{kocsis}, i.e.
\begin{equation}
    \delta \mathrm{SNR}^2 = 4 \int \frac{|\delta\tilde{h}(f)|^2}{S_n(f)} \mathrm{d}f \: ,
\end{equation}
where
\begin{align}
    \delta \tilde{h}(f) &= \tilde{h}_\mathrm{tot}(f) - \tilde{h}_\mathrm{vac}(f) \approx \tilde{h}_\mathrm{vac}(f)\left( e^{i\delta\psi(f)}-1 \right) \nonumber\\
    &\approx \tilde{h}_\mathrm{vac}(f)\cdot i\delta\psi(f) \: ,
\end{align}
where we assumed that the main contribution to the difference signal comes from the dephasing rather than the amplitude modulation. Thus
\begin{equation}
    \delta \mathrm{SNR}^2 \approx 4 \int |\delta\psi(f)|^2 \frac{|\tilde{h}(f)|^2}{S_n(f)} \mathrm{d}f \: .
\end{equation}

The SNR calculations for eccentric sources is also analogous to the circular case, only in this case the total SNR comes from a contribution from all the different harmonics:
\begin{equation}
    \mathrm{SNR}^2 = 4\int \frac{1}{S_n(f)} \Big|\sum_{\ell}\tilde{h}_\ell(f) \Big|^2 \mathrm{d}f \: .
    \label{eq:SNR_ecc}
\end{equation}
Then the difference signal is:
\begin{align}
    \delta \tilde{h}(f) &\approx \sum_\ell \tilde{h}_\mathrm{\ell,vac}(f)\cdot i\delta\psi_\ell(f) \: ,
    \label{eq:deltah_ecc}
\end{align}
and thus the $\delta$SNR:
\begin{equation}
    \delta \mathrm{SNR}^2 \approx 4\int \frac{1}{S_n(f)} \Big|\sum_{\ell} \delta \psi_\ell(f)\tilde{h}_\ell(f) \Big|^2 \mathrm{d}f \: ,
\end{equation}
where
\begin{equation}
    \delta \psi_\ell(f) = \ell \delta\Psi(f/\ell) \: .
\end{equation}
Note that the dephasing for harmonics with $\ell\gg 1$ can easily become large and even exceed $2\pi$, in which case the approximation in Eq.~\eqref{eq:deltah_ecc} clearly does not hold. Therefore the use of this formula requires some consideration and $\exp[i\delta\psi_\ell(f)]-1$ should be used instead of $i\delta\psi_\ell(f)$ in the general case.

Due to the incoherence of the phases of different harmonics at a given GW frequency $f$ (see Fig \ref{fig:ecc_waveform}), we can also assume that different harmonics give orthogonal contributions to the total SNR \cite{Barack:2003fp}, i.e. the cross terms in the sum in Eq.~\eqref{eq:SNR_ecc} can be neglected and thus the sum can be moved outside of the absolute value operator:
\begin{equation}
    \mathrm{SNR}^2 \approx 4\int \sum_{\ell}\frac{|\tilde{h}_\ell(f) |^2}{S_n(f)}  \mathrm{d}f \: .
\end{equation}
We can also make a similar assumption for the orthogonality of different harmonics that contribute to $\delta$SNR:
\begin{equation}
    \delta\mathrm{SNR}^2 \approx 4\int \sum_{\ell}\frac{ |\delta\tilde{h}_\ell(f) |^2}{S_n(f)}  \mathrm{d}f \: .
\end{equation}

We note that defining the detectability in terms of the $\delta$SNR can be useful to explore large parameter spaces efficiently, as the generation of accurate waveforms can be expensive and $\delta$SNR values instead provide fast detectability criteria. However, this way of defining detectability does not take into account possible degeneracies between parameters and can be too generous in estimating the detectability of an effect. Although potentially time-consuming, other, more accurate methods exist for estimating detectability or waveform accuracy, such as the recent update on the mismatch criterion in Ref.~\citep{Toubiana:2024car}.

\subsection{$\delta$SNR enhancement in eccentric binaries}

Higher harmonics of the GW signal can contribute significantly to the total SNR of eccentric binaries, which can make the SNR of eccentric binaries higher in some mass and initial eccentricity range \cite[e.g.][]{oleary2009}. This enhancement with eccentricity can be even more pronounced for the $\delta$SNR of dephased eccentric binaries. To understand this, let us consider the contribution of the $\ell=2$ mode, which is dominant for circular binaries, and that of the $\ell=4$ mode, taken at a given GW frequency $f$. Let us suppose that for $f_{\ell=2} = 2F = f$, the phase shift is $\delta\psi_{\ell=2}(f)=2\delta\Psi(f/2)\equiv\delta\psi$. The contribution for the same GW frequency $f$ from the $\ell=4$ comes from an earlier stage of the binary evolution. As an example, let us now consider a simple power-law scaling of $\Psi(F)\propto F^{-n}$, where $n=13/3$ corresponds to phase shifts due to Roemer-delay in the circular case. With this scaling, the phase shift of the $\ell=4$ at $f=4F$ is $\delta \psi_{\ell=4}(f) = 4\delta\Psi(f/4) = 2\cdot 2^{n} \delta\psi$. Thus we get a factor of $\sim40$ enhancement for the contribution to the $\delta$SNR at GW frequency $f$ from the $\ell=4$ mode. This huge enhancement factor can make higher harmonics dominate the $\delta$SNR even for low eccentricities, where the SNR contribution of these harmonics is normally suppressed (see Fig.~\ref{fig:ecc_harmonics}.

Let us now be more exact and calculate the $\delta$SNR from Roemer-delay for eccentric binaries. The phase shift due to line-of-sight acceleration of the center of mass of an inspiralling binary results in a phase shift \citep[e.g.][]{kai2024}:
\begin{align}
    \left|\delta\Psi(F)\right| &\approx \left|\delta\Phi(t_F)\right| \approx \frac{2\pi\Delta t}{T} \nonumber \\
    &= 2\pi \frac{a_\mathrm{CM}}{2c} t_F^2 F \: ,
\end{align}
where $a_\mathrm{CM}$ is the projection of the center-of-mass acceleration in the line-of-sight direction. The inspiral time for circular binaries, as it was shown in Sec.~\ref{sec:SPA_rev}:
\begin{equation}
    t_F^\mathrm{circ} = 5(16\pi F)^{-8/3} \mathcal{M}^{-5/3} \: .
\end{equation}
For non-zero eccentricities the inspiral time from a given orbital frequency is reduced. According to Ref.~\citep{Zwick:2019yjl} it can be estimated by the following formula:
\begin{equation}
    t_F^e \approx t_F^\mathrm{circ} \frac{R(e)}{\mathcal{F}(e)} \: ,
\end{equation}
where $\mathcal{F}(e)$ is defined by Eq.~\eqref{eq:Fe} and $R(e) = 8^{1-\sqrt{1-e}}$. Even though the extra factor reduces the inspiral time significantly, it is still $\sim0.8$ for $e=0.3$. Therefore, for eccentricities $e\lesssim0.3$ the eccentric inspiral time is approximately the circular inspiral time. Thus, for simplicity we assume a dephasing prescription of $\delta\Psi(F) = C \cdot F^{n}$ with $n=-13/3$, and investigate the change in the $\delta$SNR with increasing orbital eccentricity at some reference frequency. We leave the detailed investigation of $\delta$SNR for higher eccentricities to follow-up studies. We choose the reference frequency as the frequency of the $\ell=2$ mode at the lower cut-off frequency of a given GW detector. For the frequency-domain waveform, we use the model of Ref.~\citep{Mikoczi:2012qy} without 1PN precession.

\begin{figure}[!t]
    \centering
    \includegraphics[width=0.49\textwidth]{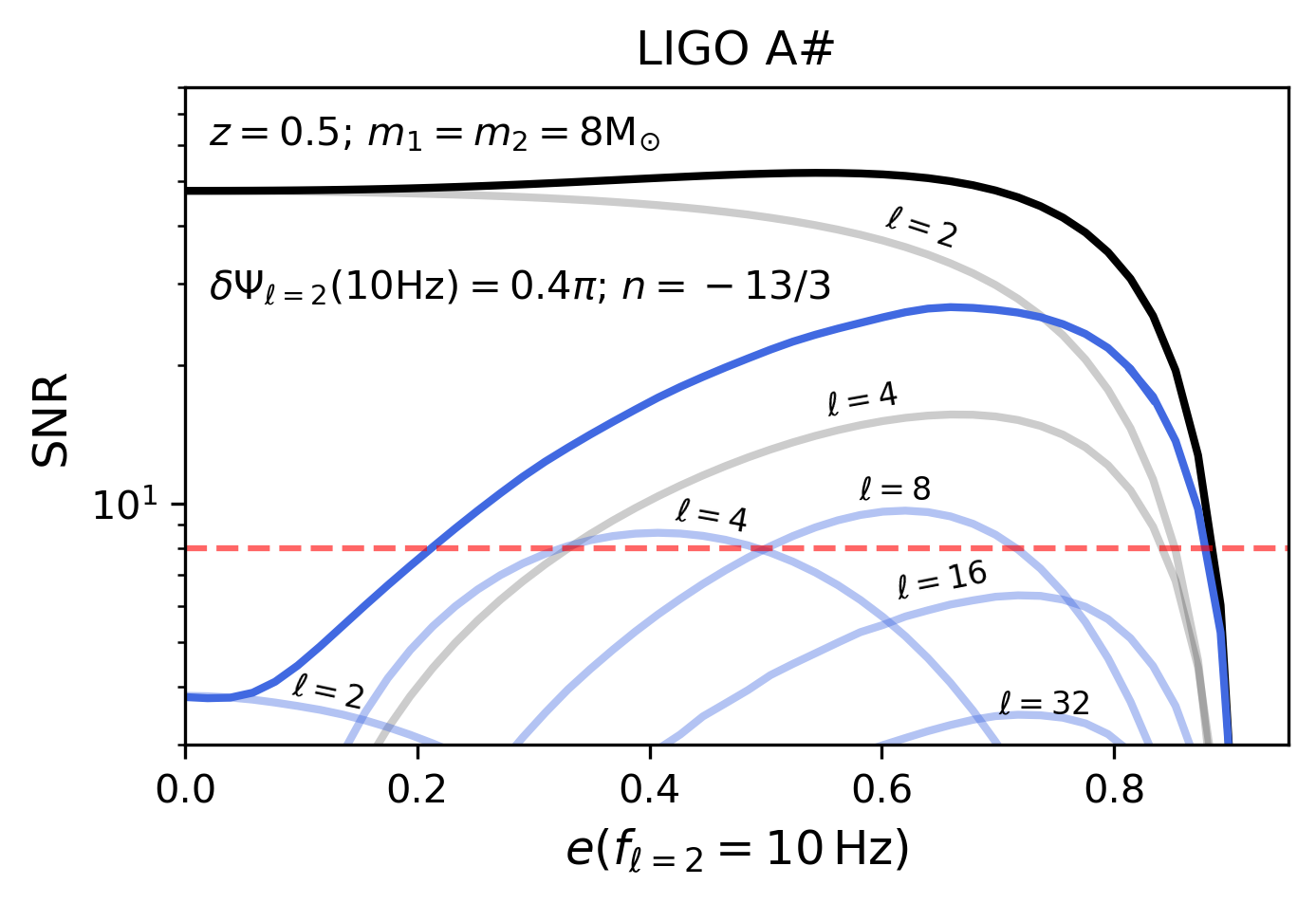}
    \includegraphics[width=0.49\textwidth]{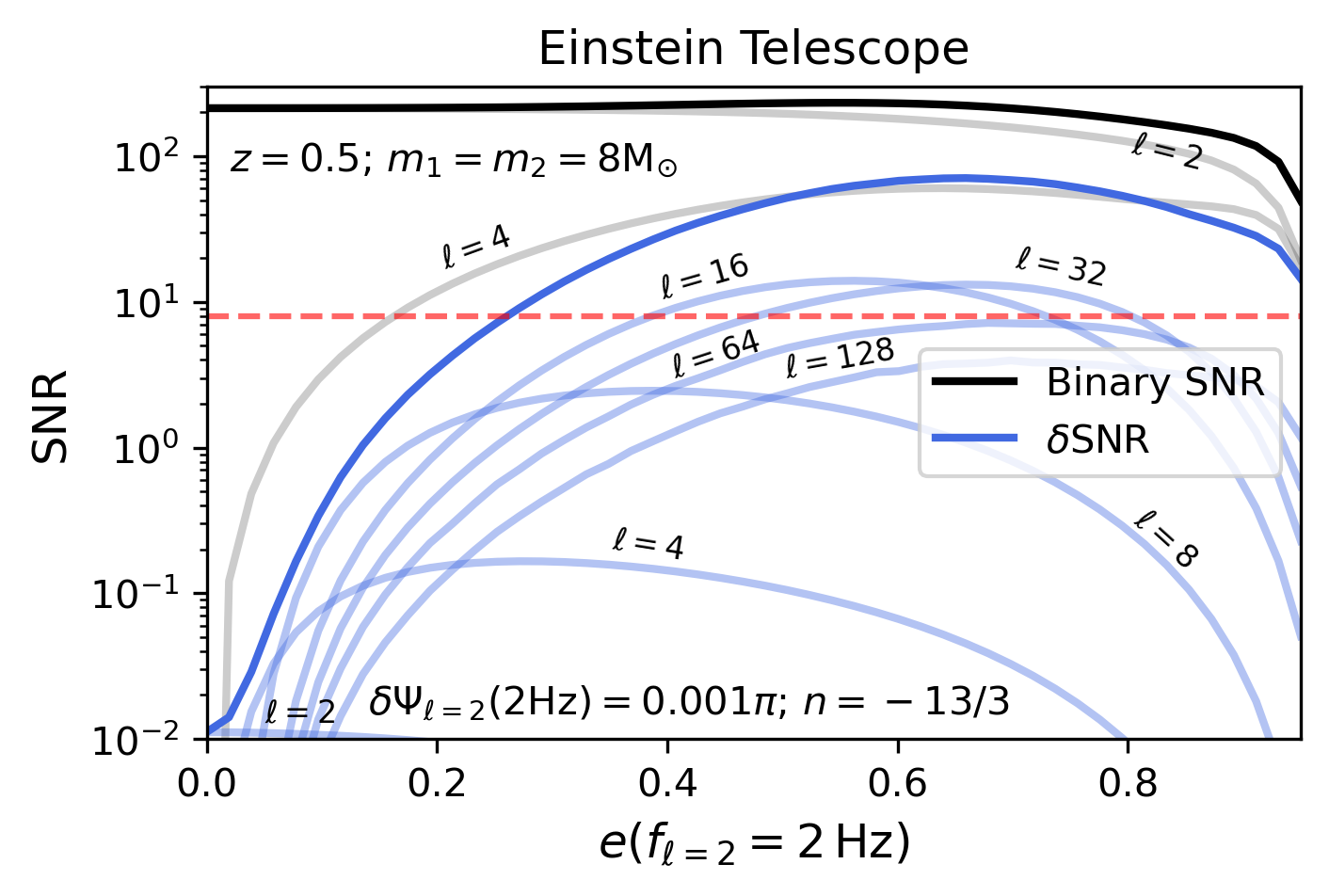}
    \caption{The total binary SNR (black) and the SNR corresponding to the difference signal $\delta h$ between the vacuum and the phase shifted signal (blue) as a function of eccentricity at a given reference frequency for the LIGO A\# (top) and the Einstein Telescope (bottom) sensitivity curves. The reference frequency was chosen as the frequency of the $\ell=2$ mode at the lower cut-off frequency of the given GW detector. The binary consists of two $8~M_\odot$ BHs at a redshift of $z=0.5$. The relative orientation of the binary and the detector was chosen to be optimal. Contributions to the SNR from a few selected harmonics are shown by the lines with lighter tones. The horizontal red line corresponds to a detection threshold of $\mathrm{SNR}=8$.}
    \label{fig:deltaSNR}
\end{figure}
The results are shown in Fig.~\ref{fig:deltaSNR}. We can see that even though the total SNR of binaries has a weak dependence on the eccentricity for moderate eccentricities, the $\delta$SNR from the phase shift increases rapidly with eccentricity and obtains a factor of $\sim5$ enhancement for $e_\mathrm{10Hz}\approx0.4$ for LIGO and a factor of $\sim10^3$ enhancement for ET at $e\approx0.3$. Interestingly but not surprisingly, apart from close to zero eccentricities, the contributions to the $\delta$SNR is dominated by harmonics $\ell>2$. This is even more pronounced for the results for ET, where already at $e_\mathrm{2Hz}\gtrsim0.2$ the contributions from $\ell>10$ harmonics dominate. Note that this is not the case for the total SNR, where, except for extreme eccentricities, the $\ell=2$ mode always dominates. Note that for $\ell\gg1$, large contributions to $\delta$SNR can come from early stages of the binary evolution, where $e\gtrsim0.9$, and thus our power-law phase shift prescription is no longer valid. Hence, our calculations here represent an idealistic scenario. For large eccentricities, the $\delta$SNR is expected to be reduced, however, the significant amplification at $e\sim0.1$ will remain valid. Note also that calculating the $\delta$SNR this way does not take into account possible degeneracies with other parameters and also assumes that these orbital parameters are known exactly. For real observations this will obviously not be the case. Nevertheless, these results underline the significance of eccentric binaries in the search for environmental effects in GW signals.

\section{Conclusion}
\label{sec:conclusion}
Here we provide a short summary of the main conclusions of this paper and emphasize the main take-aways. In Section~\ref{sec:SPA_rev}, we went through the derivation of the Fourier domain waveform using the stationary phase approximation for circular binary inspirals, based on the seminal work of Cutler \& Flanagan. In particular we highlighted how the phases in the time and Fourier domains are related. Section~\ref{sec:circ_deph} presented the relevant expressions and highlighted the key differences between various types of dephasing across the two domains. We also outlined the primary environmental effects that cause phase shifts in circular binaries, including those arising from extra energy losses, radial forces, and Doppler motion. The most important formulas are reported in Table \ref{tab:summary}. In Section~\ref{sec:circ_examp}, we explored typical astrophysical scenarios where multiple sources of dephasing occur simultaneously. We assessed their relative impacts with simple but informative toy models. In Section~\ref{sec:ecc_deph} we extended this analysis to eccentric binaries. The presence of eccentricity introduces a variety of ways in which environmental perturbations can induce dephasing. We focused here on the ways in which additional energy and angular momentum fluxes can influence the binary inspiral, deriving a novel prescription for the gravitational wave phase. Finally, in Section~\ref{sec:SNR}, we briefly addressed how signal-to-noise ratios affect the ability to detect these environmental dephasing effects, and showed in an idealized example how the SNR from the difference signal due to environmental phase shifts can be significantly amplified for non-zero eccentricities.

To conclude, we delineate again the aim of this paper is twofold. Firstly, it serves as a cohesive overview of the topic of GW dephasing, clarifying the underlying concepts and organizing essential formulas, examples, and methodological tools in a practical format. It aims to support ongoing research while also guiding newcomers through a modern and accessible entry into the field. Secondly, it highlights the interesting possibilities of analyzing dephasing prescriptions in eccentric GW signals, as the foundations worked out here already show how interesting features arise from the interplay between semi-major axis and eccentricity evolution.

\begin{acknowledgments}
We thank Rohit Chandramouli and Adrien Kuntz for useful discussions on the phase shift caused by ZKL oscillations and from gravitational redshift effects.

J.T. acknowledges support from the Horizon Europe research and innovation programs under the Marie Sk\l{}odowska-Curie grant agreement no. 101203883. This work was supported by the ERC Starting Grant No. 121817–BlackHoleMergs led by Johan Samsing, and by the Villum Fonden grant No. 29466. The Center of Gravity is a Center of Excellence funded by the Danish National Research Foundation under grant No. 184.

This work draws strong inspiration from the paper ``The construction and use of LISA sensitivity curves'' by T. Robson, N. J. Cornish and C. Liu \citep{2019robson}, which provides a similar pedagogical overview of the LISA sensitivity curve.

\end{acknowledgments}


\bibliography{stationaryphase}{}

\begin{thebibliography}{131}%
\makeatletter
\providecommand \@ifxundefined [1]{%
 \@ifx{#1\undefined}
}%
\providecommand \@ifnum [1]{%
 \ifnum #1\expandafter \@firstoftwo
 \else \expandafter \@secondoftwo
 \fi
}%
\providecommand \@ifx [1]{%
 \ifx #1\expandafter \@firstoftwo
 \else \expandafter \@secondoftwo
 \fi
}%
\providecommand \natexlab [1]{#1}%
\providecommand \enquote  [1]{``#1''}%
\providecommand \bibnamefont  [1]{#1}%
\providecommand \bibfnamefont [1]{#1}%
\providecommand \citenamefont [1]{#1}%
\providecommand \href@noop [0]{\@secondoftwo}%
\providecommand \href [0]{\begingroup \@sanitize@url \@href}%
\providecommand \@href[1]{\@@startlink{#1}\@@href}%
\providecommand \@@href[1]{\endgroup#1\@@endlink}%
\providecommand \@sanitize@url [0]{\catcode `\\12\catcode `\$12\catcode `\&12\catcode `\#12\catcode `\^12\catcode `\_12\catcode `\%12\relax}%
\providecommand \@@startlink[1]{}%
\providecommand \@@endlink[0]{}%
\providecommand \url  [0]{\begingroup\@sanitize@url \@url }%
\providecommand \@url [1]{\endgroup\@href {#1}{\urlprefix }}%
\providecommand \urlprefix  [0]{URL }%
\providecommand \Eprint [0]{\href }%
\providecommand \doibase [0]{https://doi.org/}%
\providecommand \selectlanguage [0]{\@gobble}%
\providecommand \bibinfo  [0]{\@secondoftwo}%
\providecommand \bibfield  [0]{\@secondoftwo}%
\providecommand \translation [1]{[#1]}%
\providecommand \BibitemOpen [0]{}%
\providecommand \bibitemStop [0]{}%
\providecommand \bibitemNoStop [0]{.\EOS\space}%
\providecommand \EOS [0]{\spacefactor3000\relax}%
\providecommand \BibitemShut  [1]{\csname bibitem#1\endcsname}%
\let\auto@bib@innerbib\@empty
\bibitem [{\citenamefont {{Chakrabarti}}(1993)}]{1993chakrabarti}%
  \BibitemOpen
  \bibfield  {author} {\bibinfo {author} {\bibfnamefont {S.~K.}\ \bibnamefont {{Chakrabarti}}},\ }\bibfield  {title} {\bibinfo {title} {{Binary Black Holes in Stationary Orbits and a Test of the Active Galactic Nucleus Paradigm}},\ }\href {https://doi.org/10.1086/172863} {\bibfield  {journal} {\bibinfo  {journal} {\apj}\ }\textbf {\bibinfo {volume} {411}},\ \bibinfo {pages} {610} (\bibinfo {year} {1993})}\BibitemShut {NoStop}%
\bibitem [{\citenamefont {{Ryan}}(1995)}]{1995ryan}%
  \BibitemOpen
  \bibfield  {author} {\bibinfo {author} {\bibfnamefont {F.~D.}\ \bibnamefont {{Ryan}}},\ }\bibfield  {title} {\bibinfo {title} {{Gravitational waves from the inspiral of a compact object into a massive, axisymmetric body with arbitrary multipole moments}},\ }\href {https://doi.org/10.1103/PhysRevD.52.5707} {\bibfield  {journal} {\bibinfo  {journal} {\prd}\ }\textbf {\bibinfo {volume} {52}},\ \bibinfo {pages} {5707} (\bibinfo {year} {1995})}\BibitemShut {NoStop}%
\bibitem [{\citenamefont {{Barausse}}\ and\ \citenamefont {{Rezzolla}}(2008)}]{2008barausse}%
  \BibitemOpen
  \bibfield  {author} {\bibinfo {author} {\bibfnamefont {E.}~\bibnamefont {{Barausse}}}\ and\ \bibinfo {author} {\bibfnamefont {L.}~\bibnamefont {{Rezzolla}}},\ }\bibfield  {title} {\bibinfo {title} {{Influence of the hydrodynamic drag from an accretion torus on extreme mass-ratio inspirals}},\ }\href {https://doi.org/10.1103/PhysRevD.77.104027} {\bibfield  {journal} {\bibinfo  {journal} {\prd}\ }\textbf {\bibinfo {volume} {77}},\ \bibinfo {eid} {104027} (\bibinfo {year} {2008})},\ \Eprint {https://arxiv.org/abs/0711.4558} {arXiv:0711.4558 [gr-qc]} \BibitemShut {NoStop}%
\bibitem [{\citenamefont {{Yunes}}\ \emph {et~al.}(2011)\citenamefont {{Yunes}}, \citenamefont {{Miller}},\ and\ \citenamefont {{Thornburg}}}]{2011PhRvD..83d4030Y}%
  \BibitemOpen
  \bibfield  {author} {\bibinfo {author} {\bibfnamefont {N.}~\bibnamefont {{Yunes}}}, \bibinfo {author} {\bibfnamefont {M.~C.}\ \bibnamefont {{Miller}}},\ and\ \bibinfo {author} {\bibfnamefont {J.}~\bibnamefont {{Thornburg}}},\ }\bibfield  {title} {\bibinfo {title} {{Effect of massive perturbers on extreme mass-ratio inspiral waveforms}},\ }\href {https://doi.org/10.1103/PhysRevD.83.044030} {\bibfield  {journal} {\bibinfo  {journal} {\prd}\ }\textbf {\bibinfo {volume} {83}},\ \bibinfo {eid} {044030} (\bibinfo {year} {2011})},\ \Eprint {https://arxiv.org/abs/1010.1721} {arXiv:1010.1721 [astro-ph.GA]} \BibitemShut {NoStop}%
\bibitem [{\citenamefont {{Levin}}(2007)}]{2007levin}%
  \BibitemOpen
  \bibfield  {author} {\bibinfo {author} {\bibfnamefont {Y.}~\bibnamefont {{Levin}}},\ }\bibfield  {title} {\bibinfo {title} {{Starbursts near supermassive black holes: young stars in the Galactic Centre, and gravitational waves in LISA band}},\ }\href {https://doi.org/10.1111/j.1365-2966.2006.11155.x} {\bibfield  {journal} {\bibinfo  {journal} {\mnras}\ }\textbf {\bibinfo {volume} {374}},\ \bibinfo {pages} {515} (\bibinfo {year} {2007})},\ \Eprint {https://arxiv.org/abs/astro-ph/0603583} {arXiv:astro-ph/0603583 [astro-ph]} \BibitemShut {NoStop}%
\bibitem [{\citenamefont {{Kocsis}}\ \emph {et~al.}(2011)\citenamefont {{Kocsis}}, \citenamefont {{Yunes}},\ and\ \citenamefont {{Loeb}}}]{kocsis}%
  \BibitemOpen
  \bibfield  {author} {\bibinfo {author} {\bibfnamefont {B.}~\bibnamefont {{Kocsis}}}, \bibinfo {author} {\bibfnamefont {N.}~\bibnamefont {{Yunes}}},\ and\ \bibinfo {author} {\bibfnamefont {A.}~\bibnamefont {{Loeb}}},\ }\bibfield  {title} {\bibinfo {title} {{Observable signatures of extreme mass-ratio inspiral black hole binaries embedded in thin accretion disks}},\ }\href {https://doi.org/10.1103/PhysRevD.84.024032} {\bibfield  {journal} {\bibinfo  {journal} {\prd}\ }\textbf {\bibinfo {volume} {84}},\ \bibinfo {eid} {024032} (\bibinfo {year} {2011})},\ \Eprint {https://arxiv.org/abs/1104.2322} {arXiv:1104.2322 [astro-ph.GA]} \BibitemShut {NoStop}%
\bibitem [{\citenamefont {{Barausse}}\ \emph {et~al.}(2014)\citenamefont {{Barausse}}, \citenamefont {{Cardoso}},\ and\ \citenamefont {{Pani}}}]{2014barausse}%
  \BibitemOpen
  \bibfield  {author} {\bibinfo {author} {\bibfnamefont {E.}~\bibnamefont {{Barausse}}}, \bibinfo {author} {\bibfnamefont {V.}~\bibnamefont {{Cardoso}}},\ and\ \bibinfo {author} {\bibfnamefont {P.}~\bibnamefont {{Pani}}},\ }\bibfield  {title} {\bibinfo {title} {{Can environmental effects spoil precision gravitational-wave astrophysics?}},\ }\href {https://doi.org/10.1103/PhysRevD.89.104059} {\bibfield  {journal} {\bibinfo  {journal} {\prd}\ }\textbf {\bibinfo {volume} {89}},\ \bibinfo {eid} {104059} (\bibinfo {year} {2014})},\ \Eprint {https://arxiv.org/abs/1404.7149} {arXiv:1404.7149 [gr-qc]} \BibitemShut {NoStop}%
\bibitem [{\citenamefont {{Inayoshi}}\ \emph {et~al.}(2017)\citenamefont {{Inayoshi}}, \citenamefont {{Hirai}}, \citenamefont {{Kinugawa}},\ and\ \citenamefont {{Hotokezaka}}}]{inayoshi2017}%
  \BibitemOpen
  \bibfield  {author} {\bibinfo {author} {\bibfnamefont {K.}~\bibnamefont {{Inayoshi}}}, \bibinfo {author} {\bibfnamefont {R.}~\bibnamefont {{Hirai}}}, \bibinfo {author} {\bibfnamefont {T.}~\bibnamefont {{Kinugawa}}},\ and\ \bibinfo {author} {\bibfnamefont {K.}~\bibnamefont {{Hotokezaka}}},\ }\bibfield  {title} {\bibinfo {title} {{Formation pathway of Population III coalescing binary black holes through stable mass transfer}},\ }\href {https://doi.org/10.1093/mnras/stx757} {\bibfield  {journal} {\bibinfo  {journal} {\mnras}\ }\textbf {\bibinfo {volume} {468}},\ \bibinfo {pages} {5020} (\bibinfo {year} {2017})},\ \Eprint {https://arxiv.org/abs/1701.04823} {arXiv:1701.04823 [astro-ph.HE]} \BibitemShut {NoStop}%
\bibitem [{\citenamefont {{Meiron}}\ \emph {et~al.}(2017)\citenamefont {{Meiron}}, \citenamefont {{Kocsis}},\ and\ \citenamefont {{Loeb}}}]{2017meiron}%
  \BibitemOpen
  \bibfield  {author} {\bibinfo {author} {\bibfnamefont {Y.}~\bibnamefont {{Meiron}}}, \bibinfo {author} {\bibfnamefont {B.}~\bibnamefont {{Kocsis}}},\ and\ \bibinfo {author} {\bibfnamefont {A.}~\bibnamefont {{Loeb}}},\ }\bibfield  {title} {\bibinfo {title} {{Detecting Triple Systems with Gravitational Wave Observations}},\ }\href {https://doi.org/10.3847/1538-4357/834/2/200} {\bibfield  {journal} {\bibinfo  {journal} {\apj}\ }\textbf {\bibinfo {volume} {834}},\ \bibinfo {eid} {200} (\bibinfo {year} {2017})},\ \Eprint {https://arxiv.org/abs/1604.02148} {arXiv:1604.02148 [astro-ph.HE]} \BibitemShut {NoStop}%
\bibitem [{\citenamefont {{Bonetti}}\ \emph {et~al.}(2017)\citenamefont {{Bonetti}}, \citenamefont {{Barausse}}, \citenamefont {{Faye}}, \citenamefont {{Haardt}},\ and\ \citenamefont {{Sesana}}}]{2017Bonetti}%
  \BibitemOpen
  \bibfield  {author} {\bibinfo {author} {\bibfnamefont {M.}~\bibnamefont {{Bonetti}}}, \bibinfo {author} {\bibfnamefont {E.}~\bibnamefont {{Barausse}}}, \bibinfo {author} {\bibfnamefont {G.}~\bibnamefont {{Faye}}}, \bibinfo {author} {\bibfnamefont {F.}~\bibnamefont {{Haardt}}},\ and\ \bibinfo {author} {\bibfnamefont {A.}~\bibnamefont {{Sesana}}},\ }\bibfield  {title} {\bibinfo {title} {{About gravitational-wave generation by a three-body system}},\ }\href {https://doi.org/10.1088/1361-6382/aa8da5} {\bibfield  {journal} {\bibinfo  {journal} {Classical and Quantum Gravity}\ }\textbf {\bibinfo {volume} {34}},\ \bibinfo {eid} {215004} (\bibinfo {year} {2017})},\ \Eprint {https://arxiv.org/abs/1707.04902} {arXiv:1707.04902 [gr-qc]} \BibitemShut {NoStop}%
\bibitem [{\citenamefont {{Torres-Orjuela}}\ \emph {et~al.}(2019)\citenamefont {{Torres-Orjuela}}, \citenamefont {{Chen}}, \citenamefont {{Cao}}, \citenamefont {{Amaro-Seoane}},\ and\ \citenamefont {{Peng}}}]{2019alejandro}%
  \BibitemOpen
  \bibfield  {author} {\bibinfo {author} {\bibfnamefont {A.}~\bibnamefont {{Torres-Orjuela}}}, \bibinfo {author} {\bibfnamefont {X.}~\bibnamefont {{Chen}}}, \bibinfo {author} {\bibfnamefont {Z.}~\bibnamefont {{Cao}}}, \bibinfo {author} {\bibfnamefont {P.}~\bibnamefont {{Amaro-Seoane}}},\ and\ \bibinfo {author} {\bibfnamefont {P.}~\bibnamefont {{Peng}}},\ }\bibfield  {title} {\bibinfo {title} {{Detecting the beaming effect of gravitational waves}},\ }\href {https://doi.org/10.1103/PhysRevD.100.063012} {\bibfield  {journal} {\bibinfo  {journal} {\prd}\ }\textbf {\bibinfo {volume} {100}},\ \bibinfo {eid} {063012} (\bibinfo {year} {2019})},\ \Eprint {https://arxiv.org/abs/1806.09857} {arXiv:1806.09857 [astro-ph.HE]} \BibitemShut {NoStop}%
\bibitem [{\citenamefont {{Randall}}\ and\ \citenamefont {{Xianyu}}(2019)}]{2019randall}%
  \BibitemOpen
  \bibfield  {author} {\bibinfo {author} {\bibfnamefont {L.}~\bibnamefont {{Randall}}}\ and\ \bibinfo {author} {\bibfnamefont {Z.-Z.}\ \bibnamefont {{Xianyu}}},\ }\bibfield  {title} {\bibinfo {title} {{Observing Eccentricity Oscillations of Binary Black Holes in LISA}},\ }\href {https://doi.org/10.48550/arXiv.1902.08604} {\bibfield  {journal} {\bibinfo  {journal} {arXiv e-prints}\ ,\ \bibinfo {eid} {arXiv:1902.08604}} (\bibinfo {year} {2019})},\ \Eprint {https://arxiv.org/abs/1902.08604} {arXiv:1902.08604 [astro-ph.HE]} \BibitemShut {NoStop}%
\bibitem [{\citenamefont {{Cardoso}}\ and\ \citenamefont {{Maselli}}(2020)}]{2020cardoso}%
  \BibitemOpen
  \bibfield  {author} {\bibinfo {author} {\bibfnamefont {V.}~\bibnamefont {{Cardoso}}}\ and\ \bibinfo {author} {\bibfnamefont {A.}~\bibnamefont {{Maselli}}},\ }\bibfield  {title} {\bibinfo {title} {{Constraints on the astrophysical environment of binaries with gravitational-wave observations}},\ }\href {https://doi.org/10.1051/0004-6361/202037654} {\bibfield  {journal} {\bibinfo  {journal} {\aap}\ }\textbf {\bibinfo {volume} {644}},\ \bibinfo {eid} {A147} (\bibinfo {year} {2020})},\ \Eprint {https://arxiv.org/abs/1909.05870} {arXiv:1909.05870 [astro-ph.HE]} \BibitemShut {NoStop}%
\bibitem [{\citenamefont {{D'Orazio}}\ and\ \citenamefont {{Loeb}}(2020)}]{DOrazioGWLens:2020}%
  \BibitemOpen
  \bibfield  {author} {\bibinfo {author} {\bibfnamefont {D.~J.}\ \bibnamefont {{D'Orazio}}}\ and\ \bibinfo {author} {\bibfnamefont {A.}~\bibnamefont {{Loeb}}},\ }\bibfield  {title} {\bibinfo {title} {{Repeated gravitational lensing of gravitational waves in hierarchical black hole triples}},\ }\href {https://doi.org/10.1103/PhysRevD.101.083031} {\bibfield  {journal} {\bibinfo  {journal} {\prd}\ }\textbf {\bibinfo {volume} {101}},\ \bibinfo {eid} {083031} (\bibinfo {year} {2020})},\ \Eprint {https://arxiv.org/abs/1910.02966} {arXiv:1910.02966 [astro-ph.HE]} \BibitemShut {NoStop}%
\bibitem [{\citenamefont {{Liu}}\ \emph {et~al.}(2022)\citenamefont {{Liu}}, \citenamefont {{D'Orazio}}, \citenamefont {{Vigna-G{\'o}mez}},\ and\ \citenamefont {{Samsing}}}]{2022liu}%
  \BibitemOpen
  \bibfield  {author} {\bibinfo {author} {\bibfnamefont {B.}~\bibnamefont {{Liu}}}, \bibinfo {author} {\bibfnamefont {D.~J.}\ \bibnamefont {{D'Orazio}}}, \bibinfo {author} {\bibfnamefont {A.}~\bibnamefont {{Vigna-G{\'o}mez}}},\ and\ \bibinfo {author} {\bibfnamefont {J.}~\bibnamefont {{Samsing}}},\ }\bibfield  {title} {\bibinfo {title} {{Uncovering a hidden black hole binary from secular eccentricity variations of a tertiary star}},\ }\href {https://doi.org/10.1103/PhysRevD.106.123010} {\bibfield  {journal} {\bibinfo  {journal} {\prd}\ }\textbf {\bibinfo {volume} {106}},\ \bibinfo {eid} {123010} (\bibinfo {year} {2022})},\ \Eprint {https://arxiv.org/abs/2207.10091} {arXiv:2207.10091 [astro-ph.HE]} \BibitemShut {NoStop}%
\bibitem [{\citenamefont {{Xuan}}\ \emph {et~al.}(2022)\citenamefont {{Xuan}}, \citenamefont {{Naoz}},\ and\ \citenamefont {{Chen}}}]{2022xuan}%
  \BibitemOpen
  \bibfield  {author} {\bibinfo {author} {\bibfnamefont {Z.}~\bibnamefont {{Xuan}}}, \bibinfo {author} {\bibfnamefont {S.}~\bibnamefont {{Naoz}}},\ and\ \bibinfo {author} {\bibfnamefont {X.}~\bibnamefont {{Chen}}},\ }\bibfield  {title} {\bibinfo {title} {{Detecting Accelerating Eccentric Binaries in the LISA Band}},\ }\href {https://doi.org/10.48550/arXiv.2210.03129} {\bibfield  {journal} {\bibinfo  {journal} {arXiv e-prints}\ ,\ \bibinfo {eid} {arXiv:2210.03129}} (\bibinfo {year} {2022})},\ \Eprint {https://arxiv.org/abs/2210.03129} {arXiv:2210.03129 [astro-ph.HE]} \BibitemShut {NoStop}%
\bibitem [{\citenamefont {{Garg}}\ \emph {et~al.}(2022)\citenamefont {{Garg}}, \citenamefont {{Derdzinski}}, \citenamefont {{Zwick}}, \citenamefont {{Capelo}},\ and\ \citenamefont {{Mayer}}}]{garg2022}%
  \BibitemOpen
  \bibfield  {author} {\bibinfo {author} {\bibfnamefont {M.}~\bibnamefont {{Garg}}}, \bibinfo {author} {\bibfnamefont {A.}~\bibnamefont {{Derdzinski}}}, \bibinfo {author} {\bibfnamefont {L.}~\bibnamefont {{Zwick}}}, \bibinfo {author} {\bibfnamefont {P.~R.}\ \bibnamefont {{Capelo}}},\ and\ \bibinfo {author} {\bibfnamefont {L.}~\bibnamefont {{Mayer}}},\ }\bibfield  {title} {\bibinfo {title} {{The imprint of gas on gravitational waves from LISA intermediate-mass black hole binaries}},\ }\href {https://doi.org/10.1093/mnras/stac2711} {\bibfield  {journal} {\bibinfo  {journal} {\mnras}\ }\textbf {\bibinfo {volume} {517}},\ \bibinfo {pages} {1339} (\bibinfo {year} {2022})},\ \Eprint {https://arxiv.org/abs/2206.05292} {arXiv:2206.05292 [astro-ph.GA]} \BibitemShut {NoStop}%
\bibitem [{\citenamefont {{Cole}}\ \emph {et~al.}(2022)\citenamefont {{Cole}}, \citenamefont {{Coogan}}, \citenamefont {{Kavanagh}},\ and\ \citenamefont {{Bertone}}}]{2022cole}%
  \BibitemOpen
  \bibfield  {author} {\bibinfo {author} {\bibfnamefont {P.~S.}\ \bibnamefont {{Cole}}}, \bibinfo {author} {\bibfnamefont {A.}~\bibnamefont {{Coogan}}}, \bibinfo {author} {\bibfnamefont {B.~J.}\ \bibnamefont {{Kavanagh}}},\ and\ \bibinfo {author} {\bibfnamefont {G.}~\bibnamefont {{Bertone}}},\ }\bibfield  {title} {\bibinfo {title} {{Measuring dark matter spikes around primordial black holes with Einstein Telescope and Cosmic Explorer}},\ }\href@noop {} {\bibfield  {journal} {\bibinfo  {journal} {arXiv e-prints}\ ,\ \bibinfo {eid} {arXiv:2207.07576}} (\bibinfo {year} {2022})},\ \Eprint {https://arxiv.org/abs/2207.07576} {arXiv:2207.07576 [astro-ph.CO]} \BibitemShut {NoStop}%
\bibitem [{\citenamefont {{Chandramouli}}\ and\ \citenamefont {{Yunes}}(2022)}]{2022chandramouli}%
  \BibitemOpen
  \bibfield  {author} {\bibinfo {author} {\bibfnamefont {R.~S.}\ \bibnamefont {{Chandramouli}}}\ and\ \bibinfo {author} {\bibfnamefont {N.}~\bibnamefont {{Yunes}}},\ }\bibfield  {title} {\bibinfo {title} {{Ready-to-use analytic model for gravitational waves from a hierarchical triple with Kozai-Lidov oscillations}},\ }\href {https://doi.org/10.1103/PhysRevD.105.064009} {\bibfield  {journal} {\bibinfo  {journal} {\prd}\ }\textbf {\bibinfo {volume} {105}},\ \bibinfo {eid} {064009} (\bibinfo {year} {2022})},\ \Eprint {https://arxiv.org/abs/2107.00741} {arXiv:2107.00741 [gr-qc]} \BibitemShut {NoStop}%
\bibitem [{\citenamefont {{Sberna}}\ \emph {et~al.}(2022)\citenamefont {{Sberna}}, \citenamefont {{Babak}}, \citenamefont {{Marsat}}, \citenamefont {{Caputo}}, \citenamefont {{Cusin}}, \citenamefont {{Toubiana}}, \citenamefont {{Barausse}}, \citenamefont {{Caprini}}, \citenamefont {{Dal Canton}}, \citenamefont {{Sesana}},\ and\ \citenamefont {{Tamanini}}}]{2022sberna}%
  \BibitemOpen
  \bibfield  {author} {\bibinfo {author} {\bibfnamefont {L.}~\bibnamefont {{Sberna}}}, \bibinfo {author} {\bibfnamefont {S.}~\bibnamefont {{Babak}}}, \bibinfo {author} {\bibfnamefont {S.}~\bibnamefont {{Marsat}}}, \bibinfo {author} {\bibfnamefont {A.}~\bibnamefont {{Caputo}}}, \bibinfo {author} {\bibfnamefont {G.}~\bibnamefont {{Cusin}}}, \bibinfo {author} {\bibfnamefont {A.}~\bibnamefont {{Toubiana}}}, \bibinfo {author} {\bibfnamefont {E.}~\bibnamefont {{Barausse}}}, \bibinfo {author} {\bibfnamefont {C.}~\bibnamefont {{Caprini}}}, \bibinfo {author} {\bibfnamefont {T.}~\bibnamefont {{Dal Canton}}}, \bibinfo {author} {\bibfnamefont {A.}~\bibnamefont {{Sesana}}},\ and\ \bibinfo {author} {\bibfnamefont {N.}~\bibnamefont {{Tamanini}}},\ }\bibfield  {title} {\bibinfo {title} {{Observing GW190521-like binary black holes and their environment with LISA}},\ }\href {https://doi.org/10.1103/PhysRevD.106.064056} {\bibfield  {journal} {\bibinfo  {journal} {\prd}\ }\textbf {\bibinfo {volume} {106}},\ \bibinfo {eid}
  {064056} (\bibinfo {year} {2022})},\ \Eprint {https://arxiv.org/abs/2205.08550} {arXiv:2205.08550 [gr-qc]} \BibitemShut {NoStop}%
\bibitem [{\citenamefont {{Zwick}}\ \emph {et~al.}(2023)\citenamefont {{Zwick}}, \citenamefont {{Capelo}},\ and\ \citenamefont {{Mayer}}}]{2023zwick}%
  \BibitemOpen
  \bibfield  {author} {\bibinfo {author} {\bibfnamefont {L.}~\bibnamefont {{Zwick}}}, \bibinfo {author} {\bibfnamefont {P.~R.}\ \bibnamefont {{Capelo}}},\ and\ \bibinfo {author} {\bibfnamefont {L.}~\bibnamefont {{Mayer}}},\ }\bibfield  {title} {\bibinfo {title} {{Priorities in gravitational waveforms for future space-borne detectors: vacuum accuracy or environment?}},\ }\href {https://doi.org/10.1093/mnras/stad707} {\bibfield  {journal} {\bibinfo  {journal} {\mnras}\ }\textbf {\bibinfo {volume} {521}},\ \bibinfo {pages} {4645} (\bibinfo {year} {2023})},\ \Eprint {https://arxiv.org/abs/2209.04060} {arXiv:2209.04060 [gr-qc]} \BibitemShut {NoStop}%
\bibitem [{\citenamefont {{Tiede}}\ \emph {et~al.}(2024)\citenamefont {{Tiede}}, \citenamefont {{D'Orazio}}, \citenamefont {{Zwick}},\ and\ \citenamefont {{Duffell}}}]{2023Tiede}%
  \BibitemOpen
  \bibfield  {author} {\bibinfo {author} {\bibfnamefont {C.}~\bibnamefont {{Tiede}}}, \bibinfo {author} {\bibfnamefont {D.~J.}\ \bibnamefont {{D'Orazio}}}, \bibinfo {author} {\bibfnamefont {L.}~\bibnamefont {{Zwick}}},\ and\ \bibinfo {author} {\bibfnamefont {P.~C.}\ \bibnamefont {{Duffell}}},\ }\bibfield  {title} {\bibinfo {title} {{Disk-induced Binary Precession: Implications for Dynamics and Multimessenger Observations of Black Hole Binaries}},\ }\href {https://doi.org/10.3847/1538-4357/ad2613} {\bibfield  {journal} {\bibinfo  {journal} {\apj}\ }\textbf {\bibinfo {volume} {964}},\ \bibinfo {eid} {46} (\bibinfo {year} {2024})},\ \Eprint {https://arxiv.org/abs/2312.01805} {arXiv:2312.01805 [astro-ph.HE]} \BibitemShut {NoStop}%
\bibitem [{\citenamefont {{Dyson}}\ \emph {et~al.}(2024)\citenamefont {{Dyson}}, \citenamefont {{Redondo-Yuste}}, \citenamefont {{van de Meent}},\ and\ \citenamefont {{Cardoso}}}]{2024dyson}%
  \BibitemOpen
  \bibfield  {author} {\bibinfo {author} {\bibfnamefont {C.}~\bibnamefont {{Dyson}}}, \bibinfo {author} {\bibfnamefont {J.}~\bibnamefont {{Redondo-Yuste}}}, \bibinfo {author} {\bibfnamefont {M.}~\bibnamefont {{van de Meent}}},\ and\ \bibinfo {author} {\bibfnamefont {V.}~\bibnamefont {{Cardoso}}},\ }\bibfield  {title} {\bibinfo {title} {{Relativistic aerodynamics of spinning black holes}},\ }\href {https://doi.org/10.1103/PhysRevD.109.104038} {\bibfield  {journal} {\bibinfo  {journal} {\prd}\ }\textbf {\bibinfo {volume} {109}},\ \bibinfo {eid} {104038} (\bibinfo {year} {2024})},\ \Eprint {https://arxiv.org/abs/2402.07981} {arXiv:2402.07981 [gr-qc]} \BibitemShut {NoStop}%
\bibitem [{\citenamefont {{Destounis}}\ \emph {et~al.}(2022)\citenamefont {{Destounis}}, \citenamefont {{Kulathingal}}, \citenamefont {{Kokkotas}},\ and\ \citenamefont {{Papadopoulos}}}]{2022destounis}%
  \BibitemOpen
  \bibfield  {author} {\bibinfo {author} {\bibfnamefont {K.}~\bibnamefont {{Destounis}}}, \bibinfo {author} {\bibfnamefont {A.}~\bibnamefont {{Kulathingal}}}, \bibinfo {author} {\bibfnamefont {K.~D.}\ \bibnamefont {{Kokkotas}}},\ and\ \bibinfo {author} {\bibfnamefont {G.~O.}\ \bibnamefont {{Papadopoulos}}},\ }\bibfield  {title} {\bibinfo {title} {{Gravitational-wave imprints of compact and galactic-scale environments in extreme-mass-ratio binaries}},\ }\href {https://doi.org/10.48550/arXiv.2210.09357} {\bibfield  {journal} {\bibinfo  {journal} {arXiv e-prints}\ ,\ \bibinfo {eid} {arXiv:2210.09357}} (\bibinfo {year} {2022})},\ \Eprint {https://arxiv.org/abs/2210.09357} {arXiv:2210.09357 [gr-qc]} \BibitemShut {NoStop}%
\bibitem [{\citenamefont {{Cardoso}}\ \emph {et~al.}(2022)\citenamefont {{Cardoso}}, \citenamefont {{Destounis}}, \citenamefont {{Duque}}, \citenamefont {{Macedo}},\ and\ \citenamefont {{Maselli}}}]{2022cardoso}%
  \BibitemOpen
  \bibfield  {author} {\bibinfo {author} {\bibfnamefont {V.}~\bibnamefont {{Cardoso}}}, \bibinfo {author} {\bibfnamefont {K.}~\bibnamefont {{Destounis}}}, \bibinfo {author} {\bibfnamefont {F.}~\bibnamefont {{Duque}}}, \bibinfo {author} {\bibfnamefont {R.~P.}\ \bibnamefont {{Macedo}}},\ and\ \bibinfo {author} {\bibfnamefont {A.}~\bibnamefont {{Maselli}}},\ }\bibfield  {title} {\bibinfo {title} {{Gravitational Waves from Extreme-Mass-Ratio Systems in Astrophysical Environments}},\ }\href {https://doi.org/10.1103/PhysRevLett.129.241103} {\bibfield  {journal} {\bibinfo  {journal} {\prl}\ }\textbf {\bibinfo {volume} {129}},\ \bibinfo {eid} {241103} (\bibinfo {year} {2022})},\ \Eprint {https://arxiv.org/abs/2210.01133} {arXiv:2210.01133 [gr-qc]} \BibitemShut {NoStop}%
\bibitem [{\citenamefont {{Caputo}}\ \emph {et~al.}(2020)\citenamefont {{Caputo}}, \citenamefont {{Sberna}}, \citenamefont {{Toubiana}}, \citenamefont {{Babak}}, \citenamefont {{Barausse}}, \citenamefont {{Marsat}},\ and\ \citenamefont {{Pani}}}]{2020caputo}%
  \BibitemOpen
  \bibfield  {author} {\bibinfo {author} {\bibfnamefont {A.}~\bibnamefont {{Caputo}}}, \bibinfo {author} {\bibfnamefont {L.}~\bibnamefont {{Sberna}}}, \bibinfo {author} {\bibfnamefont {A.}~\bibnamefont {{Toubiana}}}, \bibinfo {author} {\bibfnamefont {S.}~\bibnamefont {{Babak}}}, \bibinfo {author} {\bibfnamefont {E.}~\bibnamefont {{Barausse}}}, \bibinfo {author} {\bibfnamefont {S.}~\bibnamefont {{Marsat}}},\ and\ \bibinfo {author} {\bibfnamefont {P.}~\bibnamefont {{Pani}}},\ }\bibfield  {title} {\bibinfo {title} {{Gravitational-wave Detection and Parameter Estimation for Accreting Black-hole Binaries and Their Electromagnetic Counterpart}},\ }\href {https://doi.org/10.3847/1538-4357/ab7b66} {\bibfield  {journal} {\bibinfo  {journal} {\apj}\ }\textbf {\bibinfo {volume} {892}},\ \bibinfo {eid} {90} (\bibinfo {year} {2020})},\ \Eprint {https://arxiv.org/abs/2001.03620} {arXiv:2001.03620 [astro-ph.HE]} \BibitemShut {NoStop}%
\bibitem [{\citenamefont {{Zwick}}\ \emph {et~al.}(2024{\natexlab{a}})\citenamefont {{Zwick}}, \citenamefont {{Tiede}}, \citenamefont {{Trani}}, \citenamefont {{Derdzinski}}, \citenamefont {{Haiman}}, \citenamefont {{D'Orazio}},\ and\ \citenamefont {{Samsing}}}]{2024zwicknovel}%
  \BibitemOpen
  \bibfield  {author} {\bibinfo {author} {\bibfnamefont {L.}~\bibnamefont {{Zwick}}}, \bibinfo {author} {\bibfnamefont {C.}~\bibnamefont {{Tiede}}}, \bibinfo {author} {\bibfnamefont {A.~A.}\ \bibnamefont {{Trani}}}, \bibinfo {author} {\bibfnamefont {A.}~\bibnamefont {{Derdzinski}}}, \bibinfo {author} {\bibfnamefont {Z.}~\bibnamefont {{Haiman}}}, \bibinfo {author} {\bibfnamefont {D.~J.}\ \bibnamefont {{D'Orazio}}},\ and\ \bibinfo {author} {\bibfnamefont {J.}~\bibnamefont {{Samsing}}},\ }\bibfield  {title} {\bibinfo {title} {{Novel category of environmental effects on gravitational waves from binaries perturbed by periodic forces}},\ }\href {https://doi.org/10.1103/PhysRevD.110.103005} {\bibfield  {journal} {\bibinfo  {journal} {\prd}\ }\textbf {\bibinfo {volume} {110}},\ \bibinfo {eid} {103005} (\bibinfo {year} {2024}{\natexlab{a}})},\ \Eprint {https://arxiv.org/abs/2405.05698} {arXiv:2405.05698 [gr-qc]} \BibitemShut {NoStop}%
\bibitem [{\citenamefont {{Derdzinski}}\ \emph {et~al.}(2021)\citenamefont {{Derdzinski}}, \citenamefont {{D'Orazio}}, \citenamefont {{Duffell}}, \citenamefont {{Haiman}},\ and\ \citenamefont {{MacFadyen}}}]{Derdzinksi:2021}%
  \BibitemOpen
  \bibfield  {author} {\bibinfo {author} {\bibfnamefont {A.}~\bibnamefont {{Derdzinski}}}, \bibinfo {author} {\bibfnamefont {D.}~\bibnamefont {{D'Orazio}}}, \bibinfo {author} {\bibfnamefont {P.}~\bibnamefont {{Duffell}}}, \bibinfo {author} {\bibfnamefont {Z.}~\bibnamefont {{Haiman}}},\ and\ \bibinfo {author} {\bibfnamefont {A.}~\bibnamefont {{MacFadyen}}},\ }\bibfield  {title} {\bibinfo {title} {{Evolution of gas disc-embedded intermediate mass ratio inspirals in the LISA band}},\ }\href {https://doi.org/10.1093/mnras/staa3976} {\bibfield  {journal} {\bibinfo  {journal} {\mnras}\ }\textbf {\bibinfo {volume} {501}},\ \bibinfo {pages} {3540} (\bibinfo {year} {2021})},\ \Eprint {https://arxiv.org/abs/2005.11333} {arXiv:2005.11333 [astro-ph.HE]} \BibitemShut {NoStop}%
\bibitem [{\citenamefont {{Amaro-Seoane}}\ \emph {et~al.}(2017)\citenamefont {{Amaro-Seoane}}, \citenamefont {{Audley}}, \citenamefont {{Babak}}, \citenamefont {{Baker}}, \citenamefont {{Barausse}} \emph {et~al.}}]{2017lisa}%
  \BibitemOpen
  \bibfield  {author} {\bibinfo {author} {\bibfnamefont {P.}~\bibnamefont {{Amaro-Seoane}}}, \bibinfo {author} {\bibfnamefont {H.}~\bibnamefont {{Audley}}}, \bibinfo {author} {\bibfnamefont {S.}~\bibnamefont {{Babak}}}, \bibinfo {author} {\bibfnamefont {J.}~\bibnamefont {{Baker}}}, \bibinfo {author} {\bibfnamefont {E.}~\bibnamefont {{Barausse}}}, \emph {et~al.},\ }\bibfield  {title} {\bibinfo {title} {{Laser Interferometer Space Antenna}},\ }\href@noop {} {\bibfield  {journal} {\bibinfo  {journal} {arXiv e-prints}\ ,\ \bibinfo {eid} {arXiv:1702.00786}} (\bibinfo {year} {2017})},\ \Eprint {https://arxiv.org/abs/1702.00786} {arXiv:1702.00786 [astro-ph.IM]} \BibitemShut {NoStop}%
\bibitem [{\citenamefont {{Thorpe}}\ \emph {et~al.}(2019)\citenamefont {{Thorpe}}, \citenamefont {{Ziemer}}, \citenamefont {{Thorpe}}, \citenamefont {{Livas}}, \citenamefont {{Conklin}} \emph {et~al.}}]{2019lisa}%
  \BibitemOpen
  \bibfield  {author} {\bibinfo {author} {\bibfnamefont {J.~I.}\ \bibnamefont {{Thorpe}}}, \bibinfo {author} {\bibfnamefont {J.}~\bibnamefont {{Ziemer}}}, \bibinfo {author} {\bibfnamefont {I.}~\bibnamefont {{Thorpe}}}, \bibinfo {author} {\bibfnamefont {J.}~\bibnamefont {{Livas}}}, \bibinfo {author} {\bibfnamefont {J.~W.}\ \bibnamefont {{Conklin}}}, \emph {et~al.},\ }\bibfield  {title} {\bibinfo {title} {{The Laser Interferometer Space Antenna: Unveiling the Millihertz Gravitational Wave Sky}},\ }in\ \href@noop {} {\emph {\bibinfo {booktitle} {Bulletin of the American Astronomical Society}}},\ Vol.~\bibinfo {volume} {51}\ (\bibinfo {year} {2019})\ p.~\bibinfo {pages} {77},\ \Eprint {https://arxiv.org/abs/1907.06482} {arXiv:1907.06482 [astro-ph.IM]} \BibitemShut {NoStop}%
\bibitem [{\citenamefont {{Amaro-Seoane}}\ \emph {et~al.}(2022)\citenamefont {{Amaro-Seoane}}, \citenamefont {{Andrews}}, \citenamefont {{Arca Sedda}}, \citenamefont {{Askar}}, \citenamefont {{Balasov}} \emph {et~al.}}]{2022lisaastro}%
  \BibitemOpen
  \bibfield  {author} {\bibinfo {author} {\bibfnamefont {P.}~\bibnamefont {{Amaro-Seoane}}}, \bibinfo {author} {\bibfnamefont {J.}~\bibnamefont {{Andrews}}}, \bibinfo {author} {\bibfnamefont {M.}~\bibnamefont {{Arca Sedda}}}, \bibinfo {author} {\bibfnamefont {A.}~\bibnamefont {{Askar}}}, \bibinfo {author} {\bibfnamefont {R.}~\bibnamefont {{Balasov}}}, \emph {et~al.},\ }\bibfield  {title} {\bibinfo {title} {{Astrophysics with the Laser Interferometer Space Antenna}},\ }\href@noop {} {\bibfield  {journal} {\bibinfo  {journal} {arXiv e-prints}\ ,\ \bibinfo {eid} {arXiv:2203.06016}} (\bibinfo {year} {2022})},\ \Eprint {https://arxiv.org/abs/2203.06016} {arXiv:2203.06016 [gr-qc]} \BibitemShut {NoStop}%
\bibitem [{\citenamefont {{Colpi}}\ \emph {et~al.}(2024)\citenamefont {{Colpi}}, \citenamefont {{Danzmann}}, \citenamefont {{Hewitson}}, \citenamefont {{Holley-Bockelmann}}, \citenamefont {{Jetzer}} \emph {et~al.}}]{2024lisa}%
  \BibitemOpen
  \bibfield  {author} {\bibinfo {author} {\bibfnamefont {M.}~\bibnamefont {{Colpi}}}, \bibinfo {author} {\bibfnamefont {K.}~\bibnamefont {{Danzmann}}}, \bibinfo {author} {\bibfnamefont {M.}~\bibnamefont {{Hewitson}}}, \bibinfo {author} {\bibfnamefont {K.}~\bibnamefont {{Holley-Bockelmann}}}, \bibinfo {author} {\bibfnamefont {P.}~\bibnamefont {{Jetzer}}}, \emph {et~al.},\ }\bibfield  {title} {\bibinfo {title} {{LISA Definition Study Report}},\ }\href {https://doi.org/10.48550/arXiv.2402.07571} {\bibfield  {journal} {\bibinfo  {journal} {arXiv e-prints}\ ,\ \bibinfo {eid} {arXiv:2402.07571}} (\bibinfo {year} {2024})},\ \Eprint {https://arxiv.org/abs/2402.07571} {arXiv:2402.07571 [astro-ph.CO]} \BibitemShut {NoStop}%
\bibitem [{\citenamefont {{Luo}}\ \emph {et~al.}(2016)\citenamefont {{Luo}}, \citenamefont {{Chen}}, \citenamefont {{Duan}}, \citenamefont {{Gong}}, \citenamefont {{Hu}} \emph {et~al.}}]{TianQin}%
  \BibitemOpen
  \bibfield  {author} {\bibinfo {author} {\bibfnamefont {J.}~\bibnamefont {{Luo}}}, \bibinfo {author} {\bibfnamefont {L.-S.}\ \bibnamefont {{Chen}}}, \bibinfo {author} {\bibfnamefont {H.-Z.}\ \bibnamefont {{Duan}}}, \bibinfo {author} {\bibfnamefont {Y.-G.}\ \bibnamefont {{Gong}}}, \bibinfo {author} {\bibfnamefont {S.}~\bibnamefont {{Hu}}}, \emph {et~al.},\ }\bibfield  {title} {\bibinfo {title} {{TianQin: a space-borne gravitational wave detector}},\ }\href {https://doi.org/10.1088/0264-9381/33/3/035010} {\bibfield  {journal} {\bibinfo  {journal} {Classical and Quantum Gravity}\ }\textbf {\bibinfo {volume} {33}},\ \bibinfo {eid} {035010} (\bibinfo {year} {2016})},\ \Eprint {https://arxiv.org/abs/1512.02076} {arXiv:1512.02076 [astro-ph.IM]} \BibitemShut {NoStop}%
\bibitem [{\citenamefont {{Mei}}\ \emph {et~al.}(2021)\citenamefont {{Mei}}, \citenamefont {{Bai}}, \citenamefont {{Bao}}, \citenamefont {{Barausse}}, \citenamefont {{Cai}} \emph {et~al.}}]{2021tian}%
  \BibitemOpen
  \bibfield  {author} {\bibinfo {author} {\bibfnamefont {J.}~\bibnamefont {{Mei}}}, \bibinfo {author} {\bibfnamefont {Y.-Z.}\ \bibnamefont {{Bai}}}, \bibinfo {author} {\bibfnamefont {J.}~\bibnamefont {{Bao}}}, \bibinfo {author} {\bibfnamefont {E.}~\bibnamefont {{Barausse}}}, \bibinfo {author} {\bibfnamefont {L.}~\bibnamefont {{Cai}}}, \emph {et~al.},\ }\bibfield  {title} {\bibinfo {title} {{The TianQin project: Current progress on science and technology}},\ }\href {https://doi.org/10.1093/ptep/ptaa114} {\bibfield  {journal} {\bibinfo  {journal} {Progr. Theor. Exp. Phys.}\ }\textbf {\bibinfo {volume} {2021}},\ \bibinfo {eid} {05A107} (\bibinfo {year} {2021})},\ \Eprint {https://arxiv.org/abs/2008.10332} {arXiv:2008.10332 [gr-qc]} \BibitemShut {NoStop}%
\bibitem [{\citenamefont {Li}\ \emph {et~al.}(2025)\citenamefont {Li} \emph {et~al.}}]{tianqin2024}%
  \BibitemOpen
  \bibfield  {author} {\bibinfo {author} {\bibfnamefont {E.-K.}\ \bibnamefont {Li}} \emph {et~al.},\ }\bibfield  {title} {\bibinfo {title} {{Gravitational wave astronomy with TianQin}},\ }\href {https://doi.org/10.1088/1361-6633/adc9be} {\bibfield  {journal} {\bibinfo  {journal} {Rept. Prog. Phys.}\ }\textbf {\bibinfo {volume} {88}},\ \bibinfo {pages} {056901} (\bibinfo {year} {2025})},\ \Eprint {https://arxiv.org/abs/2409.19665} {arXiv:2409.19665 [astro-ph.GA]} \BibitemShut {NoStop}%
\bibitem [{\citenamefont {{Zwick}}\ \emph {et~al.}(2025)\citenamefont {{Zwick}}, \citenamefont {{Tak{\'a}tsy}}, \citenamefont {{Saini}}, \citenamefont {{Hendriks}}, \citenamefont {{Samsing}}, \citenamefont {{Tiede}}, \citenamefont {{Rowan}},\ and\ \citenamefont {{Trani}}}]{2025zwick}%
  \BibitemOpen
  \bibfield  {author} {\bibinfo {author} {\bibfnamefont {L.}~\bibnamefont {{Zwick}}}, \bibinfo {author} {\bibfnamefont {J.}~\bibnamefont {{Tak{\'a}tsy}}}, \bibinfo {author} {\bibfnamefont {P.}~\bibnamefont {{Saini}}}, \bibinfo {author} {\bibfnamefont {K.}~\bibnamefont {{Hendriks}}}, \bibinfo {author} {\bibfnamefont {J.}~\bibnamefont {{Samsing}}}, \bibinfo {author} {\bibfnamefont {C.}~\bibnamefont {{Tiede}}}, \bibinfo {author} {\bibfnamefont {C.}~\bibnamefont {{Rowan}}},\ and\ \bibinfo {author} {\bibfnamefont {A.~A.}\ \bibnamefont {{Trani}}},\ }\bibfield  {title} {\bibinfo {title} {{Environmental effects in stellar mass gravitational wave sources I: Expected fraction of signals with significant dephasing in the dynamical and AGN channels}},\ }\href {https://doi.org/10.48550/arXiv.2503.24084} {\bibfield  {journal} {\bibinfo  {journal} {arXiv e-prints}\ ,\ \bibinfo {eid} {arXiv:2503.24084}} (\bibinfo {year} {2025})},\ \Eprint {https://arxiv.org/abs/2503.24084} {arXiv:2503.24084 [astro-ph.HE]} \BibitemShut
  {NoStop}%
\bibitem [{\citenamefont {{Chen}}\ \emph {et~al.}(2020)\citenamefont {{Chen}}, \citenamefont {{Xuan}},\ and\ \citenamefont {{Peng}}}]{2020chen}%
  \BibitemOpen
  \bibfield  {author} {\bibinfo {author} {\bibfnamefont {X.}~\bibnamefont {{Chen}}}, \bibinfo {author} {\bibfnamefont {Z.-Y.}\ \bibnamefont {{Xuan}}},\ and\ \bibinfo {author} {\bibfnamefont {P.}~\bibnamefont {{Peng}}},\ }\bibfield  {title} {\bibinfo {title} {{Fake Massive Black Holes in the Milli-Hertz Gravitational-wave Band}},\ }\href {https://doi.org/10.3847/1538-4357/ab919f} {\bibfield  {journal} {\bibinfo  {journal} {\apj}\ }\textbf {\bibinfo {volume} {896}},\ \bibinfo {eid} {171} (\bibinfo {year} {2020})},\ \Eprint {https://arxiv.org/abs/2003.08639} {arXiv:2003.08639 [astro-ph.HE]} \BibitemShut {NoStop}%
\bibitem [{\citenamefont {{De Luca}}\ \emph {et~al.}(2025)\citenamefont {{De Luca}}, \citenamefont {{Del Grosso}}, \citenamefont {{Iacovelli}}, \citenamefont {{Maselli}},\ and\ \citenamefont {{Berti}}}]{2025delgrosso}%
  \BibitemOpen
  \bibfield  {author} {\bibinfo {author} {\bibfnamefont {V.}~\bibnamefont {{De Luca}}}, \bibinfo {author} {\bibfnamefont {L.}~\bibnamefont {{Del Grosso}}}, \bibinfo {author} {\bibfnamefont {F.}~\bibnamefont {{Iacovelli}}}, \bibinfo {author} {\bibfnamefont {A.}~\bibnamefont {{Maselli}}},\ and\ \bibinfo {author} {\bibfnamefont {E.}~\bibnamefont {{Berti}}},\ }\bibfield  {title} {\bibinfo {title} {{Tainted Love: Systematic biases from ignoring environmental tidal effects in gravitational wave observations}},\ }\href {https://doi.org/10.48550/arXiv.2503.10746} {\bibfield  {journal} {\bibinfo  {journal} {arXiv e-prints}\ ,\ \bibinfo {eid} {arXiv:2503.10746}} (\bibinfo {year} {2025})},\ \Eprint {https://arxiv.org/abs/2503.10746} {arXiv:2503.10746 [gr-qc]} \BibitemShut {NoStop}%
\bibitem [{\citenamefont {{Speri}}\ \emph {et~al.}(2022)\citenamefont {{Speri}}, \citenamefont {{Antonelli}}, \citenamefont {{Sberna}}, \citenamefont {{Babak}}, \citenamefont {{Barausse}}, \citenamefont {{Gair}},\ and\ \citenamefont {{Katz}}}]{2022speri}%
  \BibitemOpen
  \bibfield  {author} {\bibinfo {author} {\bibfnamefont {L.}~\bibnamefont {{Speri}}}, \bibinfo {author} {\bibfnamefont {A.}~\bibnamefont {{Antonelli}}}, \bibinfo {author} {\bibfnamefont {L.}~\bibnamefont {{Sberna}}}, \bibinfo {author} {\bibfnamefont {S.}~\bibnamefont {{Babak}}}, \bibinfo {author} {\bibfnamefont {E.}~\bibnamefont {{Barausse}}}, \bibinfo {author} {\bibfnamefont {J.~R.}\ \bibnamefont {{Gair}}},\ and\ \bibinfo {author} {\bibfnamefont {M.~L.}\ \bibnamefont {{Katz}}},\ }\bibfield  {title} {\bibinfo {title} {{Measuring accretion-disk effects with gravitational waves from extreme mass ratio inspirals}},\ }\href@noop {} {\bibfield  {journal} {\bibinfo  {journal} {arXiv e-prints}\ ,\ \bibinfo {eid} {arXiv:2207.10086}} (\bibinfo {year} {2022})},\ \Eprint {https://arxiv.org/abs/2207.10086} {arXiv:2207.10086 [gr-qc]} \BibitemShut {NoStop}%
\bibitem [{\citenamefont {{Duque}}\ \emph {et~al.}(2024)\citenamefont {{Duque}}, \citenamefont {{Kejriwal}}, \citenamefont {{Sberna}}, \citenamefont {{Speri}},\ and\ \citenamefont {{Gair}}}]{2024duque}%
  \BibitemOpen
  \bibfield  {author} {\bibinfo {author} {\bibfnamefont {F.}~\bibnamefont {{Duque}}}, \bibinfo {author} {\bibfnamefont {S.}~\bibnamefont {{Kejriwal}}}, \bibinfo {author} {\bibfnamefont {L.}~\bibnamefont {{Sberna}}}, \bibinfo {author} {\bibfnamefont {L.}~\bibnamefont {{Speri}}},\ and\ \bibinfo {author} {\bibfnamefont {J.}~\bibnamefont {{Gair}}},\ }\bibfield  {title} {\bibinfo {title} {{Constraining accretion physics with gravitational waves from eccentric extreme-mass-ratio inspirals}},\ }\href {https://doi.org/10.48550/arXiv.2411.03436} {\bibfield  {journal} {\bibinfo  {journal} {arXiv e-prints}\ ,\ \bibinfo {eid} {arXiv:2411.03436}} (\bibinfo {year} {2024})},\ \Eprint {https://arxiv.org/abs/2411.03436} {arXiv:2411.03436 [gr-qc]} \BibitemShut {NoStop}%
\bibitem [{\citenamefont {Dutta~Roy}\ \emph {et~al.}(2025)\citenamefont {Dutta~Roy}, \citenamefont {Mahapatra}, \citenamefont {Samajdar},\ and\ \citenamefont {Arun}}]{DuttaRoy:2025gnu}%
  \BibitemOpen
  \bibfield  {author} {\bibinfo {author} {\bibfnamefont {P.}~\bibnamefont {Dutta~Roy}}, \bibinfo {author} {\bibfnamefont {P.}~\bibnamefont {Mahapatra}}, \bibinfo {author} {\bibfnamefont {A.}~\bibnamefont {Samajdar}},\ and\ \bibinfo {author} {\bibfnamefont {K.~G.}\ \bibnamefont {Arun}},\ }\bibfield  {title} {\bibinfo {title} {{Identifying intermediate mass binary black hole mergers in AGN disks using LISA}},\ }\href {https://doi.org/10.1103/PhysRevD.111.104047} {\bibfield  {journal} {\bibinfo  {journal} {Phys. Rev. D}\ }\textbf {\bibinfo {volume} {111}},\ \bibinfo {pages} {104047} (\bibinfo {year} {2025})},\ \Eprint {https://arxiv.org/abs/2503.11721} {arXiv:2503.11721 [astro-ph.HE]} \BibitemShut {NoStop}%
\bibitem [{\citenamefont {{Eda}}\ \emph {et~al.}(2013)\citenamefont {{Eda}}, \citenamefont {{Itoh}}, \citenamefont {{Kuroyanagi}},\ and\ \citenamefont {{Silk}}}]{2013kazunari}%
  \BibitemOpen
  \bibfield  {author} {\bibinfo {author} {\bibfnamefont {K.}~\bibnamefont {{Eda}}}, \bibinfo {author} {\bibfnamefont {Y.}~\bibnamefont {{Itoh}}}, \bibinfo {author} {\bibfnamefont {S.}~\bibnamefont {{Kuroyanagi}}},\ and\ \bibinfo {author} {\bibfnamefont {J.}~\bibnamefont {{Silk}}},\ }\bibfield  {title} {\bibinfo {title} {{New Probe of Dark-Matter Properties: Gravitational Waves from an Intermediate-Mass Black Hole Embedded in a Dark-Matter Minispike}},\ }\href {https://doi.org/10.1103/PhysRevLett.110.221101} {\bibfield  {journal} {\bibinfo  {journal} {\prl}\ }\textbf {\bibinfo {volume} {110}},\ \bibinfo {eid} {221101} (\bibinfo {year} {2013})},\ \Eprint {https://arxiv.org/abs/1301.5971} {arXiv:1301.5971 [gr-qc]} \BibitemShut {NoStop}%
\bibitem [{\citenamefont {{Eda}}\ \emph {et~al.}(2015)\citenamefont {{Eda}}, \citenamefont {{Itoh}}, \citenamefont {{Kuroyanagi}},\ and\ \citenamefont {{Silk}}}]{2015kazunari}%
  \BibitemOpen
  \bibfield  {author} {\bibinfo {author} {\bibfnamefont {K.}~\bibnamefont {{Eda}}}, \bibinfo {author} {\bibfnamefont {Y.}~\bibnamefont {{Itoh}}}, \bibinfo {author} {\bibfnamefont {S.}~\bibnamefont {{Kuroyanagi}}},\ and\ \bibinfo {author} {\bibfnamefont {J.}~\bibnamefont {{Silk}}},\ }\bibfield  {title} {\bibinfo {title} {{Gravitational waves as a probe of dark matter minispikes}},\ }\href {https://doi.org/10.1103/PhysRevD.91.044045} {\bibfield  {journal} {\bibinfo  {journal} {\prd}\ }\textbf {\bibinfo {volume} {91}},\ \bibinfo {eid} {044045} (\bibinfo {year} {2015})},\ \Eprint {https://arxiv.org/abs/1408.3534} {arXiv:1408.3534 [gr-qc]} \BibitemShut {NoStop}%
\bibitem [{\citenamefont {{Coogan}}\ \emph {et~al.}(2022)\citenamefont {{Coogan}}, \citenamefont {{Bertone}}, \citenamefont {{Gaggero}}, \citenamefont {{Kavanagh}},\ and\ \citenamefont {{Nichols}}}]{2022coogan}%
  \BibitemOpen
  \bibfield  {author} {\bibinfo {author} {\bibfnamefont {A.}~\bibnamefont {{Coogan}}}, \bibinfo {author} {\bibfnamefont {G.}~\bibnamefont {{Bertone}}}, \bibinfo {author} {\bibfnamefont {D.}~\bibnamefont {{Gaggero}}}, \bibinfo {author} {\bibfnamefont {B.~J.}\ \bibnamefont {{Kavanagh}}},\ and\ \bibinfo {author} {\bibfnamefont {D.~A.}\ \bibnamefont {{Nichols}}},\ }\bibfield  {title} {\bibinfo {title} {{Measuring the dark matter environments of black hole binaries with gravitational waves}},\ }\href {https://doi.org/10.1103/PhysRevD.105.043009} {\bibfield  {journal} {\bibinfo  {journal} {\prd}\ }\textbf {\bibinfo {volume} {105}},\ \bibinfo {eid} {043009} (\bibinfo {year} {2022})},\ \Eprint {https://arxiv.org/abs/2108.04154} {arXiv:2108.04154 [gr-qc]} \BibitemShut {NoStop}%
\bibitem [{\citenamefont {{Figueiredo}}\ \emph {et~al.}(2023)\citenamefont {{Figueiredo}}, \citenamefont {{Maselli}},\ and\ \citenamefont {{Cardoso}}}]{2023figureido}%
  \BibitemOpen
  \bibfield  {author} {\bibinfo {author} {\bibfnamefont {E.}~\bibnamefont {{Figueiredo}}}, \bibinfo {author} {\bibfnamefont {A.}~\bibnamefont {{Maselli}}},\ and\ \bibinfo {author} {\bibfnamefont {V.}~\bibnamefont {{Cardoso}}},\ }\bibfield  {title} {\bibinfo {title} {{Black holes surrounded by generic dark matter profiles: Appearance and gravitational-wave emission}},\ }\href {https://doi.org/10.1103/PhysRevD.107.104033} {\bibfield  {journal} {\bibinfo  {journal} {\prd}\ }\textbf {\bibinfo {volume} {107}},\ \bibinfo {eid} {104033} (\bibinfo {year} {2023})},\ \Eprint {https://arxiv.org/abs/2303.08183} {arXiv:2303.08183 [gr-qc]} \BibitemShut {NoStop}%
\bibitem [{\citenamefont {Tiwari}\ \emph {et~al.}(2023)\citenamefont {Tiwari}, \citenamefont {Vijaykumar}, \citenamefont {Kapadia}, \citenamefont {Fragione},\ and\ \citenamefont {Chatterjee}}]{Tiwari:2023cpa}%
  \BibitemOpen
  \bibfield  {author} {\bibinfo {author} {\bibfnamefont {A.}~\bibnamefont {Tiwari}}, \bibinfo {author} {\bibfnamefont {A.}~\bibnamefont {Vijaykumar}}, \bibinfo {author} {\bibfnamefont {S.~J.}\ \bibnamefont {Kapadia}}, \bibinfo {author} {\bibfnamefont {G.}~\bibnamefont {Fragione}},\ and\ \bibinfo {author} {\bibfnamefont {S.}~\bibnamefont {Chatterjee}},\ }\bibfield  {title} {\bibinfo {title} {{Accelerated binary black holes in globular clusters: forecasts and detectability in the era of space-based gravitational-wave detectors}},\ }\href {https://doi.org/10.1093/mnras/stad3749} {\bibfield  {journal} {\bibinfo  {journal} {Mon. Not. Roy. Astron. Soc.}\ }\textbf {\bibinfo {volume} {527}},\ \bibinfo {pages} {8586} (\bibinfo {year} {2023})},\ \Eprint {https://arxiv.org/abs/2307.00930} {arXiv:2307.00930 [astro-ph.HE]} \BibitemShut {NoStop}%
\bibitem [{\citenamefont {Vijaykumar}\ \emph {et~al.}(2023)\citenamefont {Vijaykumar}, \citenamefont {Tiwari}, \citenamefont {Kapadia}, \citenamefont {Arun},\ and\ \citenamefont {Ajith}}]{Vijaykumar:2023tjg}%
  \BibitemOpen
  \bibfield  {author} {\bibinfo {author} {\bibfnamefont {A.}~\bibnamefont {Vijaykumar}}, \bibinfo {author} {\bibfnamefont {A.}~\bibnamefont {Tiwari}}, \bibinfo {author} {\bibfnamefont {S.~J.}\ \bibnamefont {Kapadia}}, \bibinfo {author} {\bibfnamefont {K.~G.}\ \bibnamefont {Arun}},\ and\ \bibinfo {author} {\bibfnamefont {P.}~\bibnamefont {Ajith}},\ }\bibfield  {title} {\bibinfo {title} {{Waltzing Binaries: Probing the Line-of-sight Acceleration of Merging Compact Objects with Gravitational Waves}},\ }\href {https://doi.org/10.3847/1538-4357/acd77d} {\bibfield  {journal} {\bibinfo  {journal} {Astrophys. J.}\ }\textbf {\bibinfo {volume} {954}},\ \bibinfo {pages} {105} (\bibinfo {year} {2023})},\ \Eprint {https://arxiv.org/abs/2302.09651} {arXiv:2302.09651 [astro-ph.HE]} \BibitemShut {NoStop}%
\bibitem [{\citenamefont {Tiwari}\ \emph {et~al.}(2024)\citenamefont {Tiwari}, \citenamefont {Vijaykumar}, \citenamefont {Kapadia}, \citenamefont {Chatterjee},\ and\ \citenamefont {Fragione}}]{Tiwari:2024pvb}%
  \BibitemOpen
  \bibfield  {author} {\bibinfo {author} {\bibfnamefont {A.}~\bibnamefont {Tiwari}}, \bibinfo {author} {\bibfnamefont {A.}~\bibnamefont {Vijaykumar}}, \bibinfo {author} {\bibfnamefont {S.~J.}\ \bibnamefont {Kapadia}}, \bibinfo {author} {\bibfnamefont {S.}~\bibnamefont {Chatterjee}},\ and\ \bibinfo {author} {\bibfnamefont {G.}~\bibnamefont {Fragione}},\ }\bibfield  {title} {\bibinfo {title} {{Profiling stellar environments of gravitational wave sources}},\ }\href@noop {} {\bibfield  {journal} {\bibinfo  {journal} {arXiv e-prints}\ } (\bibinfo {year} {2024})},\ \Eprint {https://arxiv.org/abs/2407.15117} {arXiv:2407.15117 [astro-ph.HE]} \BibitemShut {NoStop}%
\bibitem [{\citenamefont {{Samsing}}\ \emph {et~al.}(2024)\citenamefont {{Samsing}}, \citenamefont {{Hendriks}}, \citenamefont {{Zwick}}, \citenamefont {{D'Orazio}},\ and\ \citenamefont {{Liu}}}]{2024samsing}%
  \BibitemOpen
  \bibfield  {author} {\bibinfo {author} {\bibfnamefont {J.}~\bibnamefont {{Samsing}}}, \bibinfo {author} {\bibfnamefont {K.}~\bibnamefont {{Hendriks}}}, \bibinfo {author} {\bibfnamefont {L.}~\bibnamefont {{Zwick}}}, \bibinfo {author} {\bibfnamefont {D.~J.}\ \bibnamefont {{D'Orazio}}},\ and\ \bibinfo {author} {\bibfnamefont {B.}~\bibnamefont {{Liu}}},\ }\bibfield  {title} {\bibinfo {title} {{Gravitational Wave Phase Shifts in Eccentric Black Hole Mergers as a Probe of Dynamical Formation Environments}},\ }\href {https://doi.org/10.48550/arXiv.2403.05625} {\bibfield  {journal} {\bibinfo  {journal} {arXiv e-prints}\ ,\ \bibinfo {eid} {arXiv:2403.05625}} (\bibinfo {year} {2024})},\ \Eprint {https://arxiv.org/abs/2403.05625} {arXiv:2403.05625 [astro-ph.HE]} \BibitemShut {NoStop}%
\bibitem [{\citenamefont {{Hendriks}}\ \emph {et~al.}(2024{\natexlab{a}})\citenamefont {{Hendriks}}, \citenamefont {{Atallah}}, \citenamefont {{Martinez}}, \citenamefont {{Zevin}}, \citenamefont {{Zwick}}, \citenamefont {{Trani}}, \citenamefont {{Saini}}, \citenamefont {{Tak{\'a}tsy}},\ and\ \citenamefont {{Samsing}}}]{kai2024}%
  \BibitemOpen
  \bibfield  {author} {\bibinfo {author} {\bibfnamefont {K.}~\bibnamefont {{Hendriks}}}, \bibinfo {author} {\bibfnamefont {D.}~\bibnamefont {{Atallah}}}, \bibinfo {author} {\bibfnamefont {M.}~\bibnamefont {{Martinez}}}, \bibinfo {author} {\bibfnamefont {M.}~\bibnamefont {{Zevin}}}, \bibinfo {author} {\bibfnamefont {L.}~\bibnamefont {{Zwick}}}, \bibinfo {author} {\bibfnamefont {A.~A.}\ \bibnamefont {{Trani}}}, \bibinfo {author} {\bibfnamefont {P.}~\bibnamefont {{Saini}}}, \bibinfo {author} {\bibfnamefont {J.}~\bibnamefont {{Tak{\'a}tsy}}},\ and\ \bibinfo {author} {\bibfnamefont {J.}~\bibnamefont {{Samsing}}},\ }\bibfield  {title} {\bibinfo {title} {{Large Gravitational Wave Phase Shifts from Strong 3-body Interactions in Dense Stellar Clusters}},\ }\href {https://doi.org/10.48550/arXiv.2411.08572} {\bibfield  {journal} {\bibinfo  {journal} {arXiv e-prints}\ ,\ \bibinfo {eid} {arXiv:2411.08572}} (\bibinfo {year} {2024}{\natexlab{a}})},\ \Eprint {https://arxiv.org/abs/2411.08572} {arXiv:2411.08572 [astro-ph.HE]}
  \BibitemShut {NoStop}%
\bibitem [{\citenamefont {{Hendriks}}\ \emph {et~al.}(2024{\natexlab{b}})\citenamefont {{Hendriks}}, \citenamefont {{Zwick}},\ and\ \citenamefont {{Samsing}}}]{kai22024}%
  \BibitemOpen
  \bibfield  {author} {\bibinfo {author} {\bibfnamefont {K.}~\bibnamefont {{Hendriks}}}, \bibinfo {author} {\bibfnamefont {L.}~\bibnamefont {{Zwick}}},\ and\ \bibinfo {author} {\bibfnamefont {J.}~\bibnamefont {{Samsing}}},\ }\bibfield  {title} {\bibinfo {title} {{Eccentric features in the gravitational wave phase of dynamically formed black hole binaries}},\ }\href {https://doi.org/10.48550/arXiv.2408.04603} {\bibfield  {journal} {\bibinfo  {journal} {arXiv e-prints}\ ,\ \bibinfo {eid} {arXiv:2408.04603}} (\bibinfo {year} {2024}{\natexlab{b}})},\ \Eprint {https://arxiv.org/abs/2408.04603} {arXiv:2408.04603 [gr-qc]} \BibitemShut {NoStop}%
\bibitem [{\citenamefont {{Robson}}\ \emph {et~al.}(2018)\citenamefont {{Robson}}, \citenamefont {{Cornish}}, \citenamefont {{Tamanini}},\ and\ \citenamefont {{Toonen}}}]{2018PhRvD..98f4012R}%
  \BibitemOpen
  \bibfield  {author} {\bibinfo {author} {\bibfnamefont {T.}~\bibnamefont {{Robson}}}, \bibinfo {author} {\bibfnamefont {N.~J.}\ \bibnamefont {{Cornish}}}, \bibinfo {author} {\bibfnamefont {N.}~\bibnamefont {{Tamanini}}},\ and\ \bibinfo {author} {\bibfnamefont {S.}~\bibnamefont {{Toonen}}},\ }\bibfield  {title} {\bibinfo {title} {{Detecting hierarchical stellar systems with LISA}},\ }\href {https://doi.org/10.1103/PhysRevD.98.064012} {\bibfield  {journal} {\bibinfo  {journal} {\prd}\ }\textbf {\bibinfo {volume} {98}},\ \bibinfo {eid} {064012} (\bibinfo {year} {2018})},\ \Eprint {https://arxiv.org/abs/1806.00500} {arXiv:1806.00500 [gr-qc]} \BibitemShut {NoStop}%
\bibitem [{\citenamefont {Abbott}\ \emph {et~al.}(2019)\citenamefont {Abbott} \emph {et~al.}}]{LIGOScientific:2018hze}%
  \BibitemOpen
  \bibfield  {author} {\bibinfo {author} {\bibfnamefont {B.~P.}\ \bibnamefont {Abbott}} \emph {et~al.} (\bibinfo {collaboration} {LIGO Scientific, Virgo}),\ }\bibfield  {title} {\bibinfo {title} {{Properties of the binary neutron star merger GW170817}},\ }\href {https://doi.org/10.1103/PhysRevX.9.011001} {\bibfield  {journal} {\bibinfo  {journal} {Phys. Rev. X}\ }\textbf {\bibinfo {volume} {9}},\ \bibinfo {pages} {011001} (\bibinfo {year} {2019})},\ \Eprint {https://arxiv.org/abs/1805.11579} {arXiv:1805.11579 [gr-qc]} \BibitemShut {NoStop}%
\bibitem [{\citenamefont {Abbott}\ \emph {et~al.}(2018)\citenamefont {Abbott} \emph {et~al.}}]{LIGOScientific:2018cki}%
  \BibitemOpen
  \bibfield  {author} {\bibinfo {author} {\bibfnamefont {B.~P.}\ \bibnamefont {Abbott}} \emph {et~al.} (\bibinfo {collaboration} {LIGO Scientific, Virgo}),\ }\bibfield  {title} {\bibinfo {title} {{GW170817: Measurements of neutron star radii and equation of state}},\ }\href {https://doi.org/10.1103/PhysRevLett.121.161101} {\bibfield  {journal} {\bibinfo  {journal} {Phys. Rev. Lett.}\ }\textbf {\bibinfo {volume} {121}},\ \bibinfo {pages} {161101} (\bibinfo {year} {2018})},\ \Eprint {https://arxiv.org/abs/1805.11581} {arXiv:1805.11581 [gr-qc]} \BibitemShut {NoStop}%
\bibitem [{\citenamefont {De}\ \emph {et~al.}(2018)\citenamefont {De}, \citenamefont {Finstad}, \citenamefont {Lattimer}, \citenamefont {Brown}, \citenamefont {Berger},\ and\ \citenamefont {Biwer}}]{De:2018uhw}%
  \BibitemOpen
  \bibfield  {author} {\bibinfo {author} {\bibfnamefont {S.}~\bibnamefont {De}}, \bibinfo {author} {\bibfnamefont {D.}~\bibnamefont {Finstad}}, \bibinfo {author} {\bibfnamefont {J.~M.}\ \bibnamefont {Lattimer}}, \bibinfo {author} {\bibfnamefont {D.~A.}\ \bibnamefont {Brown}}, \bibinfo {author} {\bibfnamefont {E.}~\bibnamefont {Berger}},\ and\ \bibinfo {author} {\bibfnamefont {C.~M.}\ \bibnamefont {Biwer}},\ }\bibfield  {title} {\bibinfo {title} {{Tidal Deformabilities and Radii of Neutron Stars from the Observation of GW170817}},\ }\href {https://doi.org/10.1103/PhysRevLett.121.091102} {\bibfield  {journal} {\bibinfo  {journal} {Phys. Rev. Lett.}\ }\textbf {\bibinfo {volume} {121}},\ \bibinfo {pages} {091102} (\bibinfo {year} {2018})},\ \bibinfo {note} {[Erratum: Phys.Rev.Lett. 121, 259902 (2018)]},\ \Eprint {https://arxiv.org/abs/1804.08583} {arXiv:1804.08583 [astro-ph.HE]} \BibitemShut {NoStop}%
\bibitem [{\citenamefont {Annala}\ \emph {et~al.}(2022)\citenamefont {Annala}, \citenamefont {Gorda}, \citenamefont {Katerini}, \citenamefont {Kurkela}, \citenamefont {N\"attil\"a}, \citenamefont {Paschalidis},\ and\ \citenamefont {Vuorinen}}]{Annala:2021gom}%
  \BibitemOpen
  \bibfield  {author} {\bibinfo {author} {\bibfnamefont {E.}~\bibnamefont {Annala}}, \bibinfo {author} {\bibfnamefont {T.}~\bibnamefont {Gorda}}, \bibinfo {author} {\bibfnamefont {E.}~\bibnamefont {Katerini}}, \bibinfo {author} {\bibfnamefont {A.}~\bibnamefont {Kurkela}}, \bibinfo {author} {\bibfnamefont {J.}~\bibnamefont {N\"attil\"a}}, \bibinfo {author} {\bibfnamefont {V.}~\bibnamefont {Paschalidis}},\ and\ \bibinfo {author} {\bibfnamefont {A.}~\bibnamefont {Vuorinen}},\ }\bibfield  {title} {\bibinfo {title} {{Multimessenger Constraints for Ultradense Matter}},\ }\href {https://doi.org/10.1103/PhysRevX.12.011058} {\bibfield  {journal} {\bibinfo  {journal} {Phys. Rev. X}\ }\textbf {\bibinfo {volume} {12}},\ \bibinfo {pages} {011058} (\bibinfo {year} {2022})},\ \Eprint {https://arxiv.org/abs/2105.05132} {arXiv:2105.05132 [astro-ph.HE]} \BibitemShut {NoStop}%
\bibitem [{\citenamefont {Takatsy}\ \emph {et~al.}(2023)\citenamefont {Takatsy}, \citenamefont {Kovacs}, \citenamefont {Wolf},\ and\ \citenamefont {Schaffner-Bielich}}]{Takatsy:2023xzf}%
  \BibitemOpen
  \bibfield  {author} {\bibinfo {author} {\bibfnamefont {J.}~\bibnamefont {Takatsy}}, \bibinfo {author} {\bibfnamefont {P.}~\bibnamefont {Kovacs}}, \bibinfo {author} {\bibfnamefont {G.}~\bibnamefont {Wolf}},\ and\ \bibinfo {author} {\bibfnamefont {J.}~\bibnamefont {Schaffner-Bielich}},\ }\bibfield  {title} {\bibinfo {title} {{What neutron stars tell about the hadron-quark phase transition: A Bayesian study}},\ }\href {https://doi.org/10.1103/PhysRevD.108.043002} {\bibfield  {journal} {\bibinfo  {journal} {Phys. Rev. D}\ }\textbf {\bibinfo {volume} {108}},\ \bibinfo {pages} {043002} (\bibinfo {year} {2023})},\ \Eprint {https://arxiv.org/abs/2303.00013} {arXiv:2303.00013 [astro-ph.HE]} \BibitemShut {NoStop}%
\bibitem [{\citenamefont {Tak\'atsy}\ \emph {et~al.}(2024)\citenamefont {Tak\'atsy}, \citenamefont {Kocsis},\ and\ \citenamefont {Kov\'acs}}]{Takatsy:2024sin}%
  \BibitemOpen
  \bibfield  {author} {\bibinfo {author} {\bibfnamefont {J.}~\bibnamefont {Tak\'atsy}}, \bibinfo {author} {\bibfnamefont {B.}~\bibnamefont {Kocsis}},\ and\ \bibinfo {author} {\bibfnamefont {P.}~\bibnamefont {Kov\'acs}},\ }\bibfield  {title} {\bibinfo {title} {{Observability of dynamical tides in merging eccentric neutron star binaries}},\ }\href {https://doi.org/10.1103/PhysRevD.110.103043} {\bibfield  {journal} {\bibinfo  {journal} {Phys. Rev. D}\ }\textbf {\bibinfo {volume} {110}},\ \bibinfo {pages} {103043} (\bibinfo {year} {2024})},\ \Eprint {https://arxiv.org/abs/2407.17560} {arXiv:2407.17560 [astro-ph.HE]} \BibitemShut {NoStop}%
\bibitem [{\citenamefont {{Zwick}}\ \emph {et~al.}(2024{\natexlab{b}})\citenamefont {{Zwick}}, \citenamefont {{Tiede}}, \citenamefont {{Trani}}, \citenamefont {{Derdzinski}}, \citenamefont {{Haiman}}, \citenamefont {{D'Orazio}},\ and\ \citenamefont {{Samsing}}}]{2024zwick}%
  \BibitemOpen
  \bibfield  {author} {\bibinfo {author} {\bibfnamefont {L.}~\bibnamefont {{Zwick}}}, \bibinfo {author} {\bibfnamefont {C.}~\bibnamefont {{Tiede}}}, \bibinfo {author} {\bibfnamefont {A.~A.}\ \bibnamefont {{Trani}}}, \bibinfo {author} {\bibfnamefont {A.}~\bibnamefont {{Derdzinski}}}, \bibinfo {author} {\bibfnamefont {Z.}~\bibnamefont {{Haiman}}}, \bibinfo {author} {\bibfnamefont {D.~J.}\ \bibnamefont {{D'Orazio}}},\ and\ \bibinfo {author} {\bibfnamefont {J.}~\bibnamefont {{Samsing}}},\ }\bibfield  {title} {\bibinfo {title} {{Novel category of environmental effects on gravitational waves from binaries perturbed by periodic forces}},\ }\href {https://doi.org/10.1103/PhysRevD.110.103005} {\bibfield  {journal} {\bibinfo  {journal} {\prd}\ }\textbf {\bibinfo {volume} {110}},\ \bibinfo {eid} {103005} (\bibinfo {year} {2024}{\natexlab{b}})},\ \Eprint {https://arxiv.org/abs/2405.05698} {arXiv:2405.05698 [gr-qc]} \BibitemShut {NoStop}%
\bibitem [{\citenamefont {{Cutler}}\ and\ \citenamefont {{Flanagan}}(1994)}]{1994cutler}%
  \BibitemOpen
  \bibfield  {author} {\bibinfo {author} {\bibfnamefont {C.}~\bibnamefont {{Cutler}}}\ and\ \bibinfo {author} {\bibfnamefont {{\'E}.~E.}\ \bibnamefont {{Flanagan}}},\ }\bibfield  {title} {\bibinfo {title} {{Gravitational waves from merging compact binaries: How accurately can one extract the binary's parameters from the inspiral waveform\textbackslash?}},\ }\href {https://doi.org/10.1103/PhysRevD.49.2658} {\bibfield  {journal} {\bibinfo  {journal} {\prd}\ }\textbf {\bibinfo {volume} {49}},\ \bibinfo {pages} {2658} (\bibinfo {year} {1994})},\ \Eprint {https://arxiv.org/abs/gr-qc/9402014} {arXiv:gr-qc/9402014 [gr-qc]} \BibitemShut {NoStop}%
\bibitem [{\citenamefont {{Blanchet}}(2014)}]{blanchet2014}%
  \BibitemOpen
  \bibfield  {author} {\bibinfo {author} {\bibfnamefont {L.}~\bibnamefont {{Blanchet}}},\ }\bibfield  {title} {\bibinfo {title} {{Gravitational Radiation from Post-Newtonian Sources and Inspiralling Compact Binaries}},\ }\href {https://doi.org/10.12942/lrr-2014-2} {\bibfield  {journal} {\bibinfo  {journal} {Living Reviews in Relativity}\ }\textbf {\bibinfo {volume} {17}},\ \bibinfo {eid} {2} (\bibinfo {year} {2014})},\ \Eprint {https://arxiv.org/abs/1310.1528} {arXiv:1310.1528 [gr-qc]} \BibitemShut {NoStop}%
\bibitem [{\citenamefont {{Maggiore}}(2018)}]{2018maggiore}%
  \BibitemOpen
  \bibfield  {author} {\bibinfo {author} {\bibfnamefont {M.}~\bibnamefont {{Maggiore}}},\ }\href {https://doi.org/10.1093/oso/9780198570899.001.0001} {\emph {\bibinfo {title} {{Gravitational Waves: Volume 2: Astrophysics and Cosmology}}}}\ (\bibinfo {year} {2018})\BibitemShut {NoStop}%
\bibitem [{\citenamefont {{Damour}}\ and\ \citenamefont {{Deruelle}}(1985)}]{1985damour}%
  \BibitemOpen
  \bibfield  {author} {\bibinfo {author} {\bibfnamefont {T.}~\bibnamefont {{Damour}}}\ and\ \bibinfo {author} {\bibfnamefont {N.}~\bibnamefont {{Deruelle}}},\ }\bibfield  {title} {\bibinfo {title} {{General relativistic celestial mechanics of binary systems. I. The post-Newtonian motion.}},\ }\href@noop {} {\bibfield  {journal} {\bibinfo  {journal} {Ann. Inst. Henri Poincar{\'e} Phys. Th{\'e}or}\ }\textbf {\bibinfo {volume} {43}},\ \bibinfo {pages} {107} (\bibinfo {year} {1985})}\BibitemShut {NoStop}%
\bibitem [{\citenamefont {{Damour}}\ \emph {et~al.}(1991)\citenamefont {{Damour}}, \citenamefont {{Soffel}},\ and\ \citenamefont {{Xu}}}]{1991damour}%
  \BibitemOpen
  \bibfield  {author} {\bibinfo {author} {\bibfnamefont {T.}~\bibnamefont {{Damour}}}, \bibinfo {author} {\bibfnamefont {M.}~\bibnamefont {{Soffel}}},\ and\ \bibinfo {author} {\bibfnamefont {C.}~\bibnamefont {{Xu}}},\ }\bibfield  {title} {\bibinfo {title} {{General-relativistic celestial mechanics. I. Method and definition of reference systems}},\ }\href {https://doi.org/10.1103/PhysRevD.43.3273} {\bibfield  {journal} {\bibinfo  {journal} {\prd}\ }\textbf {\bibinfo {volume} {43}},\ \bibinfo {pages} {3273} (\bibinfo {year} {1991})}\BibitemShut {NoStop}%
\bibitem [{\citenamefont {{Jaranowski}}\ and\ \citenamefont {{Sch{\"a}fer}}(1998)}]{1998jaranowski}%
  \BibitemOpen
  \bibfield  {author} {\bibinfo {author} {\bibfnamefont {P.}~\bibnamefont {{Jaranowski}}}\ and\ \bibinfo {author} {\bibfnamefont {G.}~\bibnamefont {{Sch{\"a}fer}}},\ }\bibfield  {title} {\bibinfo {title} {{Third post-Newtonian higher order ADM Hamilton dynamics for two-body point-mass systems}},\ }\href {https://doi.org/10.1103/PhysRevD.57.7274} {\bibfield  {journal} {\bibinfo  {journal} {\prd}\ }\textbf {\bibinfo {volume} {57}},\ \bibinfo {pages} {7274} (\bibinfo {year} {1998})},\ \Eprint {https://arxiv.org/abs/gr-qc/9712075} {arXiv:gr-qc/9712075 [gr-qc]} \BibitemShut {NoStop}%
\bibitem [{\citenamefont {{Jaranowski}}\ and\ \citenamefont {{Sch{\"a}fer}}(1999)}]{1999jaranowski}%
  \BibitemOpen
  \bibfield  {author} {\bibinfo {author} {\bibfnamefont {P.}~\bibnamefont {{Jaranowski}}}\ and\ \bibinfo {author} {\bibfnamefont {G.}~\bibnamefont {{Sch{\"a}fer}}},\ }\bibfield  {title} {\bibinfo {title} {{Binary black-hole problem at the third post-Newtonian approximation in the orbital motion: Static part}},\ }\href {https://doi.org/10.1103/PhysRevD.60.124003} {\bibfield  {journal} {\bibinfo  {journal} {\prd}\ }\textbf {\bibinfo {volume} {60}},\ \bibinfo {eid} {124003} (\bibinfo {year} {1999})},\ \Eprint {https://arxiv.org/abs/gr-qc/9906092} {arXiv:gr-qc/9906092 [gr-qc]} \BibitemShut {NoStop}%
\bibitem [{\citenamefont {Flanagan}\ and\ \citenamefont {Hinderer}(2008)}]{Flanagan:2007ix}%
  \BibitemOpen
  \bibfield  {author} {\bibinfo {author} {\bibfnamefont {E.~E.}\ \bibnamefont {Flanagan}}\ and\ \bibinfo {author} {\bibfnamefont {T.}~\bibnamefont {Hinderer}},\ }\bibfield  {title} {\bibinfo {title} {{Constraining neutron star tidal Love numbers with gravitational wave detectors}},\ }\href {https://doi.org/10.1103/PhysRevD.77.021502} {\bibfield  {journal} {\bibinfo  {journal} {Phys. Rev. D}\ }\textbf {\bibinfo {volume} {77}},\ \bibinfo {pages} {021502} (\bibinfo {year} {2008})},\ \Eprint {https://arxiv.org/abs/0709.1915} {arXiv:0709.1915 [astro-ph]} \BibitemShut {NoStop}%
\bibitem [{\citenamefont {Hinderer}\ \emph {et~al.}(2016)\citenamefont {Hinderer} \emph {et~al.}}]{Hinderer:2016eia}%
  \BibitemOpen
  \bibfield  {author} {\bibinfo {author} {\bibfnamefont {T.}~\bibnamefont {Hinderer}} \emph {et~al.},\ }\bibfield  {title} {\bibinfo {title} {{Effects of neutron-star dynamic tides on gravitational waveforms within the effective-one-body approach}},\ }\href {https://doi.org/10.1103/PhysRevLett.116.181101} {\bibfield  {journal} {\bibinfo  {journal} {Phys. Rev. Lett.}\ }\textbf {\bibinfo {volume} {116}},\ \bibinfo {pages} {181101} (\bibinfo {year} {2016})},\ \Eprint {https://arxiv.org/abs/1602.00599} {arXiv:1602.00599 [gr-qc]} \BibitemShut {NoStop}%
\bibitem [{\citenamefont {Schmidt}\ and\ \citenamefont {Hinderer}(2019)}]{Schmidt:2019wrl}%
  \BibitemOpen
  \bibfield  {author} {\bibinfo {author} {\bibfnamefont {P.}~\bibnamefont {Schmidt}}\ and\ \bibinfo {author} {\bibfnamefont {T.}~\bibnamefont {Hinderer}},\ }\bibfield  {title} {\bibinfo {title} {{Frequency domain model of $f$-mode dynamic tides in gravitational waveforms from compact binary inspirals}},\ }\href {https://doi.org/10.1103/PhysRevD.100.021501} {\bibfield  {journal} {\bibinfo  {journal} {Phys. Rev. D}\ }\textbf {\bibinfo {volume} {100}},\ \bibinfo {pages} {021501} (\bibinfo {year} {2019})},\ \Eprint {https://arxiv.org/abs/1905.00818} {arXiv:1905.00818 [gr-qc]} \BibitemShut {NoStop}%
\bibitem [{\citenamefont {Steinhoff}\ \emph {et~al.}(2021)\citenamefont {Steinhoff}, \citenamefont {Hinderer}, \citenamefont {Dietrich},\ and\ \citenamefont {Foucart}}]{Steinhoff:2021dsn}%
  \BibitemOpen
  \bibfield  {author} {\bibinfo {author} {\bibfnamefont {J.}~\bibnamefont {Steinhoff}}, \bibinfo {author} {\bibfnamefont {T.}~\bibnamefont {Hinderer}}, \bibinfo {author} {\bibfnamefont {T.}~\bibnamefont {Dietrich}},\ and\ \bibinfo {author} {\bibfnamefont {F.}~\bibnamefont {Foucart}},\ }\bibfield  {title} {\bibinfo {title} {{Spin effects on neutron star fundamental-mode dynamical tides: Phenomenology and comparison to numerical simulations}},\ }\href {https://doi.org/10.1103/PhysRevResearch.3.033129} {\bibfield  {journal} {\bibinfo  {journal} {Phys. Rev. Res.}\ }\textbf {\bibinfo {volume} {3}},\ \bibinfo {pages} {033129} (\bibinfo {year} {2021})},\ \Eprint {https://arxiv.org/abs/2103.06100} {arXiv:2103.06100 [gr-qc]} \BibitemShut {NoStop}%
\bibitem [{\citenamefont {{Basu}}\ \emph {et~al.}(2024)\citenamefont {{Basu}}, \citenamefont {{Chatterjee}},\ and\ \citenamefont {{Mondal}}}]{2024basu}%
  \BibitemOpen
  \bibfield  {author} {\bibinfo {author} {\bibfnamefont {P.}~\bibnamefont {{Basu}}}, \bibinfo {author} {\bibfnamefont {S.}~\bibnamefont {{Chatterjee}}},\ and\ \bibinfo {author} {\bibfnamefont {S.}~\bibnamefont {{Mondal}}},\ }\bibfield  {title} {\bibinfo {title} {{Eccentric orbits in disc-embedded EMRIs : orbital evolution and observability trend in LISA}},\ }\href {https://doi.org/10.1093/mnras/stae1239} {\bibfield  {journal} {\bibinfo  {journal} {\mnras}\ }\textbf {\bibinfo {volume} {531}},\ \bibinfo {pages} {1506} (\bibinfo {year} {2024})}\BibitemShut {NoStop}%
\bibitem [{\citenamefont {{Caneva Santoro}}\ \emph {et~al.}(2024)\citenamefont {{Caneva Santoro}}, \citenamefont {{Roy}}, \citenamefont {{Vicente}}, \citenamefont {{Haney}}, \citenamefont {{Piccinni}}, \citenamefont {{Del Pozzo}},\ and\ \citenamefont {{Martinez}}}]{2024santoro}%
  \BibitemOpen
  \bibfield  {author} {\bibinfo {author} {\bibfnamefont {G.}~\bibnamefont {{Caneva Santoro}}}, \bibinfo {author} {\bibfnamefont {S.}~\bibnamefont {{Roy}}}, \bibinfo {author} {\bibfnamefont {R.}~\bibnamefont {{Vicente}}}, \bibinfo {author} {\bibfnamefont {M.}~\bibnamefont {{Haney}}}, \bibinfo {author} {\bibfnamefont {O.~J.}\ \bibnamefont {{Piccinni}}}, \bibinfo {author} {\bibfnamefont {W.}~\bibnamefont {{Del Pozzo}}},\ and\ \bibinfo {author} {\bibfnamefont {M.}~\bibnamefont {{Martinez}}},\ }\bibfield  {title} {\bibinfo {title} {{First Constraints on Compact Binary Environments from LIGO-Virgo Data}},\ }\href {https://doi.org/10.1103/PhysRevLett.132.251401} {\bibfield  {journal} {\bibinfo  {journal} {\prl}\ }\textbf {\bibinfo {volume} {132}},\ \bibinfo {eid} {251401} (\bibinfo {year} {2024})},\ \Eprint {https://arxiv.org/abs/2309.05061} {arXiv:2309.05061 [gr-qc]} \BibitemShut {NoStop}%
\bibitem [{\citenamefont {{Roy}}\ and\ \citenamefont {{Vicente}}(2024)}]{2024soumen}%
  \BibitemOpen
  \bibfield  {author} {\bibinfo {author} {\bibfnamefont {S.}~\bibnamefont {{Roy}}}\ and\ \bibinfo {author} {\bibfnamefont {R.}~\bibnamefont {{Vicente}}},\ }\bibfield  {title} {\bibinfo {title} {{Compact Binary Coalescences in Dense Environments Can Pose as in Vacuum}},\ }\href {https://doi.org/10.48550/arXiv.2410.16388} {\bibfield  {journal} {\bibinfo  {journal} {arXiv e-prints}\ ,\ \bibinfo {eid} {arXiv:2410.16388}} (\bibinfo {year} {2024})},\ \Eprint {https://arxiv.org/abs/2410.16388} {arXiv:2410.16388 [gr-qc]} \BibitemShut {NoStop}%
\bibitem [{\citenamefont {{Dyson}}\ \emph {et~al.}(2025)\citenamefont {{Dyson}}, \citenamefont {{Spieksma}}, \citenamefont {{Brito}}, \citenamefont {{van de Meent}},\ and\ \citenamefont {{Dolan}}}]{2025dyson}%
  \BibitemOpen
  \bibfield  {author} {\bibinfo {author} {\bibfnamefont {C.}~\bibnamefont {{Dyson}}}, \bibinfo {author} {\bibfnamefont {T.~F.~M.}\ \bibnamefont {{Spieksma}}}, \bibinfo {author} {\bibfnamefont {R.}~\bibnamefont {{Brito}}}, \bibinfo {author} {\bibfnamefont {M.}~\bibnamefont {{van de Meent}}},\ and\ \bibinfo {author} {\bibfnamefont {S.}~\bibnamefont {{Dolan}}},\ }\bibfield  {title} {\bibinfo {title} {{Environmental effects in extreme mass ratio inspirals: perturbations to the environment in Kerr}},\ }\href {https://doi.org/10.48550/arXiv.2501.09806} {\bibfield  {journal} {\bibinfo  {journal} {arXiv e-prints}\ ,\ \bibinfo {eid} {arXiv:2501.09806}} (\bibinfo {year} {2025})},\ \Eprint {https://arxiv.org/abs/2501.09806} {arXiv:2501.09806 [gr-qc]} \BibitemShut {NoStop}%
\bibitem [{\citenamefont {{Romero-Shaw}}\ \emph {et~al.}(2021)\citenamefont {{Romero-Shaw}}, \citenamefont {{Lasky}},\ and\ \citenamefont {{Thrane}}}]{2021ApJ...921L..31R}%
  \BibitemOpen
  \bibfield  {author} {\bibinfo {author} {\bibfnamefont {I.}~\bibnamefont {{Romero-Shaw}}}, \bibinfo {author} {\bibfnamefont {P.~D.}\ \bibnamefont {{Lasky}}},\ and\ \bibinfo {author} {\bibfnamefont {E.}~\bibnamefont {{Thrane}}},\ }\bibfield  {title} {\bibinfo {title} {{Signs of Eccentricity in Two Gravitational-wave Signals May Indicate a Subpopulation of Dynamically Assembled Binary Black Holes}},\ }\href {https://doi.org/10.3847/2041-8213/ac3138} {\bibfield  {journal} {\bibinfo  {journal} {\apjl}\ }\textbf {\bibinfo {volume} {921}},\ \bibinfo {eid} {L31} (\bibinfo {year} {2021})},\ \Eprint {https://arxiv.org/abs/2108.01284} {arXiv:2108.01284 [astro-ph.HE]} \BibitemShut {NoStop}%
\bibitem [{\citenamefont {{Gupte}}\ \emph {et~al.}(2024)\citenamefont {{Gupte}}, \citenamefont {{Ramos-Buades}}, \citenamefont {{Buonanno}}, \citenamefont {{Gair}}, \citenamefont {{Miller}}, \citenamefont {{Dax}}, \citenamefont {{Green}}, \citenamefont {{P{\"u}rrer}}, \citenamefont {{Wildberger}}, \citenamefont {{Macke}}, \citenamefont {{Romero-Shaw}},\ and\ \citenamefont {{Sch{\"o}lkopf}}}]{2024gupte}%
  \BibitemOpen
  \bibfield  {author} {\bibinfo {author} {\bibfnamefont {N.}~\bibnamefont {{Gupte}}}, \bibinfo {author} {\bibfnamefont {A.}~\bibnamefont {{Ramos-Buades}}}, \bibinfo {author} {\bibfnamefont {A.}~\bibnamefont {{Buonanno}}}, \bibinfo {author} {\bibfnamefont {J.}~\bibnamefont {{Gair}}}, \bibinfo {author} {\bibfnamefont {M.~C.}\ \bibnamefont {{Miller}}}, \bibinfo {author} {\bibfnamefont {M.}~\bibnamefont {{Dax}}}, \bibinfo {author} {\bibfnamefont {S.~R.}\ \bibnamefont {{Green}}}, \bibinfo {author} {\bibfnamefont {M.}~\bibnamefont {{P{\"u}rrer}}}, \bibinfo {author} {\bibfnamefont {J.}~\bibnamefont {{Wildberger}}}, \bibinfo {author} {\bibfnamefont {J.}~\bibnamefont {{Macke}}}, \bibinfo {author} {\bibfnamefont {I.~M.}\ \bibnamefont {{Romero-Shaw}}},\ and\ \bibinfo {author} {\bibfnamefont {B.}~\bibnamefont {{Sch{\"o}lkopf}}},\ }\bibfield  {title} {\bibinfo {title} {{Evidence for eccentricity in the population of binary black holes observed by LIGO-Virgo-KAGRA}},\ }\href {https://doi.org/10.48550/arXiv.2404.14286}
  {\bibfield  {journal} {\bibinfo  {journal} {arXiv e-prints}\ ,\ \bibinfo {eid} {arXiv:2404.14286}} (\bibinfo {year} {2024})},\ \Eprint {https://arxiv.org/abs/2404.14286} {arXiv:2404.14286 [gr-qc]} \BibitemShut {NoStop}%
\bibitem [{\citenamefont {G{\"u}ltekin}\ \emph {et~al.}(2006)\citenamefont {G{\"u}ltekin}, \citenamefont {Miller},\ and\ \citenamefont {Hamilton}}]{2006ApJ...640..156G}%
  \BibitemOpen
  \bibfield  {author} {\bibinfo {author} {\bibfnamefont {K.}~\bibnamefont {G{\"u}ltekin}}, \bibinfo {author} {\bibfnamefont {M.~C.}\ \bibnamefont {Miller}},\ and\ \bibinfo {author} {\bibfnamefont {D.~P.}\ \bibnamefont {Hamilton}},\ }\bibfield  {title} {\bibinfo {title} {{Three-Body Dynamics with Gravitational Wave Emission}},\ }\href@noop {} {\bibfield  {journal} {\bibinfo  {journal} {\apj}\ }\textbf {\bibinfo {volume} {640}},\ \bibinfo {pages} {156} (\bibinfo {year} {2006})}\BibitemShut {NoStop}%
\bibitem [{\citenamefont {{Samsing}}\ \emph {et~al.}(2014)\citenamefont {{Samsing}}, \citenamefont {{MacLeod}},\ and\ \citenamefont {{Ramirez-Ruiz}}}]{2014ApJ...784...71S}%
  \BibitemOpen
  \bibfield  {author} {\bibinfo {author} {\bibfnamefont {J.}~\bibnamefont {{Samsing}}}, \bibinfo {author} {\bibfnamefont {M.}~\bibnamefont {{MacLeod}}},\ and\ \bibinfo {author} {\bibfnamefont {E.}~\bibnamefont {{Ramirez-Ruiz}}},\ }\bibfield  {title} {\bibinfo {title} {{The Formation of Eccentric Compact Binary Inspirals and the Role of Gravitational Wave Emission in Binary-Single Stellar Encounters}},\ }\href {https://doi.org/10.1088/0004-637X/784/1/71} {\bibfield  {journal} {\bibinfo  {journal} {\apj}\ }\textbf {\bibinfo {volume} {784}},\ \bibinfo {eid} {71} (\bibinfo {year} {2014})},\ \Eprint {https://arxiv.org/abs/1308.2964} {arXiv:1308.2964 [astro-ph.HE]} \BibitemShut {NoStop}%
\bibitem [{\citenamefont {{Samsing}}\ and\ \citenamefont {{Ramirez-Ruiz}}(2017)}]{2017ApJ...840L..14S}%
  \BibitemOpen
  \bibfield  {author} {\bibinfo {author} {\bibfnamefont {J.}~\bibnamefont {{Samsing}}}\ and\ \bibinfo {author} {\bibfnamefont {E.}~\bibnamefont {{Ramirez-Ruiz}}},\ }\bibfield  {title} {\bibinfo {title} {{On the Assembly Rate of Highly Eccentric Binary Black Hole Mergers}},\ }\href {https://doi.org/10.3847/2041-8213/aa6f0b} {\bibfield  {journal} {\bibinfo  {journal} {\apjl}\ }\textbf {\bibinfo {volume} {840}},\ \bibinfo {eid} {L14} (\bibinfo {year} {2017})},\ \Eprint {https://arxiv.org/abs/1703.09703} {arXiv:1703.09703 [astro-ph.HE]} \BibitemShut {NoStop}%
\bibitem [{\citenamefont {{Samsing}}\ and\ \citenamefont {{Ilan}}(2018)}]{Samsing18a}%
  \BibitemOpen
  \bibfield  {author} {\bibinfo {author} {\bibfnamefont {J.}~\bibnamefont {{Samsing}}}\ and\ \bibinfo {author} {\bibfnamefont {T.}~\bibnamefont {{Ilan}}},\ }\bibfield  {title} {\bibinfo {title} {{Topology of black hole binary-single interactions}},\ }\href {https://doi.org/10.1093/mnras/sty197} {\bibfield  {journal} {\bibinfo  {journal} {\mnras}\ }\textbf {\bibinfo {volume} {476}},\ \bibinfo {pages} {1548} (\bibinfo {year} {2018})},\ \Eprint {https://arxiv.org/abs/1706.04672} {arXiv:1706.04672 [astro-ph.HE]} \BibitemShut {NoStop}%
\bibitem [{\citenamefont {{Samsing}}\ \emph {et~al.}(2018{\natexlab{a}})\citenamefont {{Samsing}}, \citenamefont {{MacLeod}},\ and\ \citenamefont {{Ramirez-Ruiz}}}]{Samsing2018}%
  \BibitemOpen
  \bibfield  {author} {\bibinfo {author} {\bibfnamefont {J.}~\bibnamefont {{Samsing}}}, \bibinfo {author} {\bibfnamefont {M.}~\bibnamefont {{MacLeod}}},\ and\ \bibinfo {author} {\bibfnamefont {E.}~\bibnamefont {{Ramirez-Ruiz}}},\ }\bibfield  {title} {\bibinfo {title} {{Dissipative Evolution of Unequal-mass Binary-single Interactions and Its Relevance to Gravitational-wave Detections}},\ }\href {https://doi.org/10.3847/1538-4357/aaa715} {\bibfield  {journal} {\bibinfo  {journal} {\apj}\ }\textbf {\bibinfo {volume} {853}},\ \bibinfo {eid} {140} (\bibinfo {year} {2018}{\natexlab{a}})},\ \Eprint {https://arxiv.org/abs/1706.03776} {arXiv:1706.03776 [astro-ph.HE]} \BibitemShut {NoStop}%
\bibitem [{\citenamefont {{Samsing}}(2018)}]{Samsing18}%
  \BibitemOpen
  \bibfield  {author} {\bibinfo {author} {\bibfnamefont {J.}~\bibnamefont {{Samsing}}},\ }\bibfield  {title} {\bibinfo {title} {{Eccentric black hole mergers forming in globular clusters}},\ }\href {https://doi.org/10.1103/PhysRevD.97.103014} {\bibfield  {journal} {\bibinfo  {journal} {\prd}\ }\textbf {\bibinfo {volume} {97}},\ \bibinfo {eid} {103014} (\bibinfo {year} {2018})},\ \Eprint {https://arxiv.org/abs/1711.07452} {arXiv:1711.07452 [astro-ph.HE]} \BibitemShut {NoStop}%
\bibitem [{\citenamefont {{Samsing}}\ \emph {et~al.}(2018{\natexlab{b}})\citenamefont {{Samsing}}, \citenamefont {{Askar}},\ and\ \citenamefont {{Giersz}}}]{2018ApJ...855..124S}%
  \BibitemOpen
  \bibfield  {author} {\bibinfo {author} {\bibfnamefont {J.}~\bibnamefont {{Samsing}}}, \bibinfo {author} {\bibfnamefont {A.}~\bibnamefont {{Askar}}},\ and\ \bibinfo {author} {\bibfnamefont {M.}~\bibnamefont {{Giersz}}},\ }\bibfield  {title} {\bibinfo {title} {{MOCCA-SURVEY Database. I. Eccentric Black Hole Mergers during Binary-Single Interactions in Globular Clusters}},\ }\href {https://doi.org/10.3847/1538-4357/aaab52} {\bibfield  {journal} {\bibinfo  {journal} {\apj}\ }\textbf {\bibinfo {volume} {855}},\ \bibinfo {eid} {124} (\bibinfo {year} {2018}{\natexlab{b}})},\ \Eprint {https://arxiv.org/abs/1712.06186} {arXiv:1712.06186 [astro-ph.HE]} \BibitemShut {NoStop}%
\bibitem [{\citenamefont {{Samsing}}\ and\ \citenamefont {{D'Orazio}}(2018)}]{2018MNRAS.tmp.2223S}%
  \BibitemOpen
  \bibfield  {author} {\bibinfo {author} {\bibfnamefont {J.}~\bibnamefont {{Samsing}}}\ and\ \bibinfo {author} {\bibfnamefont {D.~J.}\ \bibnamefont {{D'Orazio}}},\ }\bibfield  {title} {\bibinfo {title} {{Black Hole Mergers From Globular Clusters Observable by LISA I: Eccentric Sources Originating From Relativistic N-body Dynamics}},\ }\bibfield  {journal} {\bibinfo  {journal} {\mnras}\ }\href {https://doi.org/10.1093/mnras/sty2334} {10.1093/mnras/sty2334} (\bibinfo {year} {2018}),\ \Eprint {https://arxiv.org/abs/1804.06519} {arXiv:1804.06519 [astro-ph.HE]} \BibitemShut {NoStop}%
\bibitem [{\citenamefont {{Rodriguez}}\ \emph {et~al.}(2018)\citenamefont {{Rodriguez}}, \citenamefont {{Amaro-Seoane}}, \citenamefont {{Chatterjee}}, \citenamefont {{Kremer}}, \citenamefont {{Rasio}}, \citenamefont {{Samsing}}, \citenamefont {{Ye}},\ and\ \citenamefont {{Zevin}}}]{2018PhRvD..98l3005R}%
  \BibitemOpen
  \bibfield  {author} {\bibinfo {author} {\bibfnamefont {C.~L.}\ \bibnamefont {{Rodriguez}}}, \bibinfo {author} {\bibfnamefont {P.}~\bibnamefont {{Amaro-Seoane}}}, \bibinfo {author} {\bibfnamefont {S.}~\bibnamefont {{Chatterjee}}}, \bibinfo {author} {\bibfnamefont {K.}~\bibnamefont {{Kremer}}}, \bibinfo {author} {\bibfnamefont {F.~A.}\ \bibnamefont {{Rasio}}}, \bibinfo {author} {\bibfnamefont {J.}~\bibnamefont {{Samsing}}}, \bibinfo {author} {\bibfnamefont {C.~S.}\ \bibnamefont {{Ye}}},\ and\ \bibinfo {author} {\bibfnamefont {M.}~\bibnamefont {{Zevin}}},\ }\bibfield  {title} {\bibinfo {title} {{Post-Newtonian dynamics in dense star clusters: Formation, masses, and merger rates of highly-eccentric black hole binaries}},\ }\href {https://doi.org/10.1103/PhysRevD.98.123005} {\bibfield  {journal} {\bibinfo  {journal} {\prd}\ }\textbf {\bibinfo {volume} {98}},\ \bibinfo {eid} {123005} (\bibinfo {year} {2018})},\ \Eprint {https://arxiv.org/abs/1811.04926} {arXiv:1811.04926 [astro-ph.HE]} \BibitemShut {NoStop}%
\bibitem [{\citenamefont {{Liu}}\ \emph {et~al.}(2019)\citenamefont {{Liu}}, \citenamefont {{Lai}},\ and\ \citenamefont {{Wang}}}]{2019ApJ...881...41L}%
  \BibitemOpen
  \bibfield  {author} {\bibinfo {author} {\bibfnamefont {B.}~\bibnamefont {{Liu}}}, \bibinfo {author} {\bibfnamefont {D.}~\bibnamefont {{Lai}}},\ and\ \bibinfo {author} {\bibfnamefont {Y.-H.}\ \bibnamefont {{Wang}}},\ }\bibfield  {title} {\bibinfo {title} {{Black Hole and Neutron Star Binary Mergers in Triple Systems. II. Merger Eccentricity and Spin-Orbit Misalignment}},\ }\href {https://doi.org/10.3847/1538-4357/ab2dfb} {\bibfield  {journal} {\bibinfo  {journal} {\apj}\ }\textbf {\bibinfo {volume} {881}},\ \bibinfo {eid} {41} (\bibinfo {year} {2019})},\ \Eprint {https://arxiv.org/abs/1905.00427} {arXiv:1905.00427 [astro-ph.HE]} \BibitemShut {NoStop}%
\bibitem [{\citenamefont {{Zevin}}\ \emph {et~al.}(2019)\citenamefont {{Zevin}}, \citenamefont {{Samsing}}, \citenamefont {{Rodriguez}}, \citenamefont {{Haster}},\ and\ \citenamefont {{Ramirez-Ruiz}}}]{2019ApJ...871...91Z}%
  \BibitemOpen
  \bibfield  {author} {\bibinfo {author} {\bibfnamefont {M.}~\bibnamefont {{Zevin}}}, \bibinfo {author} {\bibfnamefont {J.}~\bibnamefont {{Samsing}}}, \bibinfo {author} {\bibfnamefont {C.}~\bibnamefont {{Rodriguez}}}, \bibinfo {author} {\bibfnamefont {C.-J.}\ \bibnamefont {{Haster}}},\ and\ \bibinfo {author} {\bibfnamefont {E.}~\bibnamefont {{Ramirez-Ruiz}}},\ }\bibfield  {title} {\bibinfo {title} {{Eccentric Black Hole Mergers in Dense Star Clusters: The Role of Binary-Binary Encounters}},\ }\href {https://doi.org/10.3847/1538-4357/aaf6ec} {\bibfield  {journal} {\bibinfo  {journal} {\apj}\ }\textbf {\bibinfo {volume} {871}},\ \bibinfo {eid} {91} (\bibinfo {year} {2019})},\ \Eprint {https://arxiv.org/abs/1810.00901} {arXiv:1810.00901 [astro-ph.HE]} \BibitemShut {NoStop}%
\bibitem [{\citenamefont {{Samsing}}\ \emph {et~al.}(2019{\natexlab{a}})\citenamefont {{Samsing}}, \citenamefont {{Hamers}},\ and\ \citenamefont {{Tyles}}}]{2019PhRvD.100d3010S}%
  \BibitemOpen
  \bibfield  {author} {\bibinfo {author} {\bibfnamefont {J.}~\bibnamefont {{Samsing}}}, \bibinfo {author} {\bibfnamefont {A.~S.}\ \bibnamefont {{Hamers}}},\ and\ \bibinfo {author} {\bibfnamefont {J.~G.}\ \bibnamefont {{Tyles}}},\ }\bibfield  {title} {\bibinfo {title} {{Effect of distant encounters on black hole binaries in globular clusters: Systematic increase of in-cluster mergers in the LISA band}},\ }\href {https://doi.org/10.1103/PhysRevD.100.043010} {\bibfield  {journal} {\bibinfo  {journal} {\prd}\ }\textbf {\bibinfo {volume} {100}},\ \bibinfo {eid} {043010} (\bibinfo {year} {2019}{\natexlab{a}})},\ \Eprint {https://arxiv.org/abs/1906.07189} {arXiv:1906.07189 [astro-ph.HE]} \BibitemShut {NoStop}%
\bibitem [{\citenamefont {{Samsing}}\ \emph {et~al.}(2019{\natexlab{b}})\citenamefont {{Samsing}}, \citenamefont {{D'Orazio}}, \citenamefont {{Kremer}}, \citenamefont {{Rodriguez}},\ and\ \citenamefont {{Askar}}}]{2019arXiv190711231S}%
  \BibitemOpen
  \bibfield  {author} {\bibinfo {author} {\bibfnamefont {J.}~\bibnamefont {{Samsing}}}, \bibinfo {author} {\bibfnamefont {D.~J.}\ \bibnamefont {{D'Orazio}}}, \bibinfo {author} {\bibfnamefont {K.}~\bibnamefont {{Kremer}}}, \bibinfo {author} {\bibfnamefont {C.~L.}\ \bibnamefont {{Rodriguez}}},\ and\ \bibinfo {author} {\bibfnamefont {A.}~\bibnamefont {{Askar}}},\ }\bibfield  {title} {\bibinfo {title} {{Gravitational-wave captures of single black holes in globular clusters}},\ }\href@noop {} {\bibfield  {journal} {\bibinfo  {journal} {arXiv e-prints}\ ,\ \bibinfo {eid} {arXiv:1907.11231}} (\bibinfo {year} {2019}{\natexlab{b}})},\ \Eprint {https://arxiv.org/abs/1907.11231} {arXiv:1907.11231 [astro-ph.HE]} \BibitemShut {NoStop}%
\bibitem [{\citenamefont {{Amaro-Seoane}}\ \emph {et~al.}(2007)\citenamefont {{Amaro-Seoane}}, \citenamefont {{Gair}}, \citenamefont {{Freitag}}, \citenamefont {{Miller}}, \citenamefont {{Mandel}}, \citenamefont {{Cutler}},\ and\ \citenamefont {{Babak}}}]{amaro-seoane2007}%
  \BibitemOpen
  \bibfield  {author} {\bibinfo {author} {\bibfnamefont {P.}~\bibnamefont {{Amaro-Seoane}}}, \bibinfo {author} {\bibfnamefont {J.~R.}\ \bibnamefont {{Gair}}}, \bibinfo {author} {\bibfnamefont {M.}~\bibnamefont {{Freitag}}}, \bibinfo {author} {\bibfnamefont {M.~C.}\ \bibnamefont {{Miller}}}, \bibinfo {author} {\bibfnamefont {I.}~\bibnamefont {{Mandel}}}, \bibinfo {author} {\bibfnamefont {C.~J.}\ \bibnamefont {{Cutler}}},\ and\ \bibinfo {author} {\bibfnamefont {S.}~\bibnamefont {{Babak}}},\ }\bibfield  {title} {\bibinfo {title} {{TOPICAL REVIEW: Intermediate and extreme mass-ratio inspirals{\textemdash}astrophysics, science applications and detection using LISA}},\ }\href {https://doi.org/10.1088/0264-9381/24/17/R01} {\bibfield  {journal} {\bibinfo  {journal} {Classical and Quantum Gravity}\ }\textbf {\bibinfo {volume} {24}},\ \bibinfo {pages} {R113} (\bibinfo {year} {2007})},\ \Eprint {https://arxiv.org/abs/astro-ph/0703495} {arXiv:astro-ph/0703495 [astro-ph]} \BibitemShut {NoStop}%
\bibitem [{\citenamefont {{Garg}}\ \emph {et~al.}(2024)\citenamefont {{Garg}}, \citenamefont {{Derdzinski}}, \citenamefont {{Tiwari}}, \citenamefont {{Gair}},\ and\ \citenamefont {{Mayer}}}]{2024garg}%
  \BibitemOpen
  \bibfield  {author} {\bibinfo {author} {\bibfnamefont {M.}~\bibnamefont {{Garg}}}, \bibinfo {author} {\bibfnamefont {A.}~\bibnamefont {{Derdzinski}}}, \bibinfo {author} {\bibfnamefont {S.}~\bibnamefont {{Tiwari}}}, \bibinfo {author} {\bibfnamefont {J.}~\bibnamefont {{Gair}}},\ and\ \bibinfo {author} {\bibfnamefont {L.}~\bibnamefont {{Mayer}}},\ }\bibfield  {title} {\bibinfo {title} {{Measuring eccentricity and gas-induced perturbation from gravitational waves of LISA massive black hole binaries}},\ }\href {https://doi.org/10.48550/arXiv.2402.14058} {\bibfield  {journal} {\bibinfo  {journal} {arXiv e-prints}\ ,\ \bibinfo {eid} {arXiv:2402.14058}} (\bibinfo {year} {2024})},\ \Eprint {https://arxiv.org/abs/2402.14058} {arXiv:2402.14058 [astro-ph.GA]} \BibitemShut {NoStop}%
\bibitem [{\citenamefont {{Vallisneri}}(2008)}]{2008PhRvD..77d2001V}%
  \BibitemOpen
  \bibfield  {author} {\bibinfo {author} {\bibfnamefont {M.}~\bibnamefont {{Vallisneri}}},\ }\bibfield  {title} {\bibinfo {title} {{Use and abuse of the Fisher information matrix in the assessment of gravitational-wave parameter-estimation prospects}},\ }\href {https://doi.org/10.1103/PhysRevD.77.042001} {\bibfield  {journal} {\bibinfo  {journal} {\prd}\ }\textbf {\bibinfo {volume} {77}},\ \bibinfo {eid} {042001} (\bibinfo {year} {2008})},\ \Eprint {https://arxiv.org/abs/gr-qc/0703086} {arXiv:gr-qc/0703086 [gr-qc]} \BibitemShut {NoStop}%
\bibitem [{\citenamefont {Veitch}\ and\ \citenamefont {Vecchio}(2010)}]{Veitch:2009hd}%
  \BibitemOpen
  \bibfield  {author} {\bibinfo {author} {\bibfnamefont {J.}~\bibnamefont {Veitch}}\ and\ \bibinfo {author} {\bibfnamefont {A.}~\bibnamefont {Vecchio}},\ }\bibfield  {title} {\bibinfo {title} {{Bayesian coherent analysis of in-spiral gravitational wave signals with a detector network}},\ }\href {https://doi.org/10.1103/PhysRevD.81.062003} {\bibfield  {journal} {\bibinfo  {journal} {Phys. Rev. D}\ }\textbf {\bibinfo {volume} {81}},\ \bibinfo {pages} {062003} (\bibinfo {year} {2010})},\ \Eprint {https://arxiv.org/abs/0911.3820} {arXiv:0911.3820 [astro-ph.CO]} \BibitemShut {NoStop}%
\bibitem [{\citenamefont {Veitch}\ \emph {et~al.}(2015)\citenamefont {Veitch} \emph {et~al.}}]{Veitch:2014wba}%
  \BibitemOpen
  \bibfield  {author} {\bibinfo {author} {\bibfnamefont {J.}~\bibnamefont {Veitch}} \emph {et~al.},\ }\bibfield  {title} {\bibinfo {title} {{Parameter estimation for compact binaries with ground-based gravitational-wave observations using the LALInference software library}},\ }\href {https://doi.org/10.1103/PhysRevD.91.042003} {\bibfield  {journal} {\bibinfo  {journal} {Phys. Rev. D}\ }\textbf {\bibinfo {volume} {91}},\ \bibinfo {pages} {042003} (\bibinfo {year} {2015})},\ \Eprint {https://arxiv.org/abs/1409.7215} {arXiv:1409.7215 [gr-qc]} \BibitemShut {NoStop}%
\bibitem [{\citenamefont {{Thrane}}\ and\ \citenamefont {{Talbot}}(2019)}]{2019PASA...36...10T}%
  \BibitemOpen
  \bibfield  {author} {\bibinfo {author} {\bibfnamefont {E.}~\bibnamefont {{Thrane}}}\ and\ \bibinfo {author} {\bibfnamefont {C.}~\bibnamefont {{Talbot}}},\ }\bibfield  {title} {\bibinfo {title} {{An introduction to Bayesian inference in gravitational-wave astronomy: Parameter estimation, model selection, and hierarchical models}},\ }\href {https://doi.org/10.1017/pasa.2019.2} {\bibfield  {journal} {\bibinfo  {journal} {\pasa}\ }\textbf {\bibinfo {volume} {36}},\ \bibinfo {eid} {e010} (\bibinfo {year} {2019})},\ \Eprint {https://arxiv.org/abs/1809.02293} {arXiv:1809.02293 [astro-ph.IM]} \BibitemShut {NoStop}%
\bibitem [{\citenamefont {{Owen}}\ \emph {et~al.}(2023)\citenamefont {{Owen}}, \citenamefont {{Haster}}, \citenamefont {{Perkins}}, \citenamefont {{Cornish}},\ and\ \citenamefont {{Yunes}}}]{owen23}%
  \BibitemOpen
  \bibfield  {author} {\bibinfo {author} {\bibfnamefont {C.~B.}\ \bibnamefont {{Owen}}}, \bibinfo {author} {\bibfnamefont {C.-J.}\ \bibnamefont {{Haster}}}, \bibinfo {author} {\bibfnamefont {S.}~\bibnamefont {{Perkins}}}, \bibinfo {author} {\bibfnamefont {N.~J.}\ \bibnamefont {{Cornish}}},\ and\ \bibinfo {author} {\bibfnamefont {N.}~\bibnamefont {{Yunes}}},\ }\bibfield  {title} {\bibinfo {title} {{Waveform accuracy and systematic uncertainties in current gravitational wave observations}},\ }\href@noop {} {\bibfield  {journal} {\bibinfo  {journal} {arXiv e-prints}\ ,\ \bibinfo {eid} {arXiv:2301.11941}} (\bibinfo {year} {2023})},\ \Eprint {https://arxiv.org/abs/2301.11941} {arXiv:2301.11941 [gr-qc]} \BibitemShut {NoStop}%
\bibitem [{\citenamefont {{Robson}}\ \emph {et~al.}(2019)\citenamefont {{Robson}}, \citenamefont {{Cornish}},\ and\ \citenamefont {{Liu}}}]{2019robson}%
  \BibitemOpen
  \bibfield  {author} {\bibinfo {author} {\bibfnamefont {T.}~\bibnamefont {{Robson}}}, \bibinfo {author} {\bibfnamefont {N.~J.}\ \bibnamefont {{Cornish}}},\ and\ \bibinfo {author} {\bibfnamefont {C.}~\bibnamefont {{Liu}}},\ }\bibfield  {title} {\bibinfo {title} {{The construction and use of LISA sensitivity curves}},\ }\href {https://doi.org/10.1088/1361-6382/ab1101} {\bibfield  {journal} {\bibinfo  {journal} {Class. Quantum Gravity}\ }\textbf {\bibinfo {volume} {36}},\ \bibinfo {eid} {105011} (\bibinfo {year} {2019})},\ \Eprint {https://arxiv.org/abs/1803.01944} {arXiv:1803.01944 [astro-ph.HE]} \BibitemShut {NoStop}%
\bibitem [{\citenamefont {{Gupta}}\ \emph {et~al.}(2020)\citenamefont {{Gupta}}, \citenamefont {{Suzuki}}, \citenamefont {{Okawa}},\ and\ \citenamefont {{Maeda}}}]{gupta2020}%
  \BibitemOpen
  \bibfield  {author} {\bibinfo {author} {\bibfnamefont {P.}~\bibnamefont {{Gupta}}}, \bibinfo {author} {\bibfnamefont {H.}~\bibnamefont {{Suzuki}}}, \bibinfo {author} {\bibfnamefont {H.}~\bibnamefont {{Okawa}}},\ and\ \bibinfo {author} {\bibfnamefont {K.-i.}\ \bibnamefont {{Maeda}}},\ }\bibfield  {title} {\bibinfo {title} {{Gravitational waves from hierarchical triple systems with Kozai-Lidov oscillation}},\ }\href {https://doi.org/10.1103/PhysRevD.101.104053} {\bibfield  {journal} {\bibinfo  {journal} {\prd}\ }\textbf {\bibinfo {volume} {101}},\ \bibinfo {eid} {104053} (\bibinfo {year} {2020})},\ \Eprint {https://arxiv.org/abs/1911.11318} {arXiv:1911.11318 [gr-qc]} \BibitemShut {NoStop}%
\bibitem [{\citenamefont {Deme}\ \emph {et~al.}(2020)\citenamefont {Deme}, \citenamefont {Hoang}, \citenamefont {Naoz},\ and\ \citenamefont {Kocsis}}]{Deme:2020ewx}%
  \BibitemOpen
  \bibfield  {author} {\bibinfo {author} {\bibfnamefont {B.}~\bibnamefont {Deme}}, \bibinfo {author} {\bibfnamefont {B.-M.}\ \bibnamefont {Hoang}}, \bibinfo {author} {\bibfnamefont {S.}~\bibnamefont {Naoz}},\ and\ \bibinfo {author} {\bibfnamefont {B.}~\bibnamefont {Kocsis}},\ }\bibfield  {title} {\bibinfo {title} {{Detecting Kozai{\textendash}Lidov Imprints on the Gravitational Waves of Intermediate-mass Black Holes in Galactic Nuclei}},\ }\href {https://doi.org/10.3847/1538-4357/abafa3} {\bibfield  {journal} {\bibinfo  {journal} {Astrophys. J.}\ }\textbf {\bibinfo {volume} {901}},\ \bibinfo {pages} {125} (\bibinfo {year} {2020})},\ \Eprint {https://arxiv.org/abs/2005.03677} {arXiv:2005.03677 [astro-ph.HE]} \BibitemShut {NoStop}%
\bibitem [{\citenamefont {Trani}\ and\ \citenamefont {Di~Cintio}(2025)}]{Trani:2025edb}%
  \BibitemOpen
  \bibfield  {author} {\bibinfo {author} {\bibfnamefont {A.~A.}\ \bibnamefont {Trani}}\ and\ \bibinfo {author} {\bibfnamefont {P.}~\bibnamefont {Di~Cintio}},\ }\bibfield  {title} {\bibinfo {title} {{Turbulent drag on stellar mass black holes embedded in AGN discs}},\ }\href@noop {} {\bibfield  {journal} {\bibinfo  {journal} {arXiv e-prints}\ } (\bibinfo {year} {2025})},\ \Eprint {https://arxiv.org/abs/2506.02173} {arXiv:2506.02173 [astro-ph.HE]} \BibitemShut {NoStop}%
\bibitem [{\citenamefont {Tichy}\ \emph {et~al.}(2000)\citenamefont {Tichy}, \citenamefont {Flanagan},\ and\ \citenamefont {Poisson}}]{Tichy:1999pv}%
  \BibitemOpen
  \bibfield  {author} {\bibinfo {author} {\bibfnamefont {W.}~\bibnamefont {Tichy}}, \bibinfo {author} {\bibfnamefont {E.~E.}\ \bibnamefont {Flanagan}},\ and\ \bibinfo {author} {\bibfnamefont {E.}~\bibnamefont {Poisson}},\ }\bibfield  {title} {\bibinfo {title} {{Can the postNewtonian gravitational wave form of an inspiraling binary be improved by solving the energy balance equation numerically?}},\ }\href {https://doi.org/10.1103/PhysRevD.61.104015} {\bibfield  {journal} {\bibinfo  {journal} {Phys. Rev. D}\ }\textbf {\bibinfo {volume} {61}},\ \bibinfo {pages} {104015} (\bibinfo {year} {2000})},\ \Eprint {https://arxiv.org/abs/gr-qc/9912075} {arXiv:gr-qc/9912075} \BibitemShut {NoStop}%
\bibitem [{\citenamefont {{Droz}}\ \emph {et~al.}(1999)\citenamefont {{Droz}}, \citenamefont {{Knapp}}, \citenamefont {{Poisson}},\ and\ \citenamefont {{Owen}}}]{1999PhRvD..59l4016D}%
  \BibitemOpen
  \bibfield  {author} {\bibinfo {author} {\bibfnamefont {S.}~\bibnamefont {{Droz}}}, \bibinfo {author} {\bibfnamefont {D.~J.}\ \bibnamefont {{Knapp}}}, \bibinfo {author} {\bibfnamefont {E.}~\bibnamefont {{Poisson}}},\ and\ \bibinfo {author} {\bibfnamefont {B.~J.}\ \bibnamefont {{Owen}}},\ }\bibfield  {title} {\bibinfo {title} {{Gravitational waves from inspiraling compact binaries: Validity of the stationary-phase approximation to the Fourier transform}},\ }\href {https://doi.org/10.1103/PhysRevD.59.124016} {\bibfield  {journal} {\bibinfo  {journal} {\prd}\ }\textbf {\bibinfo {volume} {59}},\ \bibinfo {eid} {124016} (\bibinfo {year} {1999})},\ \Eprint {https://arxiv.org/abs/gr-qc/9901076} {arXiv:gr-qc/9901076 [gr-qc]} \BibitemShut {NoStop}%
\bibitem [{\citenamefont {{Chatterjee}}\ \emph {et~al.}(2023)\citenamefont {{Chatterjee}}, \citenamefont {{Mondal}},\ and\ \citenamefont {{Basu}}}]{2023chatterjee}%
  \BibitemOpen
  \bibfield  {author} {\bibinfo {author} {\bibfnamefont {S.}~\bibnamefont {{Chatterjee}}}, \bibinfo {author} {\bibfnamefont {S.}~\bibnamefont {{Mondal}}},\ and\ \bibinfo {author} {\bibfnamefont {P.}~\bibnamefont {{Basu}}},\ }\bibfield  {title} {\bibinfo {title} {{Detectability of gas-rich E/IMRI's in LISA band: observable signature of transonic accretion flow}},\ }\href {https://doi.org/10.1093/mnras/stad3132} {\bibfield  {journal} {\bibinfo  {journal} {\mnras}\ }\textbf {\bibinfo {volume} {526}},\ \bibinfo {pages} {5612} (\bibinfo {year} {2023})},\ \Eprint {https://arxiv.org/abs/2307.12144} {arXiv:2307.12144 [astro-ph.HE]} \BibitemShut {NoStop}%
\bibitem [{\citenamefont {{Cole}}\ \emph {et~al.}(2023)\citenamefont {{Cole}}, \citenamefont {{Bertone}}, \citenamefont {{Coogan}}, \citenamefont {{Gaggero}}, \citenamefont {{Karydas}}, \citenamefont {{Kavanagh}}, \citenamefont {{Spieksma}},\ and\ \citenamefont {{Tomaselli}}}]{2023cole}%
  \BibitemOpen
  \bibfield  {author} {\bibinfo {author} {\bibfnamefont {P.~S.}\ \bibnamefont {{Cole}}}, \bibinfo {author} {\bibfnamefont {G.}~\bibnamefont {{Bertone}}}, \bibinfo {author} {\bibfnamefont {A.}~\bibnamefont {{Coogan}}}, \bibinfo {author} {\bibfnamefont {D.}~\bibnamefont {{Gaggero}}}, \bibinfo {author} {\bibfnamefont {T.}~\bibnamefont {{Karydas}}}, \bibinfo {author} {\bibfnamefont {B.~J.}\ \bibnamefont {{Kavanagh}}}, \bibinfo {author} {\bibfnamefont {T.~F.~M.}\ \bibnamefont {{Spieksma}}},\ and\ \bibinfo {author} {\bibfnamefont {G.~M.}\ \bibnamefont {{Tomaselli}}},\ }\bibfield  {title} {\bibinfo {title} {{Distinguishing environmental effects on binary black hole gravitational waveforms}},\ }\href {https://doi.org/10.1038/s41550-023-01990-2} {\bibfield  {journal} {\bibinfo  {journal} {Nature Astronomy}\ }\textbf {\bibinfo {volume} {7}},\ \bibinfo {pages} {943} (\bibinfo {year} {2023})},\ \Eprint {https://arxiv.org/abs/2211.01362} {arXiv:2211.01362 [gr-qc]} \BibitemShut {NoStop}%
\bibitem [{\citenamefont {{Kavanagh}}\ \emph {et~al.}(2020)\citenamefont {{Kavanagh}}, \citenamefont {{Nichols}}, \citenamefont {{Bertone}},\ and\ \citenamefont {{Gaggero}}}]{kavanagh2020}%
  \BibitemOpen
  \bibfield  {author} {\bibinfo {author} {\bibfnamefont {B.~J.}\ \bibnamefont {{Kavanagh}}}, \bibinfo {author} {\bibfnamefont {D.~A.}\ \bibnamefont {{Nichols}}}, \bibinfo {author} {\bibfnamefont {G.}~\bibnamefont {{Bertone}}},\ and\ \bibinfo {author} {\bibfnamefont {D.}~\bibnamefont {{Gaggero}}},\ }\bibfield  {title} {\bibinfo {title} {{Detecting dark matter around black holes with gravitational waves: Effects of dark-matter dynamics on the gravitational waveform}},\ }\href {https://doi.org/10.1103/PhysRevD.102.083006} {\bibfield  {journal} {\bibinfo  {journal} {\prd}\ }\textbf {\bibinfo {volume} {102}},\ \bibinfo {eid} {083006} (\bibinfo {year} {2020})},\ \Eprint {https://arxiv.org/abs/2002.12811} {arXiv:2002.12811 [gr-qc]} \BibitemShut {NoStop}%
\bibitem [{\citenamefont {{Rivera}}\ and\ \citenamefont {{Reyes}}(2024)}]{2024rivera}%
  \BibitemOpen
  \bibfield  {author} {\bibinfo {author} {\bibfnamefont {M.~I.~B.}\ \bibnamefont {{Rivera}}}\ and\ \bibinfo {author} {\bibfnamefont {R.~C.}\ \bibnamefont {{Reyes}}},\ }\bibfield  {title} {\bibinfo {title} {{Measurable parameter combinations of environmentally-dephased EMRI gravitational-wave signals}},\ }\href {https://doi.org/10.1016/j.newast.2024.102263} {\bibfield  {journal} {\bibinfo  {journal} {\na}\ }\textbf {\bibinfo {volume} {112}},\ \bibinfo {eid} {102263} (\bibinfo {year} {2024})},\ \Eprint {https://arxiv.org/abs/2406.15971} {arXiv:2406.15971 [gr-qc]} \BibitemShut {NoStop}%
\bibitem [{\citenamefont {Lindblom}\ \emph {et~al.}(2008)\citenamefont {Lindblom}, \citenamefont {Owen},\ and\ \citenamefont {Brown}}]{Lindblom:2008cm}%
  \BibitemOpen
  \bibfield  {author} {\bibinfo {author} {\bibfnamefont {L.}~\bibnamefont {Lindblom}}, \bibinfo {author} {\bibfnamefont {B.~J.}\ \bibnamefont {Owen}},\ and\ \bibinfo {author} {\bibfnamefont {D.~A.}\ \bibnamefont {Brown}},\ }\bibfield  {title} {\bibinfo {title} {{Model Waveform Accuracy Standards for Gravitational Wave Data Analysis}},\ }\href {https://doi.org/10.1103/PhysRevD.78.124020} {\bibfield  {journal} {\bibinfo  {journal} {Phys. Rev. D}\ }\textbf {\bibinfo {volume} {78}},\ \bibinfo {pages} {124020} (\bibinfo {year} {2008})},\ \Eprint {https://arxiv.org/abs/0809.3844} {arXiv:0809.3844 [gr-qc]} \BibitemShut {NoStop}%
\bibitem [{\citenamefont {{Cardoso}}\ and\ \citenamefont {{Macedo}}(2020)}]{2020cardosoself}%
  \BibitemOpen
  \bibfield  {author} {\bibinfo {author} {\bibfnamefont {V.}~\bibnamefont {{Cardoso}}}\ and\ \bibinfo {author} {\bibfnamefont {C.~F.~B.}\ \bibnamefont {{Macedo}}},\ }\bibfield  {title} {\bibinfo {title} {{Drifting through the medium: kicks and self-propulsion of binaries within accretion discs and other environments}},\ }\href {https://doi.org/10.1093/mnras/staa2396} {\bibfield  {journal} {\bibinfo  {journal} {\mnras}\ }\textbf {\bibinfo {volume} {498}},\ \bibinfo {pages} {1963} (\bibinfo {year} {2020})},\ \Eprint {https://arxiv.org/abs/2008.01091} {arXiv:2008.01091 [astro-ph.HE]} \BibitemShut {NoStop}%
\bibitem [{\citenamefont {{Kuzmin}}(1952)}]{1952kuzmin}%
  \BibitemOpen
  \bibfield  {author} {\bibinfo {author} {\bibfnamefont {G.~G.}\ \bibnamefont {{Kuzmin}}},\ }\bibfield  {title} {\bibinfo {title} {{Sobstvennye dvizheniya galaktiko-ehkvatorial'nyh A i K zvezd perpendikulyarno galakticheskoj ploskoti i dinamicheskaya plotnost' Galaktiki. Proper movements of the galactic-equatorial A and K stars of the perpendicularly galactic plane and dymanic density of the Galaxy.}},\ }\href@noop {} {\bibfield  {journal} {\bibinfo  {journal} {Publications of the Tartu Astrofizica Observatory}\ }\textbf {\bibinfo {volume} {32}},\ \bibinfo {pages} {5} (\bibinfo {year} {1952})}\BibitemShut {NoStop}%
\bibitem [{\citenamefont {{Mestel}}(1963)}]{1963mestel}%
  \BibitemOpen
  \bibfield  {author} {\bibinfo {author} {\bibfnamefont {L.}~\bibnamefont {{Mestel}}},\ }\bibfield  {title} {\bibinfo {title} {{On the galactic law of rotation}},\ }\href {https://doi.org/10.1093/mnras/126.6.553} {\bibfield  {journal} {\bibinfo  {journal} {\mnras}\ }\textbf {\bibinfo {volume} {126}},\ \bibinfo {pages} {553} (\bibinfo {year} {1963})}\BibitemShut {NoStop}%
\bibitem [{\citenamefont {{Kuzmin}}\ and\ \citenamefont {{Malasidze}}(1987)}]{1987Kuzmin}%
  \BibitemOpen
  \bibfield  {author} {\bibinfo {author} {\bibfnamefont {G.~G.}\ \bibnamefont {{Kuzmin}}}\ and\ \bibinfo {author} {\bibfnamefont {G.~A.}\ \bibnamefont {{Malasidze}}},\ }\bibfield  {title} {\bibinfo {title} {{On some mass distribution models in the theory of third quadratic integral of motion.}},\ }\href@noop {} {\bibfield  {journal} {\bibinfo  {journal} {Publications of the Tartu Astrofizica Observatory}\ }\textbf {\bibinfo {volume} {52}},\ \bibinfo {pages} {48} (\bibinfo {year} {1987})}\BibitemShut {NoStop}%
\bibitem [{\citenamefont {{Goldreich}}\ and\ \citenamefont {{Tremaine}}(1980)}]{GoldreichTremain:1980}%
  \BibitemOpen
  \bibfield  {author} {\bibinfo {author} {\bibfnamefont {P.}~\bibnamefont {{Goldreich}}}\ and\ \bibinfo {author} {\bibfnamefont {S.}~\bibnamefont {{Tremaine}}},\ }\bibfield  {title} {\bibinfo {title} {{Disk-satellite interactions.}},\ }\href {https://doi.org/10.1086/158356} {\bibfield  {journal} {\bibinfo  {journal} {\apj}\ }\textbf {\bibinfo {volume} {241}},\ \bibinfo {pages} {425} (\bibinfo {year} {1980})}\BibitemShut {NoStop}%
\bibitem [{\citenamefont {{Chandrasekhar}}(1943)}]{chandrasekhar1943}%
  \BibitemOpen
  \bibfield  {author} {\bibinfo {author} {\bibfnamefont {S.}~\bibnamefont {{Chandrasekhar}}},\ }\bibfield  {title} {\bibinfo {title} {{Dynamical Friction. I. General Considerations: the Coefficient of Dynamical Friction.}},\ }\href {https://doi.org/10.1086/144517} {\bibfield  {journal} {\bibinfo  {journal} {\apj}\ }\textbf {\bibinfo {volume} {97}},\ \bibinfo {pages} {255} (\bibinfo {year} {1943})}\BibitemShut {NoStop}%
\bibitem [{\citenamefont {Kuntz}\ and\ \citenamefont {Leyde}(2023)}]{Kuntz:2022juv}%
  \BibitemOpen
  \bibfield  {author} {\bibinfo {author} {\bibfnamefont {A.}~\bibnamefont {Kuntz}}\ and\ \bibinfo {author} {\bibfnamefont {K.}~\bibnamefont {Leyde}},\ }\bibfield  {title} {\bibinfo {title} {{Transverse Doppler effect and parameter estimation of LISA three-body systems}},\ }\href {https://doi.org/10.1103/PhysRevD.108.024002} {\bibfield  {journal} {\bibinfo  {journal} {Phys. Rev. D}\ }\textbf {\bibinfo {volume} {108}},\ \bibinfo {pages} {024002} (\bibinfo {year} {2023})},\ \Eprint {https://arxiv.org/abs/2212.09753} {arXiv:2212.09753 [gr-qc]} \BibitemShut {NoStop}%
\bibitem [{\citenamefont {{Schwarzschild}}(1916)}]{1916SPAW189S}%
  \BibitemOpen
  \bibfield  {author} {\bibinfo {author} {\bibfnamefont {K.}~\bibnamefont {{Schwarzschild}}},\ }\bibfield  {title} {\bibinfo {title} {{{\"U}ber das Gravitationsfeld eines Massenpunktes nach der Einsteinschen Theorie}},\ }\href@noop {} {\bibfield  {journal} {\bibinfo  {journal} {Sitzungsberichte der K\&ouml;niglich Preussischen Akademie der Wissenschaften}\ ,\ \bibinfo {pages} {189}} (\bibinfo {year} {1916})}\BibitemShut {NoStop}%
\bibitem [{\citenamefont {{Yunes}}\ \emph {et~al.}(2009)\citenamefont {{Yunes}}, \citenamefont {{Arun}}, \citenamefont {{Berti}},\ and\ \citenamefont {{Will}}}]{2009yunes}%
  \BibitemOpen
  \bibfield  {author} {\bibinfo {author} {\bibfnamefont {N.}~\bibnamefont {{Yunes}}}, \bibinfo {author} {\bibfnamefont {K.~G.}\ \bibnamefont {{Arun}}}, \bibinfo {author} {\bibfnamefont {E.}~\bibnamefont {{Berti}}},\ and\ \bibinfo {author} {\bibfnamefont {C.~M.}\ \bibnamefont {{Will}}},\ }\bibfield  {title} {\bibinfo {title} {{Post-circular expansion of eccentric binary inspirals: Fourier-domain waveforms in the stationary phase approximation}},\ }\href {https://doi.org/10.1103/PhysRevD.80.084001} {\bibfield  {journal} {\bibinfo  {journal} {\prd}\ }\textbf {\bibinfo {volume} {80}},\ \bibinfo {eid} {084001} (\bibinfo {year} {2009})},\ \Eprint {https://arxiv.org/abs/0906.0313} {arXiv:0906.0313 [gr-qc]} \BibitemShut {NoStop}%
\bibitem [{\citenamefont {Loutrel}\ \emph {et~al.}(2019)\citenamefont {Loutrel}, \citenamefont {Liebersbach}, \citenamefont {Yunes},\ and\ \citenamefont {Cornish}}]{Loutrel:2018ydu}%
  \BibitemOpen
  \bibfield  {author} {\bibinfo {author} {\bibfnamefont {N.}~\bibnamefont {Loutrel}}, \bibinfo {author} {\bibfnamefont {S.}~\bibnamefont {Liebersbach}}, \bibinfo {author} {\bibfnamefont {N.}~\bibnamefont {Yunes}},\ and\ \bibinfo {author} {\bibfnamefont {N.}~\bibnamefont {Cornish}},\ }\bibfield  {title} {\bibinfo {title} {{The eccentric behavior of inspiralling compact binaries}},\ }\href {https://doi.org/10.1088/1361-6382/aaf2a9} {\bibfield  {journal} {\bibinfo  {journal} {Class. Quant. Grav.}\ }\textbf {\bibinfo {volume} {36}},\ \bibinfo {pages} {025004} (\bibinfo {year} {2019})},\ \Eprint {https://arxiv.org/abs/1810.03521} {arXiv:1810.03521 [gr-qc]} \BibitemShut {NoStop}%
\bibitem [{\citenamefont {Loutrel}(2023)}]{Loutrel:2023rsl}%
  \BibitemOpen
  \bibfield  {author} {\bibinfo {author} {\bibfnamefont {N.}~\bibnamefont {Loutrel}},\ }\bibfield  {title} {\bibinfo {title} {{Eccentric catastrophes \& what to do with them}},\ }\href {https://doi.org/10.1088/1361-6382/acf9d9} {\bibfield  {journal} {\bibinfo  {journal} {Class. Quant. Grav.}\ }\textbf {\bibinfo {volume} {40}},\ \bibinfo {pages} {215004} (\bibinfo {year} {2023})},\ \Eprint {https://arxiv.org/abs/2304.00836} {arXiv:2304.00836 [gr-qc]} \BibitemShut {NoStop}%
\bibitem [{\citenamefont {Klein}\ \emph {et~al.}(2013)\citenamefont {Klein}, \citenamefont {Cornish},\ and\ \citenamefont {Yunes}}]{Klein:2013qda}%
  \BibitemOpen
  \bibfield  {author} {\bibinfo {author} {\bibfnamefont {A.}~\bibnamefont {Klein}}, \bibinfo {author} {\bibfnamefont {N.}~\bibnamefont {Cornish}},\ and\ \bibinfo {author} {\bibfnamefont {N.}~\bibnamefont {Yunes}},\ }\bibfield  {title} {\bibinfo {title} {{Gravitational waveforms for precessing, quasicircular binaries via multiple scale analysis and uniform asymptotics: The near spin alignment case}},\ }\href {https://doi.org/10.1103/PhysRevD.88.124015} {\bibfield  {journal} {\bibinfo  {journal} {Phys. Rev. D}\ }\textbf {\bibinfo {volume} {88}},\ \bibinfo {pages} {124015} (\bibinfo {year} {2013})},\ \Eprint {https://arxiv.org/abs/1305.1932} {arXiv:1305.1932 [gr-qc]} \BibitemShut {NoStop}%
\bibitem [{\citenamefont {Klein}\ \emph {et~al.}(2014)\citenamefont {Klein}, \citenamefont {Cornish},\ and\ \citenamefont {Yunes}}]{Klein:2014bua}%
  \BibitemOpen
  \bibfield  {author} {\bibinfo {author} {\bibfnamefont {A.}~\bibnamefont {Klein}}, \bibinfo {author} {\bibfnamefont {N.}~\bibnamefont {Cornish}},\ and\ \bibinfo {author} {\bibfnamefont {N.}~\bibnamefont {Yunes}},\ }\bibfield  {title} {\bibinfo {title} {{Fast Frequency-domain Waveforms for Spin-Precessing Binary Inspirals}},\ }\href {https://doi.org/10.1103/PhysRevD.90.124029} {\bibfield  {journal} {\bibinfo  {journal} {Phys. Rev. D}\ }\textbf {\bibinfo {volume} {90}},\ \bibinfo {pages} {124029} (\bibinfo {year} {2014})},\ \Eprint {https://arxiv.org/abs/1408.5158} {arXiv:1408.5158 [gr-qc]} \BibitemShut {NoStop}%
\bibitem [{\citenamefont {Berry}\ and\ \citenamefont {Upstill}(1980)}]{BERRY1980257}%
  \BibitemOpen
  \bibfield  {author} {\bibinfo {author} {\bibfnamefont {M.}~\bibnamefont {Berry}}\ and\ \bibinfo {author} {\bibfnamefont {C.}~\bibnamefont {Upstill}},\ }\bibfield  {title} {\bibinfo {title} {Iv catastrophe optics: Morphologies of caustics and their diffraction patterns}\ }(\bibinfo  {publisher} {Elsevier},\ \bibinfo {year} {1980})\ pp.\ \bibinfo {pages} {257--346}\BibitemShut {NoStop}%
\bibitem [{\citenamefont {{Peters}}(1964)}]{peters1964}%
  \BibitemOpen
  \bibfield  {author} {\bibinfo {author} {\bibfnamefont {P.~C.}\ \bibnamefont {{Peters}}},\ }\bibfield  {title} {\bibinfo {title} {{Gravitational Radiation and the Motion of Two Point Masses}},\ }\href {https://doi.org/10.1103/PhysRev.136.B1224} {\bibfield  {journal} {\bibinfo  {journal} {Physical Review}\ }\textbf {\bibinfo {volume} {136}},\ \bibinfo {pages} {1224} (\bibinfo {year} {1964})}\BibitemShut {NoStop}%
\bibitem [{\citenamefont {{von Zeipel}}(1910)}]{zeipel1910}%
  \BibitemOpen
  \bibfield  {author} {\bibinfo {author} {\bibfnamefont {H.}~\bibnamefont {{von Zeipel}}},\ }\bibfield  {title} {\bibinfo {title} {{Sur l'application des s{\'e}ries de M. Lindstedt {\`a} l'{\'e}tude du mouvement des com{\`e}tes p{\'e}riodiques}},\ }\href {https://doi.org/10.1002/asna.19091832202} {\bibfield  {journal} {\bibinfo  {journal} {Astronomische Nachrichten}\ }\textbf {\bibinfo {volume} {183}},\ \bibinfo {pages} {345} (\bibinfo {year} {1910})}\BibitemShut {NoStop}%
\bibitem [{\citenamefont {{Kozai}}(1962)}]{koz62}%
  \BibitemOpen
  \bibfield  {author} {\bibinfo {author} {\bibfnamefont {Y.}~\bibnamefont {{Kozai}}},\ }\bibfield  {title} {\bibinfo {title} {{Secular perturbations of asteroids with high inclination and eccentricity}},\ }\href {https://doi.org/10.1086/108790} {\bibfield  {journal} {\bibinfo  {journal} {\aj}\ }\textbf {\bibinfo {volume} {67}},\ \bibinfo {pages} {591} (\bibinfo {year} {1962})}\BibitemShut {NoStop}%
\bibitem [{\citenamefont {{Lidov}}(1962)}]{lid62}%
  \BibitemOpen
  \bibfield  {author} {\bibinfo {author} {\bibfnamefont {M.~L.}\ \bibnamefont {{Lidov}}},\ }\bibfield  {title} {\bibinfo {title} {{The evolution of orbits of artificial satellites of planets under the action of gravitational perturbations of external bodies}},\ }\href {https://doi.org/10.1016/0032-0633(62)90129-0} {\bibfield  {journal} {\bibinfo  {journal} {\planss}\ }\textbf {\bibinfo {volume} {9}},\ \bibinfo {pages} {719} (\bibinfo {year} {1962})}\BibitemShut {NoStop}%
\bibitem [{\citenamefont {{Naoz}}\ \emph {et~al.}(2013)\citenamefont {{Naoz}}, \citenamefont {{Farr}}, \citenamefont {{Lithwick}}, \citenamefont {{Rasio}},\ and\ \citenamefont {{Teyssandier}}}]{2013MNRAS.431.2155N}%
  \BibitemOpen
  \bibfield  {author} {\bibinfo {author} {\bibfnamefont {S.}~\bibnamefont {{Naoz}}}, \bibinfo {author} {\bibfnamefont {W.~M.}\ \bibnamefont {{Farr}}}, \bibinfo {author} {\bibfnamefont {Y.}~\bibnamefont {{Lithwick}}}, \bibinfo {author} {\bibfnamefont {F.~A.}\ \bibnamefont {{Rasio}}},\ and\ \bibinfo {author} {\bibfnamefont {J.}~\bibnamefont {{Teyssandier}}},\ }\bibfield  {title} {\bibinfo {title} {{Secular dynamics in hierarchical three-body systems}},\ }\href {https://doi.org/10.1093/mnras/stt302} {\bibfield  {journal} {\bibinfo  {journal} {\mnras}\ }\textbf {\bibinfo {volume} {431}},\ \bibinfo {pages} {2155} (\bibinfo {year} {2013})},\ \Eprint {https://arxiv.org/abs/1107.2414} {arXiv:1107.2414 [astro-ph.EP]} \BibitemShut {NoStop}%
\bibitem [{\citenamefont {Barack}\ and\ \citenamefont {Cutler}(2004)}]{Barack:2003fp}%
  \BibitemOpen
  \bibfield  {author} {\bibinfo {author} {\bibfnamefont {L.}~\bibnamefont {Barack}}\ and\ \bibinfo {author} {\bibfnamefont {C.}~\bibnamefont {Cutler}},\ }\bibfield  {title} {\bibinfo {title} {{LISA capture sources: Approximate waveforms, signal-to-noise ratios, and parameter estimation accuracy}},\ }\href {https://doi.org/10.1103/PhysRevD.69.082005} {\bibfield  {journal} {\bibinfo  {journal} {Phys. Rev. D}\ }\textbf {\bibinfo {volume} {69}},\ \bibinfo {pages} {082005} (\bibinfo {year} {2004})},\ \Eprint {https://arxiv.org/abs/gr-qc/0310125} {arXiv:gr-qc/0310125} \BibitemShut {NoStop}%
\bibitem [{\citenamefont {Toubiana}\ and\ \citenamefont {Gair}(2024)}]{Toubiana:2024car}%
  \BibitemOpen
  \bibfield  {author} {\bibinfo {author} {\bibfnamefont {A.}~\bibnamefont {Toubiana}}\ and\ \bibinfo {author} {\bibfnamefont {J.~R.}\ \bibnamefont {Gair}},\ }\bibfield  {title} {\bibinfo {title} {{Indistinguishability criterion and estimating the presence of biases}},\ }\href@noop {} {\bibfield  {journal} {\bibinfo  {journal} {arXiv e-prints}\ } (\bibinfo {year} {2024})},\ \Eprint {https://arxiv.org/abs/2401.06845} {arXiv:2401.06845 [gr-qc]} \BibitemShut {NoStop}%
\bibitem [{\citenamefont {{O'Leary}}\ \emph {et~al.}(2009)\citenamefont {{O'Leary}}, \citenamefont {{Kocsis}},\ and\ \citenamefont {{Loeb}}}]{oleary2009}%
  \BibitemOpen
  \bibfield  {author} {\bibinfo {author} {\bibfnamefont {R.~M.}\ \bibnamefont {{O'Leary}}}, \bibinfo {author} {\bibfnamefont {B.}~\bibnamefont {{Kocsis}}},\ and\ \bibinfo {author} {\bibfnamefont {A.}~\bibnamefont {{Loeb}}},\ }\bibfield  {title} {\bibinfo {title} {{Gravitational waves from scattering of stellar-mass black holes in galactic nuclei}},\ }\href {https://doi.org/10.1111/j.1365-2966.2009.14653.x} {\bibfield  {journal} {\bibinfo  {journal} {\mnras}\ }\textbf {\bibinfo {volume} {395}},\ \bibinfo {pages} {2127} (\bibinfo {year} {2009})},\ \Eprint {https://arxiv.org/abs/0807.2638} {arXiv:0807.2638 [astro-ph]} \BibitemShut {NoStop}%
\bibitem [{\citenamefont {Zwick}\ \emph {et~al.}(2020)\citenamefont {Zwick}, \citenamefont {Capelo}, \citenamefont {Bortolas}, \citenamefont {Mayer},\ and\ \citenamefont {Amaro-Seoane}}]{Zwick:2019yjl}%
  \BibitemOpen
  \bibfield  {author} {\bibinfo {author} {\bibfnamefont {L.}~\bibnamefont {Zwick}}, \bibinfo {author} {\bibfnamefont {P.~R.}\ \bibnamefont {Capelo}}, \bibinfo {author} {\bibfnamefont {E.}~\bibnamefont {Bortolas}}, \bibinfo {author} {\bibfnamefont {L.}~\bibnamefont {Mayer}},\ and\ \bibinfo {author} {\bibfnamefont {P.}~\bibnamefont {Amaro-Seoane}},\ }\bibfield  {title} {\bibinfo {title} {{Improved gravitational radiation time-scales: significance for LISA and LIGO-Virgo sources}},\ }\href {https://doi.org/10.1093/mnras/staa1314} {\bibfield  {journal} {\bibinfo  {journal} {Mon. Not. Roy. Astron. Soc.}\ }\textbf {\bibinfo {volume} {495}},\ \bibinfo {pages} {2321} (\bibinfo {year} {2020})},\ \Eprint {https://arxiv.org/abs/1911.06024} {arXiv:1911.06024 [astro-ph.GA]} \BibitemShut {NoStop}%
\bibitem [{\citenamefont {Mikoczi}\ \emph {et~al.}(2012)\citenamefont {Mikoczi}, \citenamefont {Kocsis}, \citenamefont {Forgacs},\ and\ \citenamefont {Vasuth}}]{Mikoczi:2012qy}%
  \BibitemOpen
  \bibfield  {author} {\bibinfo {author} {\bibfnamefont {B.}~\bibnamefont {Mikoczi}}, \bibinfo {author} {\bibfnamefont {B.}~\bibnamefont {Kocsis}}, \bibinfo {author} {\bibfnamefont {P.}~\bibnamefont {Forgacs}},\ and\ \bibinfo {author} {\bibfnamefont {M.}~\bibnamefont {Vasuth}},\ }\bibfield  {title} {\bibinfo {title} {{Parameter estimation for inspiraling eccentric compact binaries including pericenter precession}},\ }\href {https://doi.org/10.1103/PhysRevD.86.104027} {\bibfield  {journal} {\bibinfo  {journal} {Phys. Rev. D}\ }\textbf {\bibinfo {volume} {86}},\ \bibinfo {pages} {104027} (\bibinfo {year} {2012})},\ \Eprint {https://arxiv.org/abs/1206.5786} {arXiv:1206.5786 [gr-qc]} \BibitemShut {NoStop}%
\end{thebibliography}%


\end{document}